\documentclass[useAMS,usenatbib]{mnras}
\usepackage{times}
\usepackage{./aas_macros}
\usepackage{amssymb}
\usepackage{amsmath}
\usepackage{textcomp}
\usepackage{multirow}
\usepackage[dvipdfmx]{graphicx} 
\graphicspath{{images/}}

\begin{document}%
\title[The infancy of core-collapse SN remnants]{The infancy of 
core-collapse supernova remnants}
\author[Michael Gabler, Annop Wongwathanarat, and Hans-Thomas Janka]
{Michael Gabler$^{1}$, 
Annop Wongwathanarat$^2$
and
Hans-Thomas Janka$^2$, 
\\
$^1$Departamento de Astronom\'{\i}a y Astrof\'{\i}sica,
  Universidad de Valencia, 46100 Burjassot (Valencia), Spain\\
$^2$Max-Planck-Institut f\"ur Astrophysik,
  Karl-Schwarzschild-Str.~1, 85748 Garching, Germany 
}
\date{\today}
\maketitle
\begin{abstract}
We present 3D hydrodynamic simulations of neutrino-driven supernovae (SNe) with
the \textsc{Prometheus-HotB} code, evolving the asymmetrically expanding ejecta
from shock breakout until they reach the homologous expansion phase after 
roughly one year. Our calculations continue the simulations for two red 
supergiant (RSG) 
and two blue supergiant (BSG) progenitors by Wongwathanarat et al., who 
investigated the growth of explosion asymmetries produced by
hydrodynamic instabilities during the first second of the explosion and their 
later fragmentation by Rayleigh-Taylor instabilities.
We focus on the late time acceleration and inflation of the ejecta caused by the 
heating due to the radioactive decay of $^{56}$Ni to $^{56}$Fe and by a new 
outward-moving shock, which forms when the reverse shock from the He/H-shell 
interface compresses the central part of the ejecta. The mean 
velocities of the iron-rich ejecta increase between 100\,km/s and 350\,km/s 
($\sim$8--30\%), and the fastest one percent of the iron accelerates by up to 
$\sim$1000\,km/s ($\sim$20--25\%). This `Ni-bubble effect', known from 1D 
models, accelerates the bulk of the nickel in our 3D models and 
causes an inflation of the initially overdense Ni-rich clumps, which leads 
to underdense, extended fingers, enveloped by overdense skins of compressed 
surrounding matter. 
We also provide volume and surface filling factors as well as a spherical 
harmonics analysis to characterize the spectrum of Ni-clump sizes 
quantitatively. Three of our four models give volume filling factors larger 
than $0.3$, consistent with what is suggested for SN~1987A by observations.
\end{abstract}
%
\section{Introduction}

Core-collapse supernova (CCSN) explosions are the most violent phenomena that 
happen at the end of the lifetime of massive stars. They shed light onto 
extreme physical conditions and processes inside the exploding star, 
which otherwise are inaccessible by observations in the electromagnetic 
spectrum. Despite the 
significant progress of our theoretical understanding of these events due to 
the feasibility of three-dimensional (3D) simulations, answering the question 
whether the explosion is driven by the delayed neutrino-heating mechanism still
requires further studies and, in particular, observational assessment in direct 
comparison to 3D model 
predictions. Therefore, it is of great importance to determine possibilities of 
testing the consequences of the explosion mechanism with detailed observations. 
Promising objects for this kind of observations are young SN remnants 
(SNR), which still carry the imprints of explosion asymmetries reflected by the
3D spatial distributions of different chemical elements synthesized 
during the SN outburst.

In particular, observations of SN~1987A in the Large Magellanic Cloud and 
Cassiopeia A (Cas A), a $\sim340$ year old galactic SNR, offer possibilities to 
indirectly probe the CCSN mechanism. The explosions producing these two 
fascinating objects must have been of genuine 3D nature, as already expected by 
extensive theoretical studies, and as suggested by abundant observational 
evidence
gathered over the past decades. For instance, 
\cite{Larsson2016} inferred the 3D distribution of the ejecta of SN~1987A by 
using Doppler shift information of the velocities of different elements 
obtained from spectroscopic observations. The 
analysis showed global large-scale asymmetries of the SN ejecta extending along 
the northeast and the southwest directions. \cite{DeLaney2010} reconstructed 
the 3D ejecta structure of Cas A using observational data obtained in 
infrared by the {\it Spitzer Space Telescope} \citep{Isensee2010}, in X-ray by 
the {\it Chandra} satellite \citep{Lazendic2006}, and in optical 
\citep{Fesen1996,Fesen2001}. The reconstruction revealed that the remnant can 
be characterized by a spherical component illuminated by the reverse shock, 
a flattened ejecta structure seen as a tilted thick disk, two opposing 
wide-angle, jet-like ejecta pistons, and numerous optical fast-moving knots 
lying in the thick disk plane. Spectroscopic observations of SN light echoes 
from SN~1987A \citep{Sinnott2013} and Cas A \citep{Rest2011} provide
evidences that the observed large-scale ejecta asymmetries in these objects 
originate from very early phases of the explosions. In addition, direct 
observations of spectra and the lightcurve of SN~1987A show 
the presence of large scale anisotropies \citep{Utrobin2015}. 
\cite{Grefenstette2014} and 
\cite{Grefenstette2017} directly imaged the spatial distribution of radioactive 
$^{44}$Ti in Cas A. They found strong hints that there must have been 
significant asymmetries during the explosion. \cite{Milisavljevic2013} and 
\cite{Milisavljevic2015} showed that the shocked ejecta strongly emitting in 
optical light are organized in ring-like structures that connect to the borders 
of seemingly empty bubbles or cavities in the interior of unshocked 
sulfur-rich ejecta. A 
comparison of recent Very Large Telescope/SINFONI observations of HII emission 
regions \citep{Larsson2019b} and {\it Atacama Large Millimeter/Submillimeter 
Array (ALMA)} observations of CO and SiO molecules and dust \citep{Abellan2017, 
Cigan2019} shows that these molecules reside in different regions of the young 
supernova remnant. 

On the theoretical side, there have been first successful 
attempts to model the observed structures of the ejecta in Cas A 
\citep{Orlando2016} and of SN~1987A \citep{Ono2020, Orlando2020}. However, the 
former models relied on a particular choice of the initial conditions at the 
shock breakout and the latter on parameterized initial explosion asphericities, 
both of which are not compatible with or would have to be checked against 
self-consistent calculations \citep{Wongwathanarat2017}.

In the context of neutrino-driven explosions, which we consider here, 
large-scale asymmetries originate from the nonlinear growth of hydrodynamic 
instabilities, as for example the convective instability \citep{Bethe1990, 
Herant1992, Herant1994, Burrows1995, Janka1995, Janka1996} and the standing 
accretion shock instability \citep[SASI;][]{Foglizzo2002, Blondin2003, 
Blondin2006, Ohnishi2006, Foglizzo2007, Scheck2008, Fernandez2010}, during the 
revival of the stalled SN shock wave. These asymmetries particularly manifest 
themselves in the distribution of the heavy elements, freshly synthesized 
during 
the explosion. After the revival of the shock wave, which takes less than 
$1$\,s, 
the initial asymmetries get shaped further by the growth of secondary 
Rayleigh-Taylor instabilities (RTIs) that develop due to the propagation of the 
SN shock through the non-monotonically varying density gradients of the mantle 
and the envelope 
of the exploding progenitor star 
\citep{Chevalier1976, Chevalier1978}. Inspired by the SN~1987A event a large 
number of multi-D 
simulations studying the growth of RTIs at 
different composition shell interfaces (e.g. C+O/He and He/H interfaces) inside 
the progenitor star have been performed
\citep[e.g.,][]{Arnett1989, Mueller1991, Fryxell1991, 
Hachisu1990, Hachisu1992, Hachisu1994, Iwamoto1997, Nagataki1998, 
Hungerford2003, Hungerford2005, Joggerst2009, Joggerst2010, 
Joggerst2010b, Couch2009, Couch2011, Ono2013, Ellinger2012, Ellinger2013, 
Mao2015}. However, these studies did not consistently model the 
development of explosion asymmetries introduced by convection and SASI during 
the first second of the explosion. To circumvent the underlying problem of a 
still uncertain CCSN explosion mechanism, these previous simulations either 
assumed spherical explosions or relied on asymmetric explosions with global, 
low-mode asymmetries imposed artificially.
More recently, simulations of supernova explosions have been achieved with 
fully self-consistent calculations, where the shock revival was computed 
with detailed neutrino transport \citep{Takiwaki2014, Melson2015a, Melson2015b, 
Lentz2015, Roberts2016, Summa2016, Mueller2017, Mueller2018, Vartanyan2018, 
Ott2018, OConnor2018, Melson2020, Vartanyan2019, Burrows2019}. Typically, these 
simulations stop after the shock wave is revived and starts to expand through 
the progenitor.

Long-time CCSN simulations which consider the explosion engine in multi-D and 
follow the time evolution of explosion asymmetries from the initiation of 
neutrino-driven explosions until late phases were 
carried out first in 2D by \cite{Kifonidis2003,Kifonidis2006} and 
\cite{Gawryszczak2010}, and more recently in 3D by \cite{Hammer2010}, 
\cite{Wongwathanarat2013, Wongwathanarat2015, Wongwathanarat2017}, and 
\cite{Stockinger2020}. In these calculations, the emission of neutrinos by the 
nascent proto-neutron star (PNS) is parameterized and the interactions of these 
neutrinos with the post-shock matter are calculated by solving 
neutrino-transport equations with a grey approximation in a ray-by-ray manner 
\citep{Scheck2006}. The neutrino-matter interactions play a crucial role in 
reviving the stalled SN shock and in depositing the energy of the SN blast. 
With this approach it is not possible to determine all the properties of 
the involved neutrinos, whereas the growth of the hydrodynamic instabilities 
in the post-shock layer can be studied in most aspects realistically. These 
long-time CCSN simulations typically follow the propagation of the SN 
ejecta until hours or a day after the onset of the explosion. This is roughly 
the time at which the SN shock wave breaks out from the surface of the 
progenitor star. \cite{Mueller2018} studied the explosion of an 
ultrastripped supernova and evolved their model until shock breakout.

After the SN shock breakout additional energy input from the radioactive decay 
of $^{56}$Ni continues to drive inflation of $^{56}$Ni-rich structures and 
facilitates mixing between ejecta components. This late time expansion can 
still lead to substantial modifications of the overall SN ejecta morphology on 
timescales of weeks or months \citep{Benz1994}. In 2D calculations and in 
calculations in 
a $30^\circ$ wedge of a 3D domain, \cite{Herant1991,Herant1992} found that the 
energy input by radioactive $\beta$ decays can boost the ejected velocity of 
$^{56}$Ni-rich clumps from $900\,$km/s to $1300\,$km/s and from $1400\,$km/s to 
$1900\,$km/s, corresponding to about a $30\%$ increase. A similar magnitude of 
the velocity increase was found by \cite{Basko1994}, who studied the growth of 
RTI at the surface of an inflating $^{56}$Ni-rich bubble. With artificial 
initial setups, \cite{Blondin2001} studied how $^{56}$Ni-rich clumps are heated 
and inflated by the radioactive decay energy and how they interact with the 
surrounding SN ejecta and the reverse shock. They confirmed previous 
expectations that the density along the borders of the $^{56}$Ni-bubbles 
increases. The density contrast between these structures of overdense 
filaments, and the matter inside the $^{56}$Ni-rich bubbles increases, 
because the latter reduces its density due to an additional expansion. The 
corresponding study was motivated by an analysis of observational data of 
SN~1987A carried out by \cite{Li1993}, who provided an estimate of the filling 
factor of $^{56}$Ni clumps of $f\gtrsim0.3$ in the SN ejecta. In a 1D model 
considering either pure hydrodynamical or coupled radiation-hydrodynamical 
evolution, \cite{wang2005} found that during the inflation of a central 
spherical $^{56}$Ni-bubble a dense shell of up to $1\,$M$_\odot$ is swept 
up, resulting in a maximal density enhancement of a factor of $100$ with 
respect to the ambient medium density. Such a  `Ni bubble' effect was also 
observed in recent 1D SN models with $^{56}$Ni decay analyzed by 
\cite{Jerkstrand2018}

To follow the creation of the early-time SN ejecta asymmetries and their
continuous transformation by secondary instabilities and by inflation 
caused by $\beta$-decay energy input, it is indispensable to perform 3D 
computer simulations. To capture the initial asymmetries consistently, these 
simulations have to start before the onset of the explosion and continue until 
the 
ejecta evolve into its gaseous remnant state. Such multi-physics, multi-scale 
simulations are computationally challenging and expensive.
In this work, we continue the efforts by \cite{Wongwathanarat2015, 
Wongwathanarat2017} to model the long-time evolution of CCSNe beyond the SN 
shock breakout until the early SNR phase roughly 1 year after the 
shock formation. We employ the models calculated by \cite{Wongwathanarat2015} 
as our initial data. In order to investigate the effect of radioactive heating 
on the SN ejecta asymmetries on a long timescale, we extend previous work by 
implementing a simplified treatment of the energy input due to the $\beta$ 
decay of 
$^{56}$Ni and $^{44}$Ti. Our approach is different from the one typically 
employed in other calculations \citep[e.g.,][]{Herant1991, Herant1992}, 
where energy deposition by the radioactive decay of $^{56}$Ni is assumed to be 
local regardless of the optical depth of the $^{56}$Ni-rich ejecta. Results 
from our 3D hydrodynamic calculations we present here have already been used in 
comparisons to 3D distributions of CO and SiO molecular emission in SN~1987A 
obtained recently by ALMA \citep{Abellan2017, Cigan2019}, for more 
realistic estimates of the X-ray absorption and emission in young CCSN 
remnants like Cas A and SN~1987A \citep{Alp2018a, Alp2018b, Alp2019, 
Jerkstrand2020}, and in a geometrical analysis of the Fe distribution and 
neutron star kick in SN~1987A by \cite{Janka2017}.

This paper is organized as follows. In Section 2, we briefly 
describe the numerical methods employed in our code. In addition, we explain in 
detail our approach to model the radioactive $\beta$-decay energy deposition and 
provide a brief overview of the properties of the considered progenitor models. 
In Section\,\ref{sec_long-time}, we present results from our numerical models, 
beginning with the dynamics of a self-reflected reverse shock, the effect of 
$\beta$ decays on the global properties of the ejecta, a detailed view on 
the properties of the ejecta structures such as the velocity and density 
distributions, and finally the inflation of $^{56}$Ni-rich clumps and their 
properties. We conclude and discuss our findings 
in Section\,\ref{sec_conclusions}. In Appendix\,\ref{app_beta}, we provide more 
details about our treatment of the $\beta$ decay.

\section{Theoretical framework}
\subsection{Numerics} \label{sec_numerics}
For our simulations we use the 3D, explicit finite-volume hydrodynamics code  
{\sc Prometheus} \citep{Fryxell1991,Mueller1991,Mueller1991b}  
in its version {\sc Prometheus-HotB} 
\citep{Janka1996, Kifonidis2003, Kifonidis2006, Scheck2006, Arcones2007, 
Wongwathanarat2013, Wongwathanarat2015, Wongwathanarat2017, Gessner2018, 
Stockinger2020}, which includes neutrino physics, a general equation of state 
applicable above and below nuclear statistical equilibrium, and a treatment of 
nuclear burning via a small alpha network. The hydrodynamics equations are 
solved with the piecewise parabolic method 
\citep[PPM;][]{Colella1984} employing an exact Riemann solver for real gases 
\citep{Colella1985} and treating a multi-fluid system with the consistent 
multi-fluid advection (CMA) scheme by \cite{Plewa1999}. In our simulations, we 
consider a stellar fluid consisting of 19 nuclear species: protons, alpha nuclei 
from $^{4}$He to $^{56}$Ni, $^{56}$Co, $^{56}$Fe, $^{44}$Sc, $^{44}$Ca, and a 
neutronization tracer X which traces production of neutron rich 
nuclear species when the electron fraction $Y_e<0.49$.
The multi-dimensional Euler equations are solved in one-dimensional sweeps 
following the splitting technique of \cite{Strang1968}. 

Spatial discretization of the computational sphere is done using an axis-free 
overlapping `Yin-Yang' grid technique \citep{Kageyama2004} implemented into 
{\sc Prometheus-HotB} by \cite{Wongwathanarat2010}. The Yin-Yang overset grid 
avoids numerical artefacts which can arise near the polar axis of a spherical 
polar grid. In addition, it also alleviates time step constraints imposed by 
the Courant-Friedrich-Levy (CFL) condition, which in the case of a spherical 
polar grid is very restrictive due to small azimuthal grid cells in the polar 
regions. Thus, the use of the Yin-Yang grid allows the simulations to advance 
with larger time steps. 

As in \cite{Wongwathanarat2015}, Newtonian self-gravity is taken into 
account by solving the integral form of Poisson’s equation with a multipole 
expansion method as described in \cite{Mueller1995} and we omit the 
local relativistic corrections to the potential, which were included in 
\cite{Wongwathanarat2013}, who calculated the early time evolution of 
the explosion of our models. The gravitational potential of the 
central point mass, which accounts for the gravitating effects of the neutron 
star (sitting far interior to our inner grid boundary) and includes monopole 
general-relativistic corrections, is treated continuously to avoid numerical 
transients, and it is updated during the simulation for mass leaving the inner 
grid boundary and assumed to be accreted by the neutron star. 

At the late times considered here, the only 
remaining effect of neutrinos on the expanding ejecta is the waning 
neutrino-driven wind (i.e., a mass outflow from the nascent neutron star driven 
by neutrino-energy deposition), which is taken into account in the long-time 
simulations in a parametrized functional form that is prescribed as a boundary 
condition at the inner grid boundary. This is described in detail in
\cite{Wongwathanarat2015}, and already there the influence at the end of 
the simulation was negligible. For the details of the grey, ray-by-ray  
treatment of neutrinos applied in the early phases of the explosion (but not of 
relevance for the long-time simulations discussed in the present paper), we 
refer to \cite{Scheck2006} and \cite{Wongwathanarat2013, Wongwathanarat2015}.

At late phases, when the ejecta expand almost homologously, we move the grid 
radially as the SN ejecta expand. This moving mesh further relaxes the CFL 
condition imposed by grid cells with smallest radial extension, which are found 
at the smallest radii. The grid velocity is set to be linearly proportional to 
the radius, with the velocity of the outermost grid cell being set to 
$\sim120\%$ of the maximal fluid velocity. The grid velocity of the inner radial 
grid boundary is forced to be zero, i.e. it remains at a fixed radius at all 
times. The shock may still expand faster than the maximum grid velocity. To 
avoid that the shock leaves the numerical grid during the simulations, we 
remove the innermost cell in 
radial direction and add a new cell in the exterior whenever the shock gets 
closer than 10 grid cells to the outer boundary of the computational domain. 
The physical conditions of the new grid cell are determined by the assumed
stellar wind in the exterior. All other cell indices are shifted by minus one 
in radial direction, such that the previously second cell is now the first one.
Since $|v_r|\gg\{|v_\theta|,|v_\varphi|\}$ the grid movement is quasi-Lagrangian 
and this treatment thus minimizes the numerical diffusion associated with 
the expansion of the SN ejecta over many orders of magnitude of the initial 
radial scale.\footnote{\label{footnote_problem} One of our models 
(B15) had to be rerun at an advanced stage of the project because of a 
numerical problem that occurred with the equation of state in a few cells with 
very low densities behind the shock front. In order to save computer resources 
and to repeat the model calculation within a shorter period of time, we decided 
to increase the central volume that is cut out and to choose it larger than in 
the other models. We made sure (by comparison with the original run) that this 
volume still contained a negligible amount of mass and the larger cut radius had 
no noticeable influence on the simulation results.}

While, in general, we use an exact Riemann solver for ideal gases, 
we employ either the HLLE \citep{Einfeldt1988} or the AUSM+ solver 
\citep{Liou1996} inside grid cells where strong shocks are present in order to 
suppress 
numerical artefacts that can arise due to odd-even decoupling 
\citep{Quirk1994}. The more diffusive HLLE solver is used when the 
computational grid is expanding radially, while the AUSM+ solver is employed in 
the case of a static grid.

A previous version of the {\sc Prometheus-HotB} code has already been 
applied to compute the propagation of the shock and the ejecta during a 
neutrino-driven supernova explosion up to the shock breakout in three 
dimensions and to study the production of $^{44}$Ti and $^{56}$Ni in Cas A 
\citep{Wongwathanarat2013,Wongwathanarat2015,Wongwathanarat2017}. It 
was further used to study light curves of different progenitors and to compare 
them to SN~1987A \citep{Utrobin2015, Utrobin2017, Utrobin2019}.

\subsection{Radioactive  {\protect $\beta$} decay}\label{sec_beta}
As in \cite{Stockinger2020}, we use an extension of {\sc Prometheus-HotB} 
\citep{Wongwathanarat2015,Wongwathanarat2017} that includes the effects of 
$\beta$ decay, which cause additional heating of the $^{56}$Ni-rich ejecta. 
$^{56}$Ni has a half-life time 
of $\tau_{1/2}^\mathrm{Ni}=6.08\,$d to $^{56}$Co, which in turn decays to the 
stable $^{56}$Fe with $\tau_{1/2}^\mathrm{Co}=77.23\,$d:
\begin{eqnarray}
 ^{56} \mathrm{Ni} + e^- &\xrightarrow{6.08\,\mathrm{d}}& {^{56}\mathrm{Co}} + 
\gamma + \nu_e\,.\\
 \left. \begin{aligned}
^{56}\text{Co} + e^- \\
^{56}\text{Co}
 \end{aligned}\right\rbrace
&\xrightarrow{77.23\,\mathrm{d}}& \begin{cases}
                            ^{56}\text{Fe}+\gamma+\nu_e\,,\\
                            ^{56}\text{Fe}+e^++\gamma+\nu_e\,.
                           \end{cases}
\end{eqnarray} 
Considering the relative probabilities of the two decay channels of $^{56}$Co, 
the mean energies carried away by the $\gamma$ photons are 
$Q_\mathrm{Ni}=1.72\,$MeV and
$Q_{\text{Co,}\gamma}=3.61\,$MeV. In the case that $^{56}$Co 
decays via $\beta^+$ emission, the positron obtains an energy of about 
$Q_{\text{Co,}e^+}=0.125\,$MeV per decay on average \citep{Junde2011}.
The total energy emitted in photons and positrons of the $^{56}$Co decay is 
thus $Q_\text{Co}=3.735\,$MeV$=Q_{\text{Co,}\gamma}+Q_{\text{Co,}e^+}$.
If this energy per decay is deposited locally, the specific energy (per unit 
mass) increases in a time interval $\Delta t$ by 
\begin{equation}
 \Delta \varepsilon^\mathrm{release}= \sum_i 
\frac{Q_{i}  X_i}{m_i} \left[1 - 
\exp\left({-\frac{\Delta t \ln{2}}{ \tau_{1/2}^i}}\right)\right]\,,
\end{equation}
where $X_i$ and $m_i$ are the mass fraction and atomic mass of the 
respective element $i\in\{\text{Ni,Co}\}$. 

At early times, when the matter is still optically thick, all this energy is 
deposited locally close to where the radioactive decay proceeds. However, the 
longer the ejecta expand, the more transparent they become with respect to the 
$\gamma$ radiation. A self-consistent treatment of 
the non-local deposition and the escape of the $\gamma$-photons would require a 
detailed radiation transport coupled to the hydrodynamic calculation and is far 
beyond the scope of this paper. Thus, we approximate the energy deposition in 
the following way.

First, we find the maximal radial extent of $^{56}$Ni-rich ejecta in each 
angular direction given as the outermost radial point where the mass fraction 
of $^{56}$Ni and its decay products is greater than $10^{-3}$. We denote the 
radius of the 
corresponding grid cell $R^{56}_\mathrm{max}(\theta,\varphi)$ and the maximal 
index in the radial grid as $i^{56}_\mathrm{max}(\theta,\varphi)$. For each 
grid 
cell with radius $r<R^{56}_\mathrm{max}(\theta,\varphi)$, we calculate the 
optical depth up to $R^{56}_\mathrm{max}(\theta,\varphi)$ in the radial 
direction
\begin{equation}
\tau_\gamma^r=\int_{r}^{R^{56}_\mathrm{max}}\kappa_\gamma\rho ~dr, 
\label{eq_taur}
\end{equation} 
and the respective optical depths in the angular directions 
\begin{equation}
\tau_\gamma^{\{\theta,\varphi\}}=\int\kappa_\gamma\rho~dl.
\label{eq_tauthetaphi}
\end{equation} 
Here, $\kappa_\gamma=0.06 Y_e$\,cm$^2$/g is an effective, grey absorption 
coefficient describing the interaction of $\gamma$ rays with 
the cool supernova gas \citep{Swartz1995}, $\rho$ is the density, $dl$ 
the differential length along the photon path, 
and $Y_e$ the electron fraction per baryon. The minimum 
$\tau_\gamma^\mathrm{min}=\min{(\tau_\gamma^r,\tau_\gamma^\theta,
\tau_\gamma^\varphi)}$ is 
used to determine the amount of energy we deposit locally in each respective 
cell of our numerical grid. For $\tau_\gamma^\theta$ and 
$\tau_\gamma^\varphi$, we limit the integration to a maximum of three 
neighbouring cells in the angular 
directions and the photon path length $l=\int dl$, must not exceed the distance 
a photon can 
travel during one hydrodynamic time step $\Delta t$: $l\leq 
l_\mathrm{max}= c \Delta t$, where $c$ is the speed of light.  
These limits are motivated by the fact that we expect that the 
optical depth usually decreases faster in the radial direction and that at a 
given radius mainly the cells located close to the lateral boundaries of the 
$^{56}$Ni-rich RT fingers lose significant radioactively generated energy to 
the 
surrounding $^{56}$Ni-poor ejecta. Given $\tau_\gamma^\mathrm{min}$,
the specific energy (per unit mass) $\varepsilon$ is increased in a 
time interval $\Delta t$ by
\begin{equation}
\Delta \varepsilon^\mathrm{deposit}= 
\Delta 
\varepsilon^\mathrm{release}\left[1-\exp\left({-\tau_\gamma^\mathrm{min}}
\right)\right ] ~.
\label{eq_decay}
\end{equation}
The energy input from positrons 
produced by the $^{56}$Co decay is assumed to be local always.

The sum of escaping energy from all cells
\begin{equation}
\Delta\varepsilon^\mathrm{escape}=\sum_\mathrm{cells} \Delta 
\varepsilon^\mathrm{release}\exp\left({-\tau_\gamma^\mathrm{min}}\right)
~,
\end{equation}
is not deposited locally within the grid. However, this radiation can still
interact with the ejected matter further away from the $\beta$-decay sites. 
To take this non-local deposition into account, we deposit parts of 
$\Delta\varepsilon^\mathrm{escape}$ homogeneously within the ejecta. To this 
end, we define a mean optical depth:
\begin{equation}
\tau_\gamma^\mathrm{mean}(r)=\int^{r}_{R_\mathrm{mean}^\mathrm{56} } 
\kappa_\gamma\bar\rho ~dr^\prime\,,
\label{eq_taumean}
\end{equation} 
where, $\bar\rho$ is the angular average of the density, 
$R_\mathrm{mean}^{56}$ the radius at $i_\mathrm{mean}$, and the latter is the 
mean radial grid index of the outermost cells where $X_\mathrm{Ni}>10^{-3}$:
\begin{equation}
 i_\mathrm{mean} = \frac{1}{N} \sum_{j,k} i^{56}_\mathrm{max}\,.
\end{equation}
Here, $N$ is the total number of cells in the angular directions.
We use this approach for $R_\mathrm{mean}^{56}$ rather than taking the mean 
radius, because we want to reduce the influence of very extended RT fingers, 
which may extend to very large radii. 
Now we can define the radius at which $\tau_\gamma^\mathrm{mean}=1.0$. 
Since $1-\exp(-1)$ of all photons interact until reaching this optical depth, we
deposit $2/3$ ($\sim 1-\exp(-1)$) of $\Delta\varepsilon^\mathrm{escape}$ 
isotropically within the 
sphere determined by this radius. For simplicity, and because the 
influence in the huge affected volume is expected to be very small, the
remaining one third of the escaping energy is deposited homogeneously in the 
spherical shell limited by the radii where $\tau_\gamma^\mathrm{mean}(r)=1$ and 
$\tau_\gamma^\mathrm{mean}(r)=2$.
If the optical depth to the outer grid boundary is less than 
$\tau_\gamma^\mathrm{mean}(r)<2$ or $\tau_\gamma^\mathrm{mean}(r)<1$ the 
corresponding energy of $ 1/3~ \Delta\varepsilon^\mathrm{escape}$ or 
$\Delta\varepsilon^\mathrm{escape}$ is allowed to escape completely from 
the ejecta. 

In addition to the decay chain of $^{56}$Ni, we also implemented the 
radioactive decay of $^{44}$Ti to $^{44}$Sc and then to $^{44}$Ca.
\begin{eqnarray}
 ^{44}\mathrm{Ti} &\xrightarrow{60.25\,\mathrm{y}}& {^{44}\mathrm{Sc}} + 
Q_\mathrm{Ti}\,,\\
 ^{44}\mathrm{Sc} &\xrightarrow{3.972\,\mathrm{h}}& {^{44}\mathrm{Ca}} + 
Q_\mathrm{Sc}\,,
\end{eqnarray}
with $Q_\mathrm{Ti}= 0.143\,$MeV and $Q_\mathrm{Sc}= 2.73\,$MeV.
This decay happens at very late times, when the ejecta are expected to be 
effectively optically thin. In addition, there is much less $^{44}$Ti than 
$^{56}$Ni. Thus, 
we only expect a negligible influence on the overall dynamics of the $^{44}$Ti 
decay. We show some tests of our implementation in  
Appendix\,\ref{app_beta}.

\begin{table*}
\begin{tabular}{c c c c c c c c c c c c c c c}
 Model&\multicolumn{4}{c}{Progenitor}&Mapping
&Shock &\multicolumn{2}{c}{Wind}&$\beta$ decay & 
$M_\mathrm{Ni}^\mathrm{initial}$&Explosion\\
&Name in&Type&M$_\mathrm{ZAMS}$&Radius&Time&Breakout ($t_\mathrm{out}$)&Mass 
loss& Speed&&&Energy\\
&Literature&&[M$_\odot$]&[$10^6\,$km]&[$10^3\,$s]&[$10^3\,$s]&[ 
M$_\odot$yr$^{-1}$ ] & [km s$^{-1}$] &&[M$_\odot$]&[B]\\\hline
 W15&W15-2-cw&RSG&15&339&5.8&85&$10^{-5}$&10&standard&0.056&1.47\\
 L15&L15-1-cw&RSG&15&434&5.0&95&$10^{-5}$&10&standard&0.034&1.75\\
 N20&N20-4-cw&BSG&20&33.8&1.4&5.6&$4\times10^{-6}$&550&standard&0.044&1.65\\
 B15&B15-1-pw&BSG&15&39.0&3.2&7.3&$4\times10^{-6}$&550&standard&0.034&1.39\\
 B15$_0$&B15-1-pw&BSG&15&39.0&3.2&7.3&$4\times10^{-6}$&550&no&-&1.39\\
 B15$_\mathrm{X}$&B15-1-pw&BSG&15&39.0&3.2&7.3&$4\times10^{-6}$&550&with 
X&0.103&1.39\\
 \end{tabular}
\caption{Properties of the six models considered in this work (see text for 
details). The model names in the second column are those used in  
{\protect\cite{Wongwathanarat2015}}. We further provide the type of the 
progenitor, its Zero Age Main Sequence mass M$_\mathrm{ZAMS}$, progenitor 
radius, time of mapping when we continue the simulations of 
{\protect\cite{Wongwathanarat2015}}, the time $t_\mathrm{out}$ when 
the shock leaves the 
progenitor, the assumed mass loss rate of the 
progenitor model and the 
corresponding wind velocity. For the $\beta$ decay, we also provide the kind of 
treatment we apply: i) standard: all synthezised $^{56}$Ni decays, ii) no: 
nothing decays and iii) with X: enhanced $\beta$ decay, where we add the entire 
mass of the tracer nucleus X to the mass of $^{56}$Ni for the calculation of 
the $\beta$ decay. The corresponding $^{56}$Ni masses that are actually used as 
the basis for the $\beta$ decay are given in the following column. Finally, we 
provide the explosion energy in $1\mathrm{B}=10^{51}$ as given 
in table 2 of {\protect\cite{Wongwathanarat2015}}.}
\label{tab_models}
\end{table*}

\subsection{Stellar models}

We investigate four stellar progenitor models: two red supergiant (RSG) and two 
blue supergiant (BSG) stars. The two RSGs are the model s15s7b2 computed by 
\cite{Woosley1995}, and a 15\,M$_\odot$ star evolved by \cite{Limongi2000}. The 
two BSGs are a $20\,$M$_\odot$ progenitor model for SN~1987A from 
\cite{Shigeyama1990}, and a 15\,M$_\odot$ star by \cite{Woosley1988}. A 
detailed description of these progenitor models can be found in 
\cite{Wongwathanarat2015}, and a summary of their properties is given in 
Table\,\ref{tab_models}. The four models were computed from a time shortly 
($\sim$15\,ms) after core bounce through the onset of the explosions by 
\cite{Wongwathanarat2013}, and were followed until the SN shock breaks out from 
the surface of the respective progenitor star by \cite{Wongwathanarat2015}. 
In this work, we selected the more extensively studied model for each of the 
considered progenitors  computed by \cite{Wongwathanarat2015}: W15-2-cw, 
L15-1-cw, N20-4-cw, and 
B15-1-pw.
These initial models are mapped onto our computational domain at times between 
$\sim1000-6000\,$s after the onset of the the explosion depending on the 
respective model. The 
mapping time for each model is given in Table\,\ref{tab_models}. The models are 
then followed until approximately 1 year after the explosion began, taking into 
account the energy deposition by radioactive $\beta$ decay as described in 
Section\,\ref{sec_beta}. Since we calculate only one model for each of the 
progenitor stars we discard the suffixes from the model names in this work, and 
denote our models W15, L15, N20, and B15. 

To study in detail the influence of the energy input due to the $\beta$ decay 
on the SN ejecta morphology at late times, we calculate two additional variants 
of model B15. On the one hand, we carry out one simulation without the 
radioactive decay, which we denote B15$_0$. On the other hand, we compute
another model B15$_\mathrm{X}$, in which we assume that all of the 
tracer nucleus X radioactively decays as $^{56}$Ni. Therefore, the amount of 
$^{56}$Ni given in Table\,\ref{tab_models} is the $^{56}$Ni produced by the 
burning network for the standard models B15, N20, L15, and W15, while for model 
B15$_\mathrm{X}$ we add the entire mass of the tracer $X$ to $^{56}$Ni 
\citep[see also][for a similar treatment]{Utrobin2015,Utrobin2017}. 
During this work, we denote our treatment of the $\beta$ decay of the other 
models as {\it 
standard}, while we say that the $\beta$ decay is {\it enhanced} in the case of 
model B15$_\mathrm{X}$. We consider in particular this latter case because the 
synthesized yields of $^{56}$Ni may be underestimated in our 
simulations due to uncertainties of the electron fraction $Y_e$ of 
neutrino-processed ejecta caused by the use of a simplified neutrino treatment 
during the shock revival phase \citep[see][]{Wongwathanarat2013}. A significant 
fraction of the tracer element X is expected to actually be $^{56}$Ni. 
Therefore, models B15$_\mathrm{0}$ and B15$_\mathrm{X}$ provide a lower and an 
upper limit for the effect of the $\beta$-decay energy input in the SN ejecta. 

The $^{56}$Ni masses of our models are given in the 
penultimate column of Table\,\ref{tab_models}. These masses are well compatible 
with, for example, SN~1999em, for which Hillier \& Dessart 
(2019) found that a 15\,M$_\odot$explosion with a kinetic energy
of 1.2\,B and an ejected $^{56}$Ni mass of 0.036--0.043\,M$_\odot$ 
yields a good match of the observed multiband light curves and spectra.

To follow the propagation of the SN shock into regions beyond the surface of 
the progenitor star, we assume a stellar wind environment with prescribed 
properties. 
Following \cite{Lundquist1991} who provide estimated properties of the potential 
BSG wind of SN~1987A, we assume that the BSG progenitors lose their material 
at a rate $\dot M_\mathrm{wind}\sim4\times10^{-6}\,$ M$_\odot$/yr. The 
estimated temperature and velocity are $T_\mathrm{wind}\sim2.5\times10^4\,$K 
and $v_\mathrm{wind}\sim550\,$km/s, respectively. The properties of the RSG 
wind are $v_\mathrm{wind}\sim10\,$km/s, 
$T_\mathrm{wind}\sim10^5\,$K, and $\dot M_\mathrm{wind}\sim10^{-5}\,$ 
M$_\odot$/yr. For both BSG and RSG progenitors, we assume a wind
density profile that is proportional to $r^{-2}$. However, to make a smooth 
transition between the steep density gradient at the surface of the 
progenitor star and the density profile of the corresponding stellar wind we 
assume a $r^{-4}$ density dependence in between these two regions. 

\subsection{Terminology}

In this work we mainly focus on discussing differences of the 
morphological structures of the SN ejecta resulting from explosions of 
different 
stellar progenitor models. We often use terms like bubble, clumps, and 
fingers. The former are used to describe the central ejecta, which is rich in 
heavy nuclei like $^{56}$Ni. It is often used in the literature in the context 
of the `Ni-bubble effect' to describe the inflation of the central ejecta due 
to $\beta$ decay, which was first noted by \cite{Chevalier1976} and 
\cite{Woosley1988}. 
'Clump' and 'finger' are used to describe extended or isolated structures and 
often can be used interchangeably. The term `finger' is used to denote 
elongated structures that 
arise due to the growth of RTIs \citep[see][and references therein for a 
detailed description]{Wongwathanarat2015} after the propagation of the SN shock 
through shell interfaces of different chemical compositions inside the 
progenitor. The expression `clump' is usually used when referring to a 
disconnected finger-like 
structure or just a fast-moving blob of matter that cannot be associated to a 
finger. A clump is essentially any structure that does not connect to the central 
bubble.

Since $^{56}$Ni decays to $^{56}$Co and subsequently to the stable $^{56}$Fe 
isotope at late times, we introduce an abbreviation to denote the mixture of 
these three isotopes in a consistent way throughout the entire time evolution in 
our simulations. From here on, we refer to the mixture of 
$^{56}$Ni+$^{56}$Co+$^{56}$Fe as NiCoFe, and, if we additionally include the 
tracer X in the list, we denote this as NiCoFeX. Therefore, we define the 
corresponding mass 
fractions as $X_\mathrm{NiCoFe(X)}\equiv 
X_\mathrm{Ni}+X_\mathrm{Co}+X_\mathrm{Fe}(+X_\mathrm{X})$.

At late times the ejecta are expected to expand homologously. After the 
breakout from the progenitor at $t=t_\mathrm{out}$ (see 
Table\,\ref{tab_models}), only $\beta$ decay 
leads to an additional 
inflation of the NiCoFe-rich clumps/bubbles/fingers. To differentiate from the 
homologous expansion, we thus use the term bubble/clump/finger {\it inflation} 
to denote this additional expansion.

\subsection{Definition of clumps and corresponding filling factors}
In Section\,\ref{sec_clumps}, we discuss the properties of the clumps 
containing the largest amounts of NiCoFeX quantitatively. Since these clumps 
are characterized exclusively by NiCoFeX, the discussion in this section is 
based exclusively on the density and mass of these nuclei. We assume that one 
can observe only the densest of the NiCoFeX-rich material. Thus, to define the 
clumps, we take a certain fraction $F_\rho$ of the total mass 
$M^{\mathrm{tot}}_\mathrm{NiCoFeX}$, 
\begin{equation}
F_\rho\equiv\frac{M_\mathrm{NiCoFeX}^{>\rho^\mathrm {min }_\mathrm{NiCoFeX} } } 
 { M_\mathrm { NiCoFeX } ^\mathrm{tot}} \,,
\end{equation}
 which contains the densest NiCoFeX-rich material. Prescribing $F_\rho$, we can 
calculate the corresponding `visible' matter in the clumps which has the mass 
\begin{equation}\label{eq_def_M_1}
M_\mathrm{NiCoFeX}^{>\rho^\mathrm {min }_\mathrm{NiCoFeX}} =  
F_\rho\times  M_\mathrm {NiCoFeX } ^\mathrm{tot}\,.
\end{equation}
This mass can be obtained by integrating the mass of the densest NiCoFeX 
material
\begin{equation}\label{eq_def_M_2}
 M_\mathrm{NiCoFeX}^{>\rho^\mathrm {min}_\mathrm{NiCoFeX}} =  
\int_{\rho^\mathrm {min}_\mathrm{NiCoFeX}}^{\rho^\mathrm 
{max}_\mathrm{NiCoFeX}} V_\rho d\rho^\prime_\mathrm{NiCoFeX}\,,
 \end{equation}
where $\rho^\mathrm {max }_\mathrm{NiCoFeX}$ is the maximal density of NiCoFeX 
at a given time, and $V_\rho$ is the volume which is occupied by the 
NiCoFeX-rich ejecta with densities $\rho^\mathrm 
{min}_\mathrm{NiCoFeX}<\rho_\mathrm{NiCoFeX}<\rho^\mathrm 
{max}_\mathrm{NiCoFeX}$. With Eqs.\,(\ref{eq_def_M_1}) and 
(\ref{eq_def_M_2}) we can 
now determine the minimal density of NiCoFeX $\rho^\mathrm {min 
}_\mathrm{NiCoFeX}$, which we still assume to be part of the clumps.

We further will use a volume filling fraction 
$V^\mathrm{x}_\mathrm{NiCoFeX}\equiv{V_\rho}/{V_\mathrm{x} }$ 
 which we define as the ratio of the volume occupied by the NiCoFeX-rich 
matter above the minimal density $\rho^\mathrm {min}_\mathrm{NiCoFeX}$ and
the volume $V_\mathrm{x}$ of the sphere defined by the mean radius, where the 
ejecta move with a given mean velocity $\bar v=\mathrm{x}\,$km/s.
To have a measure to describe the clumpiness of the ejecta when 3D 
information in observations is not available, we further provide the surface 
filling factors 
$A^\mathrm{x}_\mathrm{NiCoFeX}\equiv{A_\mathrm{NiCoFeX}}/{A_\mathrm{x}}$.
The $A^\mathrm{x}_\mathrm{NiCoFeX}$ are defined as 
the fraction of a plane perpendicular to the line of sight which is covered by 
NiCoFeX clumps. For the extension of the plane we choose a square with the 
side length of the twice the corresponding radius $r_\mathrm{x}$, which we 
define as the 
radius at which the ejecta move with $\bar v=\mathrm{x}\,$km/s.

\section{Long-time evolution}\label{sec_long-time}

We continue the simulations of some models of \cite{Wongwathanarat2015} 
at the mapping times given in Table\,\ref{tab_models} and follow the evolution 
of the SN ejecta for all models until a time $t\gtrsim1\,$yr. The numerical 
grid consists of $1200$ cells in radial direction and has an  angular resolution 
of 2 degrees in $\theta$ and $\varphi$. The radius of the inner grid 
boundary of the BSGs is set to $r_\mathrm{IB}\lesssim 3\times10^{10}$\,cm and 
for the RSGs to $r_\mathrm{IB}\sim 10^{11}$\,cm. The radial grid is 
logarithmically spaced, and the outer grid boundary is placed just outside of 
the surface of the progenitor (see Table\,\ref{tab_models}) at the beginning of 
the simulations. 
In this section, we first discuss the two main processes that further modify 
the structures 
of the ejecta separately: a `self reflection' of the reverse shock 
which forms at the He/H-interface as it travels back to the stellar 
centre,  
and the energy input due to $\beta$ decay. Then, we show how 
their common action affects the structures of the NiCoFe-rich ejecta and 
we analyse properties of the ejecta clumps quantitatively.

\subsection{Self-reflected reverse shock}\label{sec_rev_shock}

In our 3D simulations, we confirm that the reverse shock from the He/H 
interface experiences a self-reflection at the stellar centre.  This 
reflection was first discussed in 1D simulations by \cite{Ertl2016err}. They 
showed that 
during the inward motion of the reverse-shock heated matter, the latter is 
decelerated because the flow gets 
geometrically focussed, leading to a negative pressure gradient in the radial 
direction. The deceleration produces an outward moving wave that steepens into 
a shock front when the expansion velocity exceeds the local sound speed. 

\begin{figure}  
\includegraphics[width=.49\textwidth]{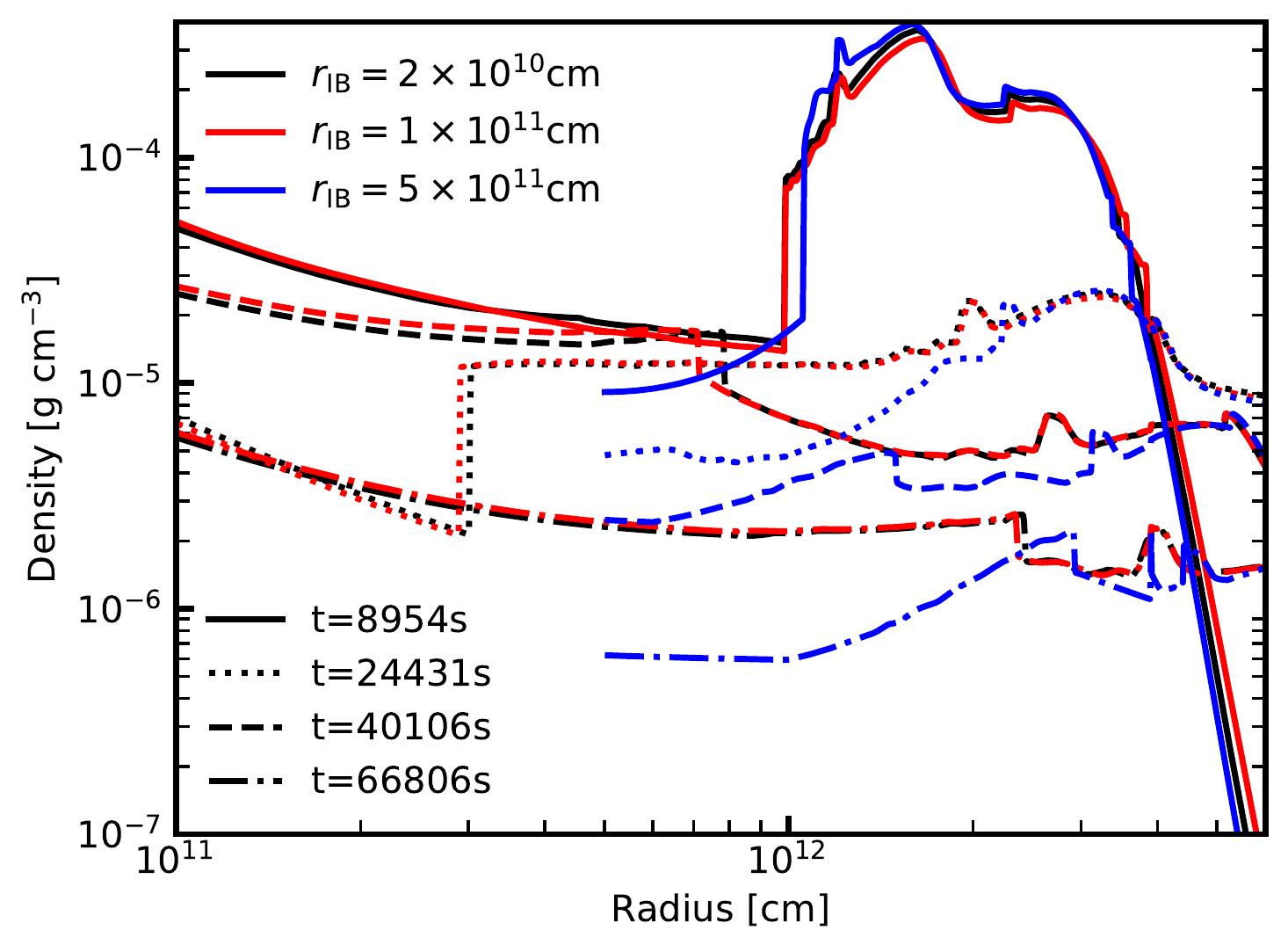}\\
\includegraphics[width=.49\textwidth]{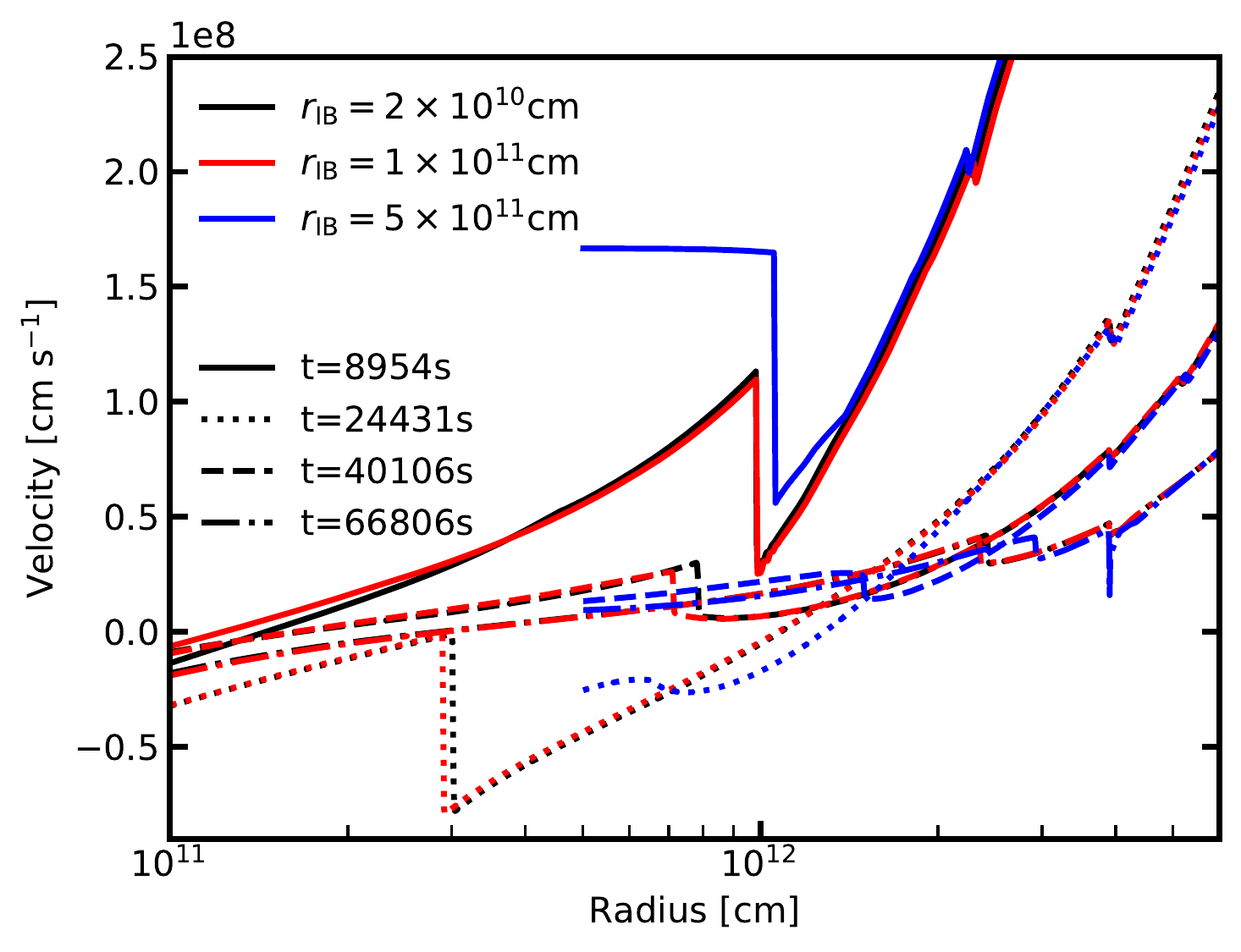}
\caption{Density (top panel) and velocity (bottom panel) profiles for 1D 
simulations of model B15 performed  
with different choices of the radius of the inner grid boundary 
$r_\mathrm{IB}=2\times10^{10},1\times 10^{11}$ and $5\times10^{11}\,$cm shown by 
black, red, and blue lines, respectively. The profiles are plotted at times 
$t\approx8950,24430,40100,66800\,$s indicated by different line styles (solid, 
dotted, dashed, and dash-dotted, respectively). Note that the times plotted are 
approximate because we output the simulation data at fixed interval numbers of 
time steps rather than fixed intervals in time. Depending on the choice of the 
placing of the inner boundary, the inward moving, reverse shock self-reflects at 
different times. The larger $r_\mathrm{IB}$ is, the farther out the 
self-reflection occurs, and the weaker the self-reflected shocks become. For 
$r_\mathrm{IB}=2\times10^{10}\,$cm and $r_\mathrm{IB}=1\times10^{10}\,$cm we 
obtain similar results.}
\label{fig_rev_shock}
\end{figure}

We tested that our choice of the radius of the inner grid boundary does not 
influence the strength of this self-reflected shock significantly by performing 
1D simulations with the inner boundary placed at different radii 
$r_\mathrm{IB}$. We computed three simulations with 
$r_\mathrm{IB}=2\times10^{10}$\,cm, $1\times10^{11}$\,cm and 
$5\times10^{11}$\,cm using model B15 as initial model. Angle-averaged 
profiles of hydrodynamic quantities of model B15 are mapped onto a 1D radial 
grid at $t\sim3150\,$s. Profiles of the density and velocity of the three 
simulations at different snapshots are displayed in the top and bottom panels of
Figure\,\ref{fig_rev_shock}, respectively.
The density profiles of the simulations with $r_\mathrm{IB}$ at 
$2\times10^{10}$\,cm and $1\times10^{11}$\,cm (black and red lines) are very 
similar at all times. In both cases, the reverse shock is visible at 
$r\sim10^{12}$\,cm for $t\sim8950\,$s (solid lines), and at 
$r\sim3\times10^{11}$\,cm for $t\sim24430\,$s (dotted lines). 
However, when the inner grid boundary is placed at 
$r_\mathrm{IB}=5\times10^{11}$ the resulting density profiles at the given 
times are significantly different from the other two cases. 
Already at $t\sim8950\,$s a low-density region is present close to the inner 
grid 
boundary. The outflow boundary condition we apply there first leads to a 
faster expansion (see higher velocities in bottom 
panel of Fig.\,\ref{fig_rev_shock} for $r_\mathrm{IB}=5\times10^{11}$), and at 
later times, when the velocities become negative, it allows more of the ejecta 
to leave the grid. Both effects lead to lower densities in the central region. 
In addition, the self-reflected shock forms 
only at a slightly larger radius. Thus, we conclude that our choice of 
$r_\mathrm{IB}\lesssim3\times10^{10}\,$cm used in the 3D simulations of the 
BSGs has a negligible impact on the formation of the newly formed
outward moving shock. The same holds true for $r_\mathrm{IB}\sim10^{11}\,$cm 
for the RSGs, because these progenitors are more extended by a factor of 10 and, 
hence, the reverse shock also forms much farther out than in the case of the 
BSGs. The self-reflected reverse shock impacts the SN ejecta dynamics by 
driving additional acceleration of the slow ejecta in our simulations, and we 
discuss this effect in more detail in Section\,\ref{sec_ejecta}.

\subsection{Beta decay}
\label{sec_volume_increase}
At early times when the SN ejecta are still optically thick, we expect the 
$\gamma$-rays produced in the decays of $^{56}$Ni and $^{56}$Co to heat up the 
matter in regions with high concentration of these two radioactive isotopes. 
This heating should lead to a non-homologous expansion (inflation) of these 
regions due to $pdV$-work. The total energy available from the decay of 
$^{56}$Ni to $^{56}$Co and the subsequent decay to $^{56}$Fe is 
\begin{equation}
E_\mathrm{\beta}\sim(Q_\mathrm{Ni}+Q_\mathrm{Co}) \times 
\frac{m_\mathrm{Ni}}{56m_u}  = 1.9\times10^{50} 
\frac{m_\mathrm{Ni}}{M_\odot} \mathrm{erg}\,,
\end{equation}
where $m_u$ is the atomic mass unit. For model B15 $m_\mathrm{Ni}=0.034 
M_\odot$ and we find $E_\mathrm{\beta}^\mathrm{B15}=6.5\times10^{48}\,$ erg.
Comparing this with the total kinetic energy at about $1\,$yr, 
$E_\mathrm{kin}^{1\,\mathrm{yr}}=1.4\times10^{51}\,$erg, we see that 
$E_\mathrm{kin}^{1\,\mathrm{yr}}\gg E_\mathrm{beta}$, 
and, hence, one would not expect a huge change of structures in the ejecta due 
to the $\beta$ decay. We can test how much of the decay energy is transformed 
to kinetic energy by comparing results from simulations computed with (B15) 
and without (B15$_0$) $\beta$-decay energy input. At $1\,$yr after the 
explosions we find a difference in the total kinetic energy of the SN ejecta 
between the two models of $\Delta 
E^{1\,\mathrm{yr}}_\mathrm{kin}\sim3\times10^{48}\,$erg, 
which is approximately half of the total available $\beta$-decay energy 
$E_\mathrm{\beta}^\mathrm{B15}$.

\begin{figure}
\includegraphics[width=.49\textwidth]{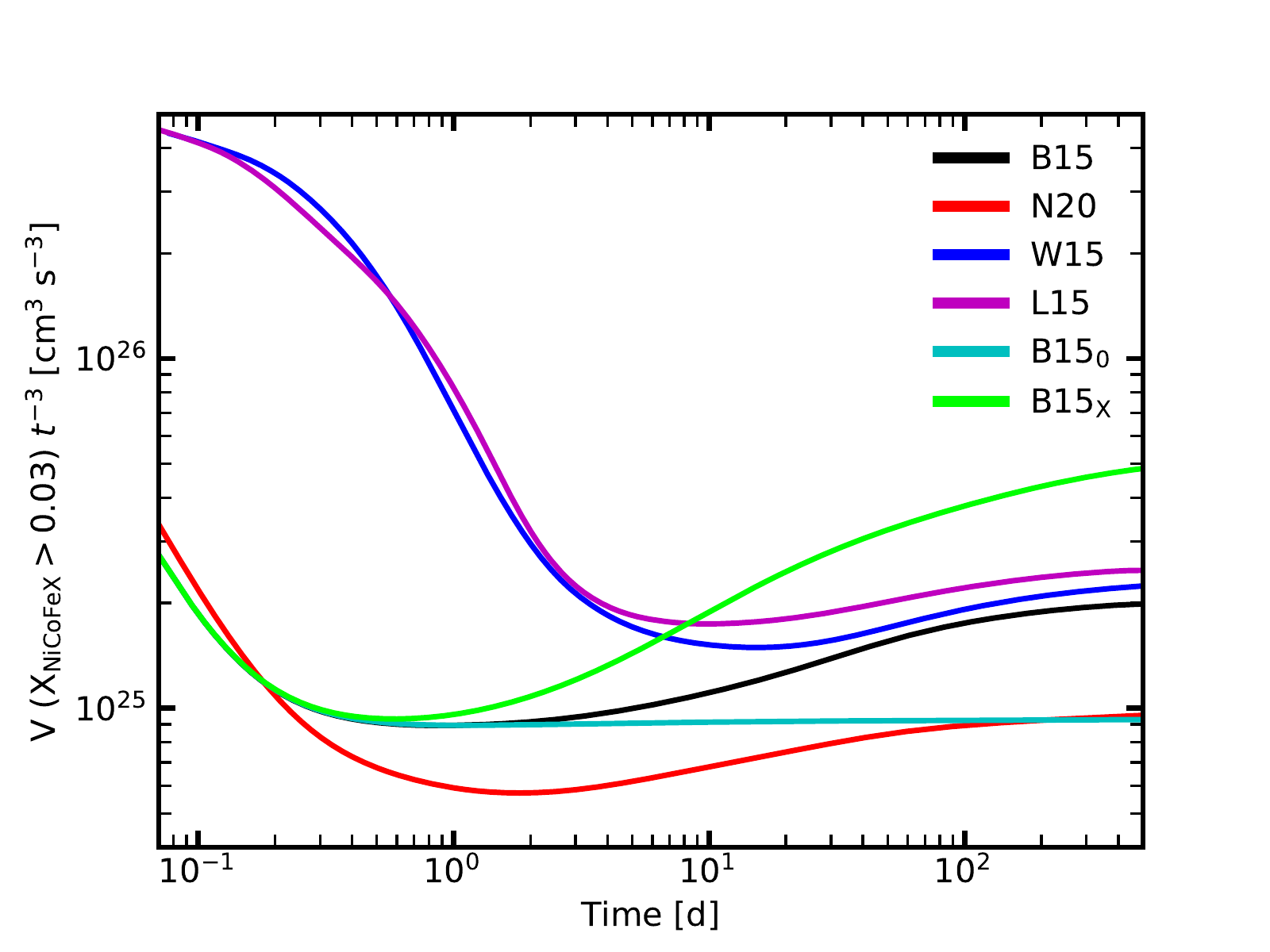}
\caption{Volume of the structures enclosed by the $3\%$-surface 
of $X_\mathrm{NiCoFeX}$ ($X_\mathrm{Ni}+X_\mathrm{Co}+X_\mathrm{Fe}+X_\mathrm{X}
>3\%$) rescaled by $t^{-3}$ as a function of time. Homologous expansion would be 
represented by a horizontal line like for model B15$_0$ for $t\gtrsim5\,$h. 
The expansion in the BSG models B15 and N20 
is slower than homologous until $t\lesssim1\,$d and later it is faster due to 
the energy input of the $\beta$-radiation causing the inflation of NiCoFe-rich 
structures. L15 and W15 expand more slowly than homologous until later and 
their rescaled volumes have their minima around $t\sim10\,$d. The bubble and 
finger inflation comes to rest after a few hundred days in all cases.}
\label{fig_expansion}
\end{figure}

For freely or homologously expanding ejecta one would expect the volume of any 
clump or bubble to expand proportional to $r^3$ or $t^3$. Since the 
$\beta$ decay provides an additional energy source that may inflate the 
$^{56}$Ni-rich structures, we expect deviations from this scaling in our 
simulations. To quantify this influence of the radioactive decay, we 
plot the volume enclosed by an isosurface of a constant mass fraction 
$X_\mathrm{NiCoFeX}=0.03$ multiplied by $t^{-3}$ as a function of time in 
Fig.\,\ref{fig_expansion}. As expected, the rescaled volume $Vt^{-3}$ of model 
B15$_0$ performed without $\beta$-decay expands homologously after about 
$t\sim6\,$h, as can be seen by the horizontal cyan line in 
Fig.\,\ref{fig_expansion}. For all other models the rescaled volumes of 
ejecta structures initially containing high Ni mass fractions increase after 
several hours for the BSG progenitors and after a few days for the RSG 
progenitors, i.e. the ejecta expand faster than homologous because they 
inflate. The initial decrease of the rescaled volume is related to the 
deceleration of the expansion due to the swept up masses in the outer layers 
still inside the progenitor and the deceleration due to the interaction 
of the ejecta with the reverse shock formed at the He/H shell interface.
As expected, the rescaled volume of model B15$_\mathrm{X}$ increases more than 
that of model B15. The additional energy from the decay of the tracer X leads 
to the production of more kinetic energy inflating the initially 
$^{56}$Ni-rich structures even further. We discuss these differences 
quantitatively in Section\,\ref{sec_clumps}.

After about $150\,$days, the inflation of the $^{56}$Ni clumps stagnates
because the ejecta become optically thin for $\gamma$-ray photons and only a  
small amount of energy associated with the $e^+$, which are released during the 
$\beta$ decay of $^{56}$Co, contributes to the heating of the bubbles and 
clumps. Additionally, a significant fraction of the radioactive material has 
already decayed after $\tau_{1/2}^\mathrm{Co}=77.23\,$d. Similar trends can be 
observed for all models.

However, since the BSG models B15 and N20 are more compact, the deceleration 
and interaction with the reverse shock occur and terminate earlier ($t<1\,$d) 
than for the RSG models ($t\sim10\,$d). Furthermore, because the liberated 
energy 
is deposited within a smaller volume the effect of the bubble and finger 
inflation sets in also at earlier times, and the inflation is relatively 
stronger as indicated by faster rises of the rescaled volume in the BSG models. 
At $t<0.1\,$d the rescaled volume of model W15 (or L15) is more than a factor 
of $10$ larger than that of B15, and at the end of the simulations around 
$t\sim1\,$yr both volumes are almost equal.

\subsection{Ejecta structures}\label{sec_ejecta}

\subsubsection{Radial velocities}\label{sec_radial_vel}

\begin{figure}
\includegraphics[width=.49\textwidth]{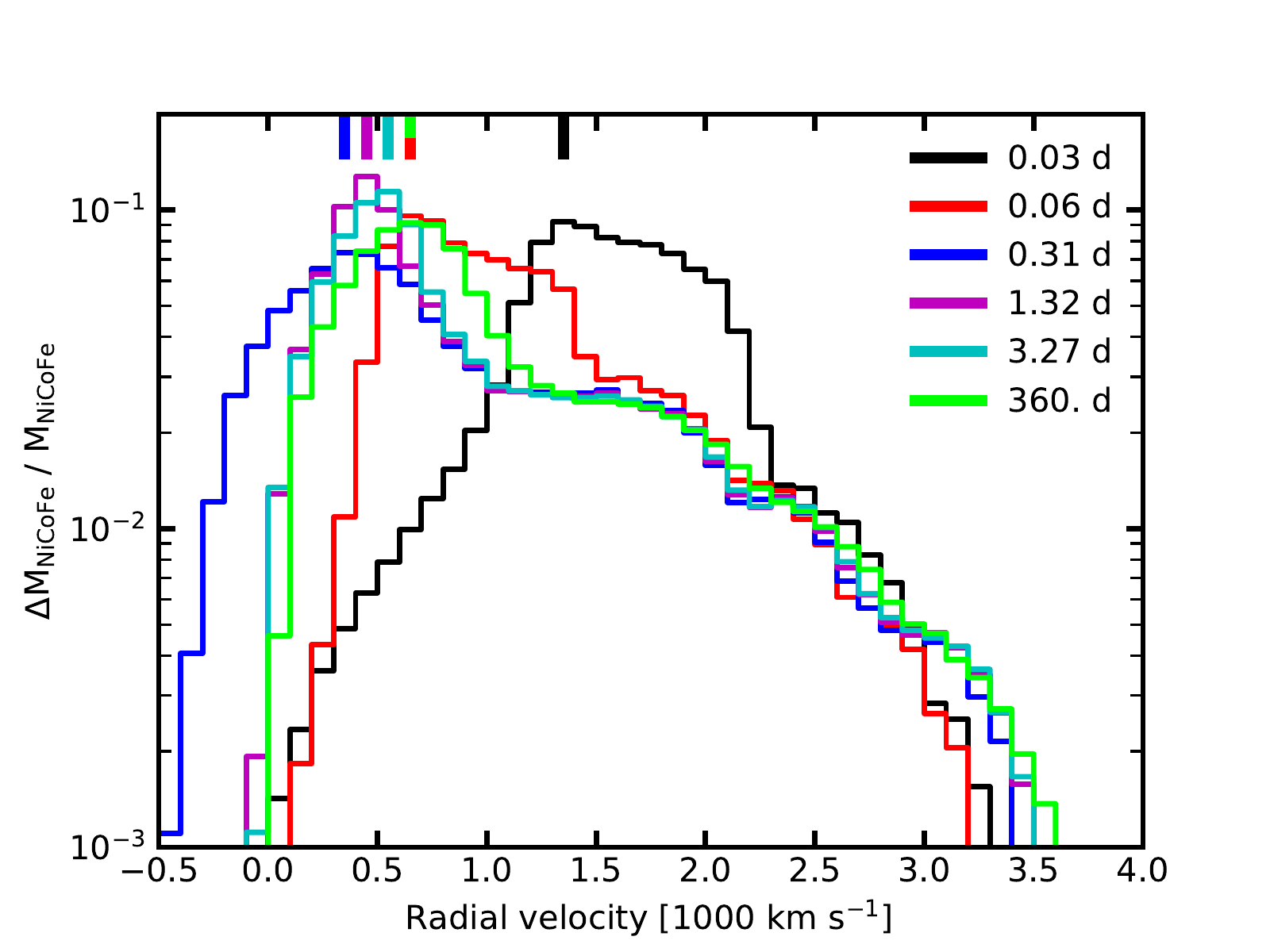}
\caption{Fractional mass of the $^{56}$Ni decay chain elements for model B15. 
The initially rapidly moving ejecta (black line) slow down during the 
propagation through the progenitor mainly due to the interaction with the 
reverse shock. After about $t\sim1\,$d  (magenta line) the self-reflected shock 
and the $\beta$-decay energy input reaccelerate the ejecta until 
they reach their final velocities at $t\lesssim1\,$yr (green line). Short 
lines at the top of the panel indicate the location of the maximum of the 
corresponding distribution and the bin width of the velocity is $100\,$km/s.}
\label{fig_vel_B15_1d} 
\end{figure}

\begin{figure*}  
\includegraphics[width=\textwidth]{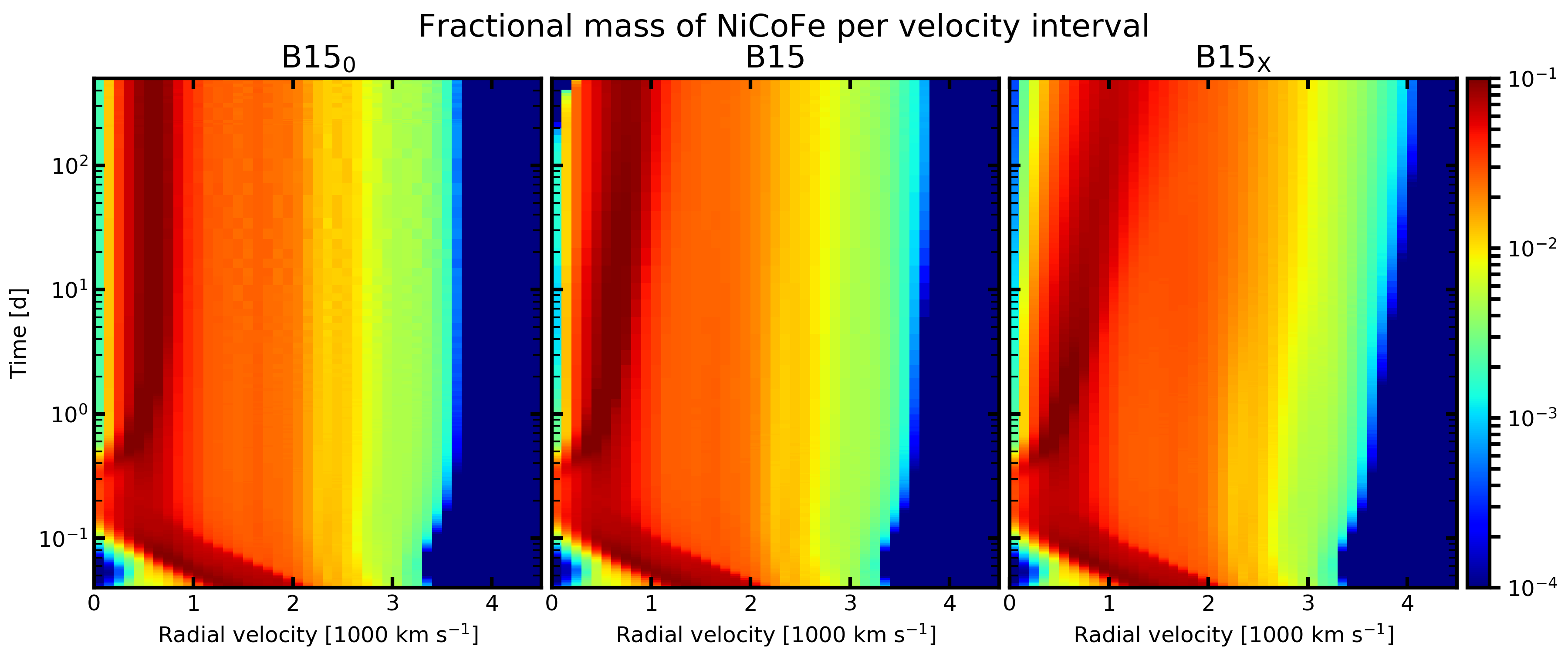}\\
\includegraphics[width=\textwidth]{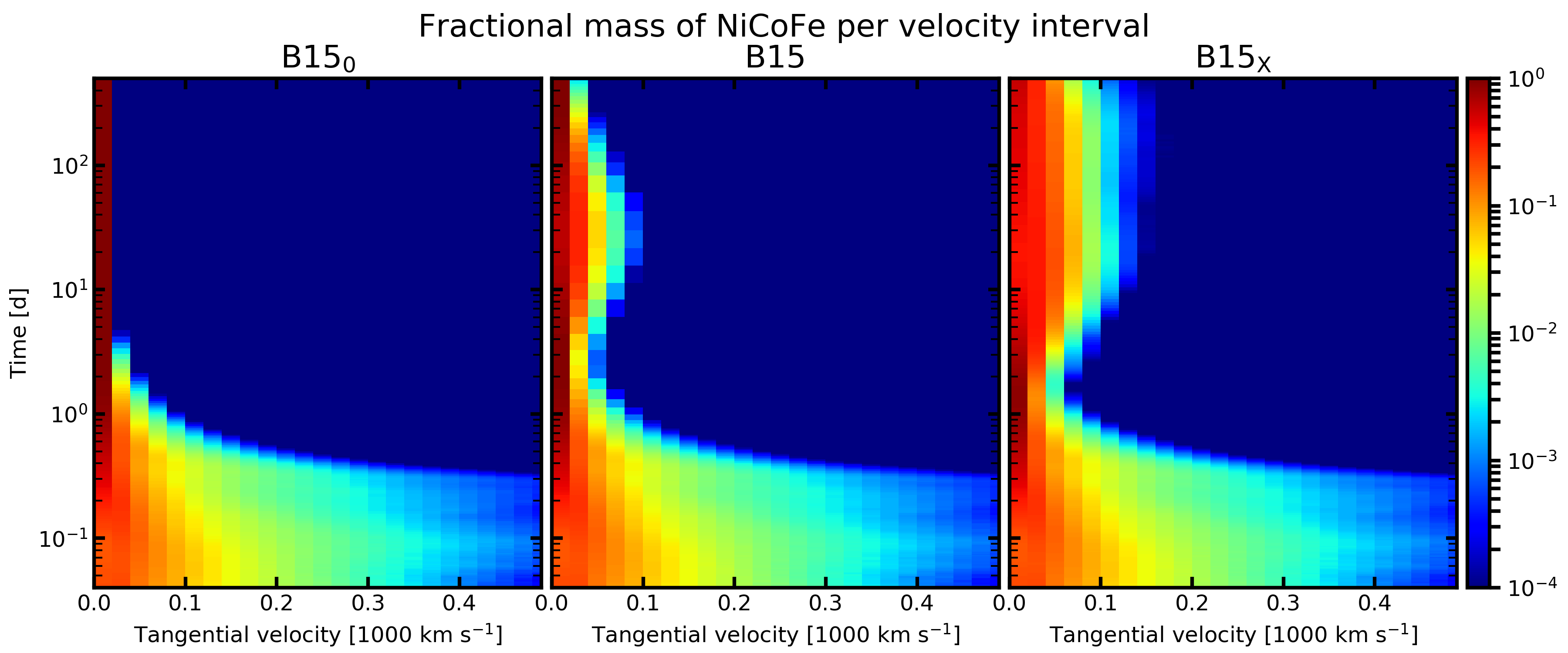}
\caption{Colour coded is the fractional mass of NiCoFe per velocity bin 
(x-axis) as a function of time (y-axis) for the first year of the evolution of 
models computed with the B15 progenitor. {\it Top row}: radial velocity. {\it 
Bottom row}: tangential velocity. Different columns show the models without 
$\beta$ decay (B15$_0$), with standard $\beta$ decay (B15), and with enhanced 
$\beta$ decay (B15$_\mathrm{X}$), respectively. After the initial deceleration 
due to 
the interaction with the reverse shock up to $t\lesssim0.5\,$d, the ejecta of 
models with $\beta$ decay accelerate again until $t\sim1\,$yr when the 
expansion becomes homologous. In model B15$_0$, only the effects of the 
self-reflected reverse shock can be witnessed.}
\label{fig_vel_B15}
\end{figure*}

Both effects discussed in the two preceding sections, namely the $\beta$ 
decay and 
the self-reflection of the reverse shock, have a similar impact on the 
NiCoFe-rich ejecta: they accelerate in particular the innermost slow material. 
To study their combined action, we investigate the mass distribution of 
NiCoFe-rich ejecta in the radial velocity space. In Fig. \ref{fig_vel_B15_1d}, 
we plot exemplarily the mass fractions per velocity bin of model B15 at 
different times. At early times ($t<0.5\,$d) the material gets decelerated as 
can be seen by comparing peaks of the distributions shown with the black, red, 
and blue curves. 

This deceleration is caused by the interaction of the NiCoFe-rich ejecta with 
the reverse shock formed after the forward shock of the SN crosses the He/H 
interface \citep{Wongwathanarat2015}. There is even a significant amount of 
material falling back towards the centre with negative velocities (blue curve 
at $0.31\,$d). When the reverse shock gets self reflected, it reaccelerates the 
innermost  (and slowest) material such that only a negligible amount of matter 
has negative velocities at $t=1.32\,$d (magenta curve). Within a few days, the 
outward-moving, self-reflected shock runs into denser material, and 
transfers all its energy so that it cannot accelerate the NiCoFe-rich 
ejecta any longer (cyan curve). At this epoch, the $\beta$ decay of $^{56}$Ni 
provides an additional energy source that leads to further acceleration of the 
NiCoFe-rich ejecta by about $150\,$km/s (see difference in the maxima between 
the cyan and green curves). The propagation of the fastest moving NiCoFe-rich 
ejecta can neither be influenced significantly by the self-reflected shock, 
because it loses its power before reaching them, nor by the $\beta$ decay, 
because there is not sufficient energy deposition due to the very low mass 
fraction of $^{56}$Ni in the fastest ejecta.

In the top panels of Fig.\,\ref{fig_vel_B15}, we plot the fractional mass of 
NiCoFe in given velocity bins in 2D plots as function of time and radial 
velocity for model B15$_0$ (left column), B15 
(central column), and B15$_\mathrm{X}$ (right column). The first few hours 
proceed nearly identically in all three cases: the reverse shock decelerates the 
NiCoFe-rich ejecta, causing parts of them to fall back with negative velocities. 
Then, around $t\sim0.2\,$d the reverse shock self-reflects and turns outward, 
accelerating the innermost material to positive velocities. For model B15$_0$ 
this acceleration terminates at around 4 days. In contrast, the low-velocity 
NiCoFe-rich ejecta of models B15 and B15$_\mathrm{X}$ continue to accelerate 
until approximately $150\,$d. Consequently, the mean velocity of the 
NiCoFe-rich ejecta increases by about $200\,$km/s and even up to $\sim350\,$km/s 
for model B15 and B15$_\mathrm{X}$, respectively. In the bottom row of 
Fig.\,\ref{fig_vel_B15}, we plot the fractional mass according to their 
tangential velocities $v_t= \left(v_\theta^2+v_\varphi^2\right)^{1/2}$. 
As for the radial velocity, the tangential velocities of the NiCoFe-rich 
ejecta, which initially arise mainly due to the growth of RTIs at composition 
shell interfaces inside the progenitor star \citep{Wongwathanarat2017}, 
decrease in all models up to about $4\,$d. At later times, $v_t$ of the models 
including $\beta$ decay increases up to a maximum of $100\,$km/s (B15, central 
panel) and $160\,$km/s (B15$_\mathrm{X}$, right panel). Note that in the latter 
case much more of the NiCoFe-rich ejecta gets accelerated to $v_t>100\,$km/s 
and the tangential velocities need longer time to decline to very low 
values.
The maximal velocities in model B15 are reached after $20\,$d and then $v_t$ 
decreases until it becomes negligible again at around $1\,$yr.
At this point the SN ejecta have attained homology for this model, and we do 
not expect further strong effects of the $\beta$ decay of $^{56}$Ni and 
$^{56}$Co on the structure of the NiCoFe-rich ejecta. In model B15$_X$, even 
after more than $1\,$yr  the tangential velocity is still significant. 

%
\begin{figure*}  
\includegraphics[width=\textwidth]{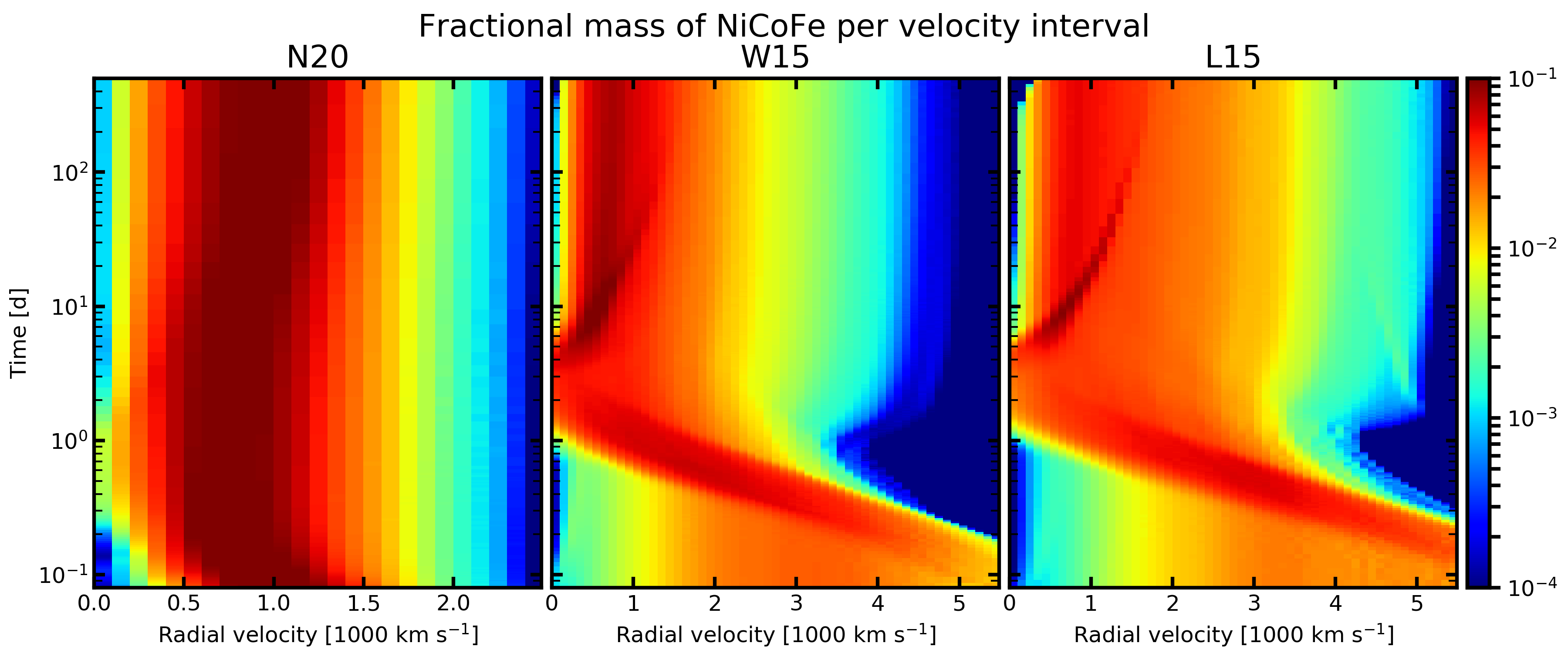}
\caption{Same as top row of Fig.\,\ref{fig_vel_B15} but for models N20, 
W15, and L15. Model N20 has the weakest reverse shock and, hence, the weakest 
acceleration due to the self-reflected shock. The two RSGs have
higher initial velocities (because of more efficient radial mixing of heavy 
elements) and the reverse shock reaches the central (slow) part 
of the ejecta at later times ($t\sim5\,$d). For all models the acceleration 
stalls after a few hundred days. Note the different scale of the $x$-axis 
for model N20.} 
\label{fig_vel_models}
\end{figure*}

\begin{figure}  
\includegraphics[width=.49\textwidth]{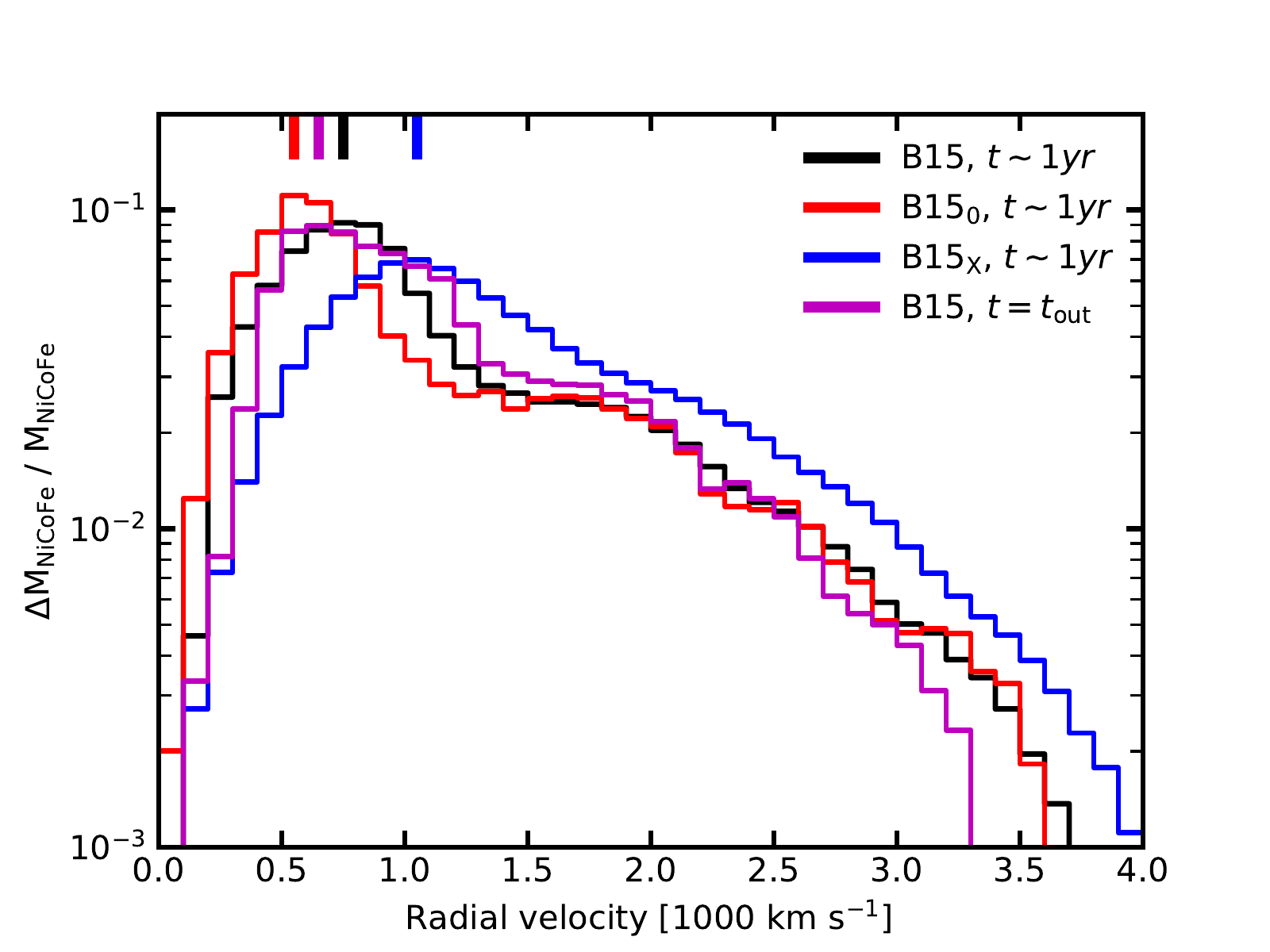}\\
\includegraphics[width=.49\textwidth]{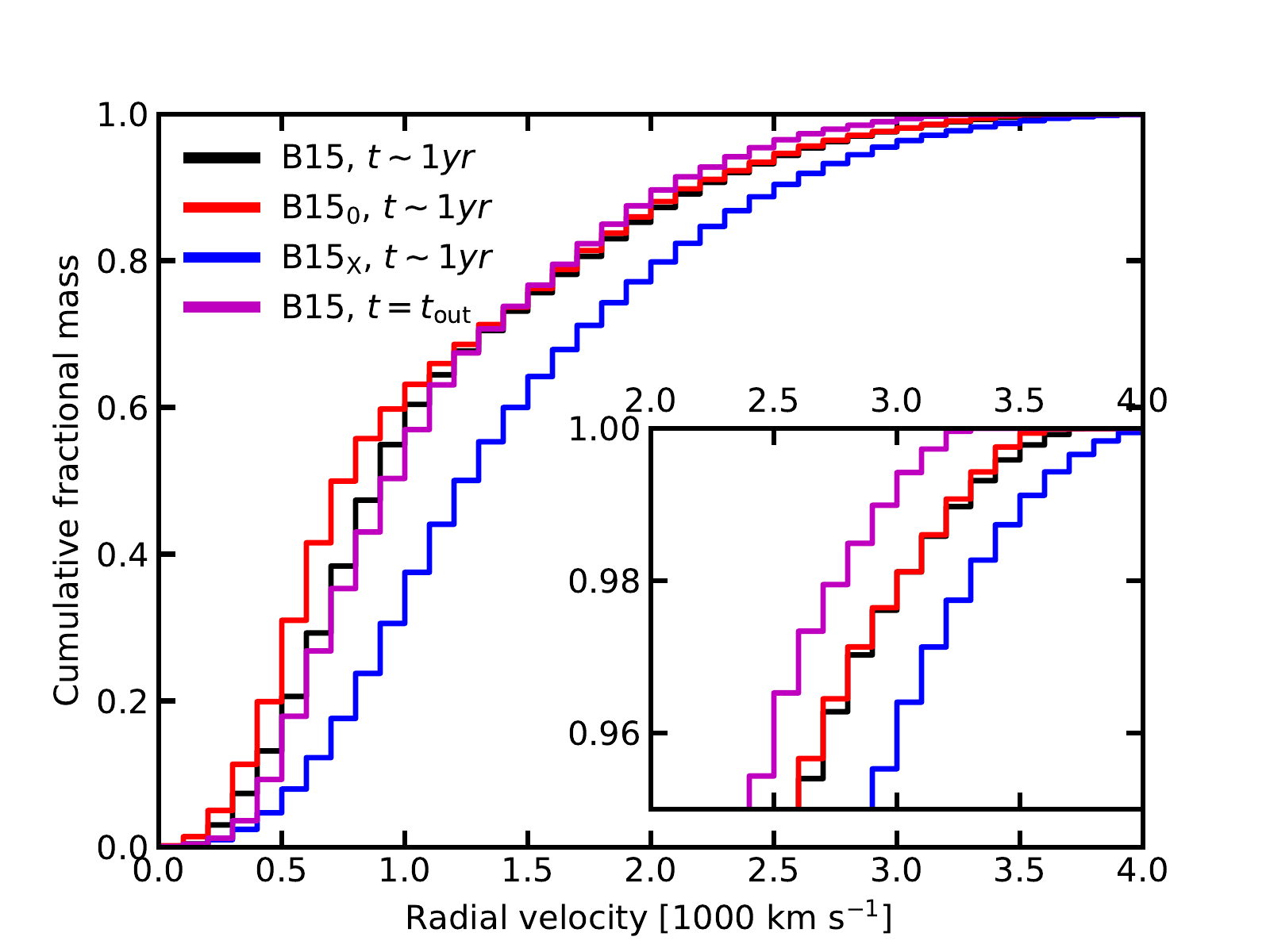}
\caption{Fractional mass (top panel) and cumulative fractional mass 
(bottom panel) of NiCoFe in radial velocity bins of $100\,$km width 
at shock breakout (magenta) and after $t\sim1\,$yr of the evolution of the 
models B15, B15$_0$, and B15$_\mathrm{x}$, respectively. At shock breakout all 
models have an almost identical velocity distribution and we only show that of 
model B15. Short lines at the top of the panel indicate the location of the 
maximum of the corresponding distribution. The inset in the lower panel shows 
the distributions of the fastest 5\% of the NiCoFe material.
}
\label{fig_vel_1y_B15}
\end{figure}

\begin{figure}
\includegraphics[width=.49\textwidth]{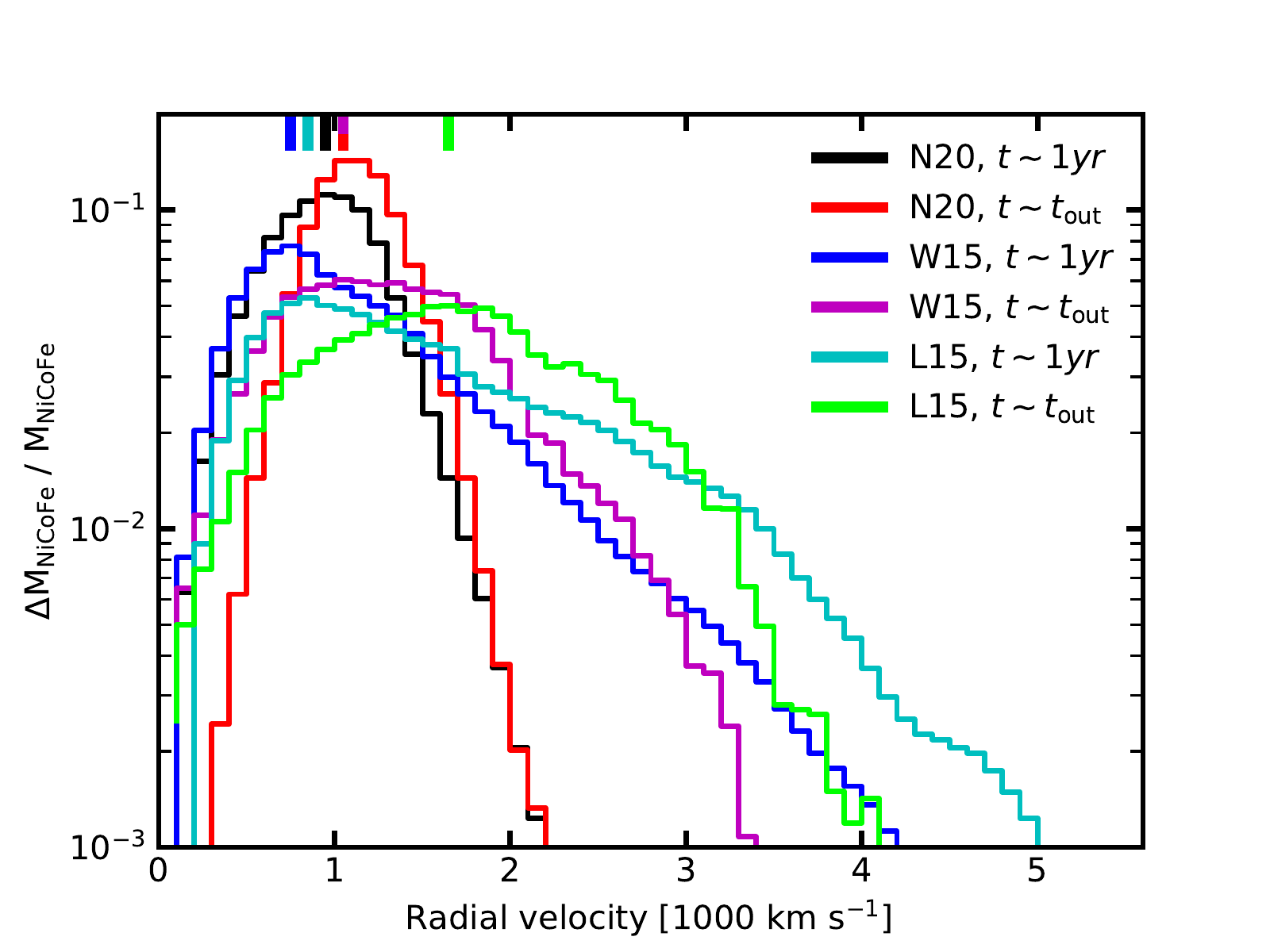}\\
\includegraphics[width=.49\textwidth]{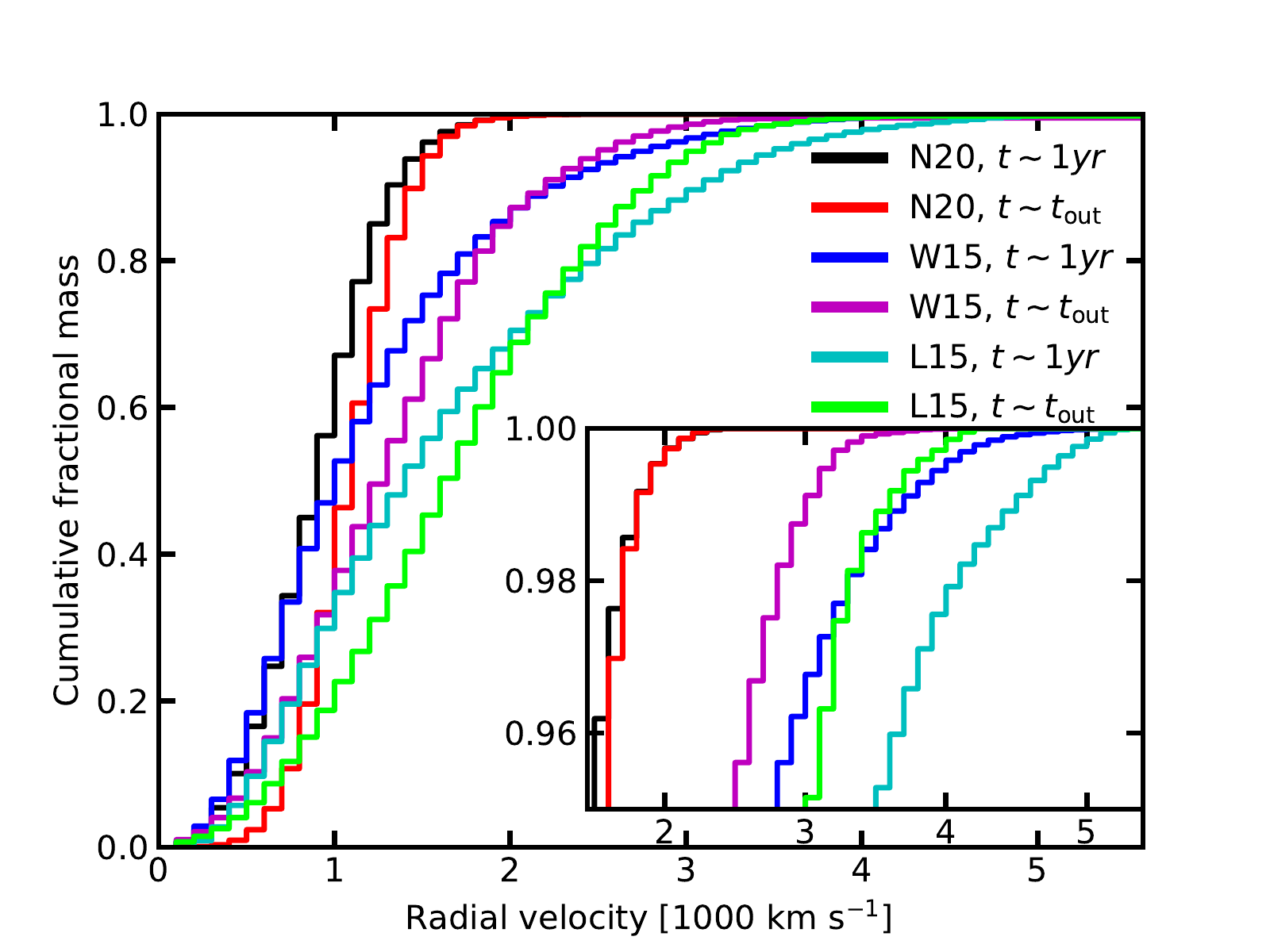}
\caption{Fractional mass  (top panel) and cumulative fractional mass 
(bottom panel) of NiCoFe in radial velocity bins of $100\,$km width 
at  shock breakout and after $t\sim1\,$yr of the evolution of the models N20, 
W15, and L15, respectively. Short lines at the top of the panel indicate the 
location of the maximum of the corresponding distribution. The inset in the 
lower panel shows the distributions of the fastest 5\% of the NiCoFe material.
}
\label{fig_vel_1y}
\end{figure}

The time evolutions of the NiCoFe mass distributions in the radial velocity 
space of the other three models, N20, W15, and L15, are given in 
Fig.\,\ref{fig_vel_models}. In model N20, the 
deceleration of the NiCoFe-rich material is not as strong as in the other  
two models, i.e. there is almost no NiCoFe-rich material with negative radial 
velocities. However, since the reverse shock is very weak in this model, the 
self-reflected shock is also very weak. Acceleration of NiCoFe-rich ejecta after 
$t\sim1\,$d is almost entirely due to the $\beta$-decay energy deposition. The 
two RSG models, W15 and L15, have a strong reverse shock that decelerates the 
NiCoFe-rich ejecta drastically. Consequently, after a few days there is some 
NiCoFe-rich material with negative velocities in these two models. As for model 
B15, the reverse shock then self-reflects and accelerates the innermost ejecta 
outward. Around the same time, a significant amount of the radioactive 
$^{56}$Ni 
has decayed, and the deposited $\beta$-decay energy heats up the ejecta and 
contributes to the reacceleration of NiCoFe-rich material. Around $t\sim1\,$yr,
the acceleration stagnates and homologous expansion follows. The reverse shock 
in the BSGs is generally weaker than in the RSGs, because the density drop at 
the He/H-interface is less steep in the BSGs \citep{Wongwathanarat2015}. 
The shallower density gradient leads to slower acceleration of the shock when 
crossing this interface and, consequently, the following deceleration inside 
the 
H-shell is less drastic. Therefore, the reverse shock, forming as a consequence 
of this deceleration, is weaker in models N20 and B15.

Next we will study how the mass distribution of the NiCoFe-rich ejecta 
changes from the shock breakout to $t\sim1\,$yr after the onset of 
the explosions. The 
graphs for models B15, B15$_0$, and B15$_\mathrm{X}$ are given in 
Fig.~\ref{fig_vel_1y_B15}. Without $\beta$ decay, the peak of the mass 
distribution shifts to lower values of the radial velocity (compare magenta and 
red curves in Fig.~\ref{fig_vel_1y_B15}). In contrast, the peaks of the 
distributions shift towards higher velocities at late times for models computed 
with $\beta$ decay. The highest velocities of the NiCoFe-rich ejecta are larger 
at late times in all cases, regardless of whether the $\beta$ decay is included 
or not. Even without $\beta$ decay, the ejecta still did not expand homologously 
at shock break out for model B15 (see also Fig.\,\ref{fig_vel_B15}).

In  Fig.\,\ref{fig_vel_1y}, we show the fractional mass distribution versus 
radial velocity for models N20,
W15, and L15. In all cases, the maxima of the distributions at $t\sim1\,$yr are 
at lower velocities than during the shock breakout. At the breakout time, the 
bulk of the NiCoFe-rich matter is still decelerating due to the interaction with 
the reverse shock. In contrast to model B15, the acceleration due to the 
self-reflected reverse shock and the 
$\beta$ decay is not sufficient to reach bulk velocities larger than during the 
shock breakout. However, similar to model B15, there is sufficient acceleration 
of the high-velocity tail of the mass distributions for models W15 and L15 
that the tails extend to higher velocities at late times $t\sim1\,$yr, even 
though the bulk of the matter is moving slower than before. This acceleration 
of the high-velocity component happens still at early times until about 
several days (see 
Fig.\,\ref{fig_vel_models}) and the effects of the $\beta$ decay are 
subdominant. The fractional mass 
distribution of model N20 has different characteristics. The fastest moving 
NiCoFe-rich ejecta of this model are almost at the same velocities at 
$t=t_\mathrm{out}$ and $t\sim1\,$yr, i.e. the fastest material was expanding 
homologously almost since it left the progenitor star, and the $\beta$-decay 
energy input was not able to accelerate this material significantly.

We can study this behaviour also by looking at the mean velocities 
of the NiCoFe-rich ejecta at $t\sim[t_\mathrm{out},1\,$d$,10\,$d$,1\,$yr$]$ in 
Table\,\ref{tab_vel_final}. During and shortly after the breakout, all models 
have a strong decrease of the mean velocity. Note that for model L15 
$t_\mathrm{out}\sim1.1\,$d, such that also in this model the mean velocity 
decreases after the breakout time. After this period, the models show different 
behaviours. The B15$_0$ model without $\beta$ decay has only a very mild 
acceleration of the mean velocity between $t=1\,$d and $t=10\,$d and remains 
constant afterward, indicating that it reaches homologous expansion after a few 
days. In model B15 with standard $\beta$ decay, there is significantly more 
acceleration even after $t=10\,$d, and model B15$_\mathrm{X}$ has the strongest
velocity increase of up to $30\%$.

From $t=1\,$d until the end of our simulations, the mean velocities in model 
B15 and B15$_\mathrm{X}$ increase by about $100\,$km/s and $400\,$km/s, 
respectively. This shows that the $\beta$ decay has a significant imprint on 
the final velocities and, thus, has to be considered in the simulations. 
The other three models show similar trends as model B15 or B15$_\mathrm{X}$.
Around the time of the shock breakout, the mean velocity of the NiCoFe-rich 
ejecta decreases, while at latest after a few days, an acceleration occurs. In 
contrast to model B15 or B15$_\mathrm{X}$, the mean velocities of all 
other models during shock 
break out are not reached again until the end of the acceleration phase. 
In addition to the mean velocity, we give the velocity of the fastest one 
percent of the NiCoFe-rich ejecta in Table\,\ref{tab_vel_final_1percent}. Most 
of the acceleration is finished after ten days and only the velocity of model 
B15$_\mathrm{X}$ increases significantly by $200$ km/s until one year.
Comparing the different models, L15 has the fastest moving 
NiCoFeX-rich ejecta. The higher velocities are a direct consequence of the  
high explosion energy of that model, see Table\,\ref{tab_models}.
\begin{table}
\centering
 \begin{tabular}{c | c c c c c c}
time&B15$_0$&B15&B15$_\mathrm{X}$&N20&W15&L15\\\hline
$t_\mathrm{out}$&1.22&1.22&1.22&1.18&1.40&1.79\\
$1\,$d&1.12&1.12&1.17&0.89&1.39&1.90\\
$10\,$d&1.14&1.16&1.33&0.93&1.15&1.60\\
$1\,$yr&1.14&1.22&1.52&1.00&1.29&1.73
\end{tabular}
\caption{Mean velocities of NiCoFe at different times in $1000$\,km/s.
Note that $t_\mathrm{out}$ for Model L15 is $t_\mathrm{out}\sim1.1\,$d, while 
all other times of the breakout are shorter than one day (see 
Table\,\ref{tab_models}). }
\label{tab_vel_final}
 \begin{tabular}{c | c c c c c c}
time&B15$_0$&B15&B15$_\mathrm{X}$&N20&W15&L15\\\hline
$t_\mathrm{out}$&3.19&3.18&3.19&2.08&3.31&3.92\\
$1\,$d&3.46&3.45&3.47&2.05&3.32&3.92\\
$10\,$d&3.48&3.48&3.60&2.06&4.09&4.81\\
$1\,$yr&3.48&3.54&3.80&2.08&4.16&4.89
\end{tabular}
\caption{ Mean velocities for the fastest one percent of the 
NiCoFe-rich ejecta in $1000\,$km/s.}
\label{tab_vel_final_1percent}
\end{table}

\begin{figure*}  
\includegraphics[width=.325\textwidth]{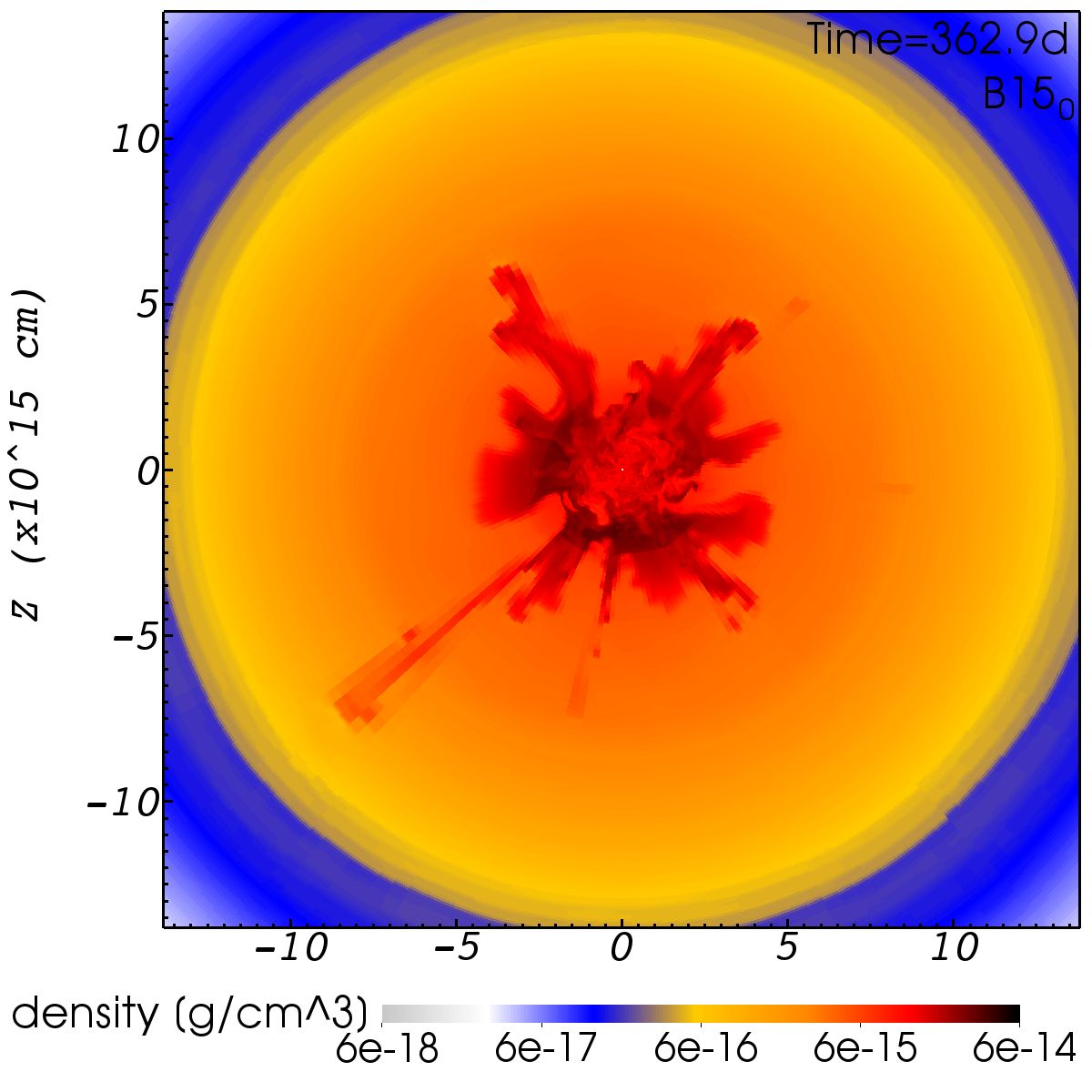}
\includegraphics[width=.325\textwidth]{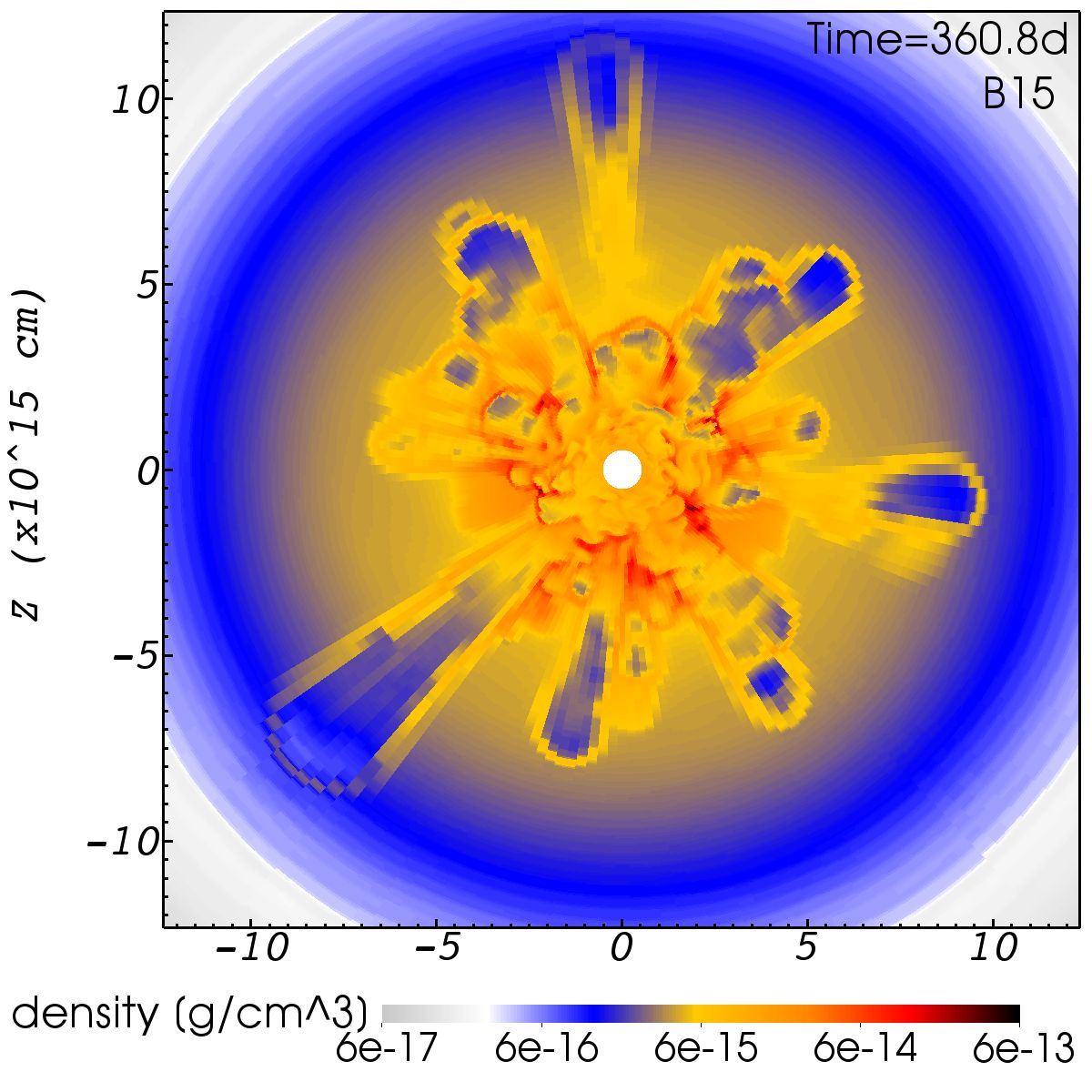}
\includegraphics[width=.325\textwidth]{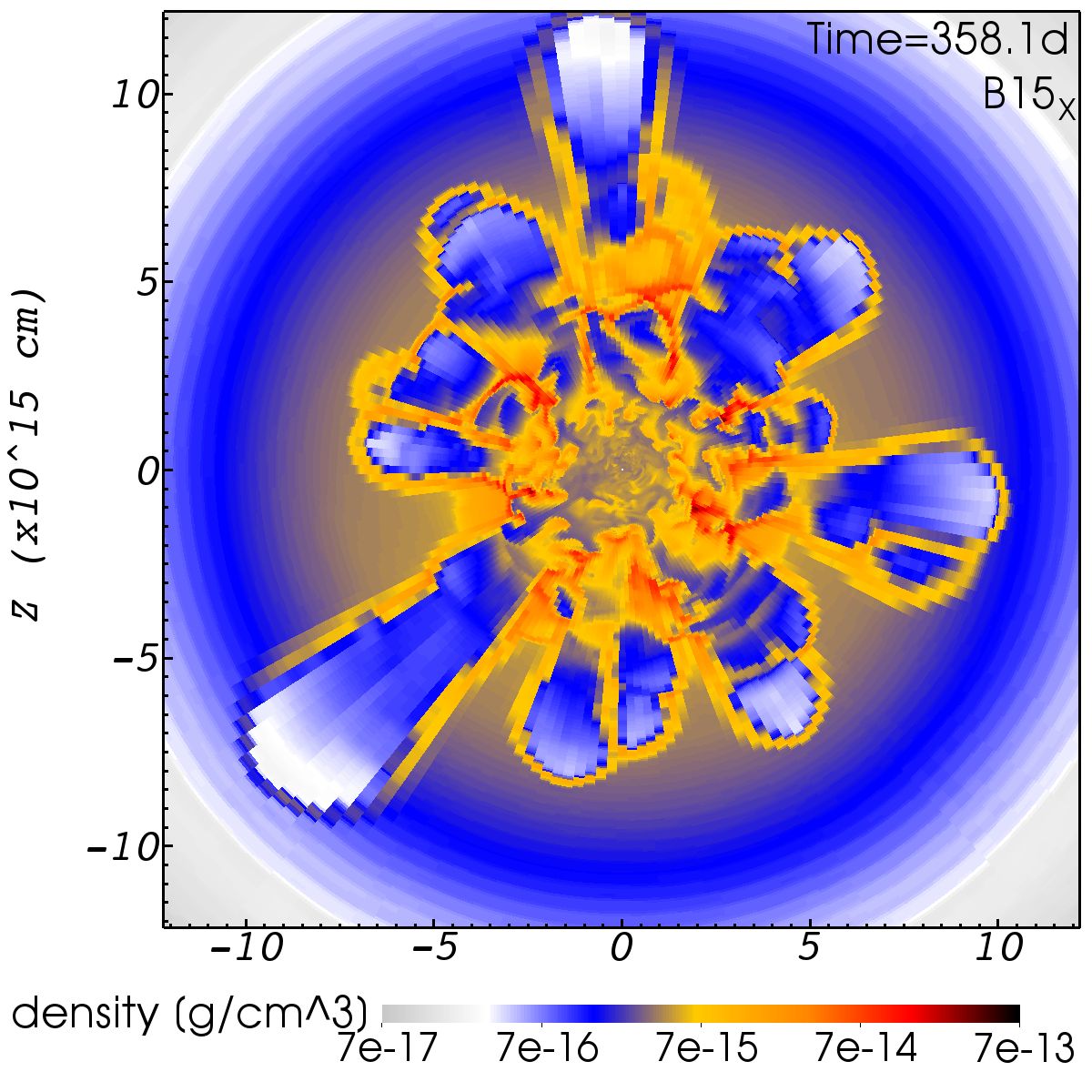}\\
\includegraphics[width=.325\textwidth]{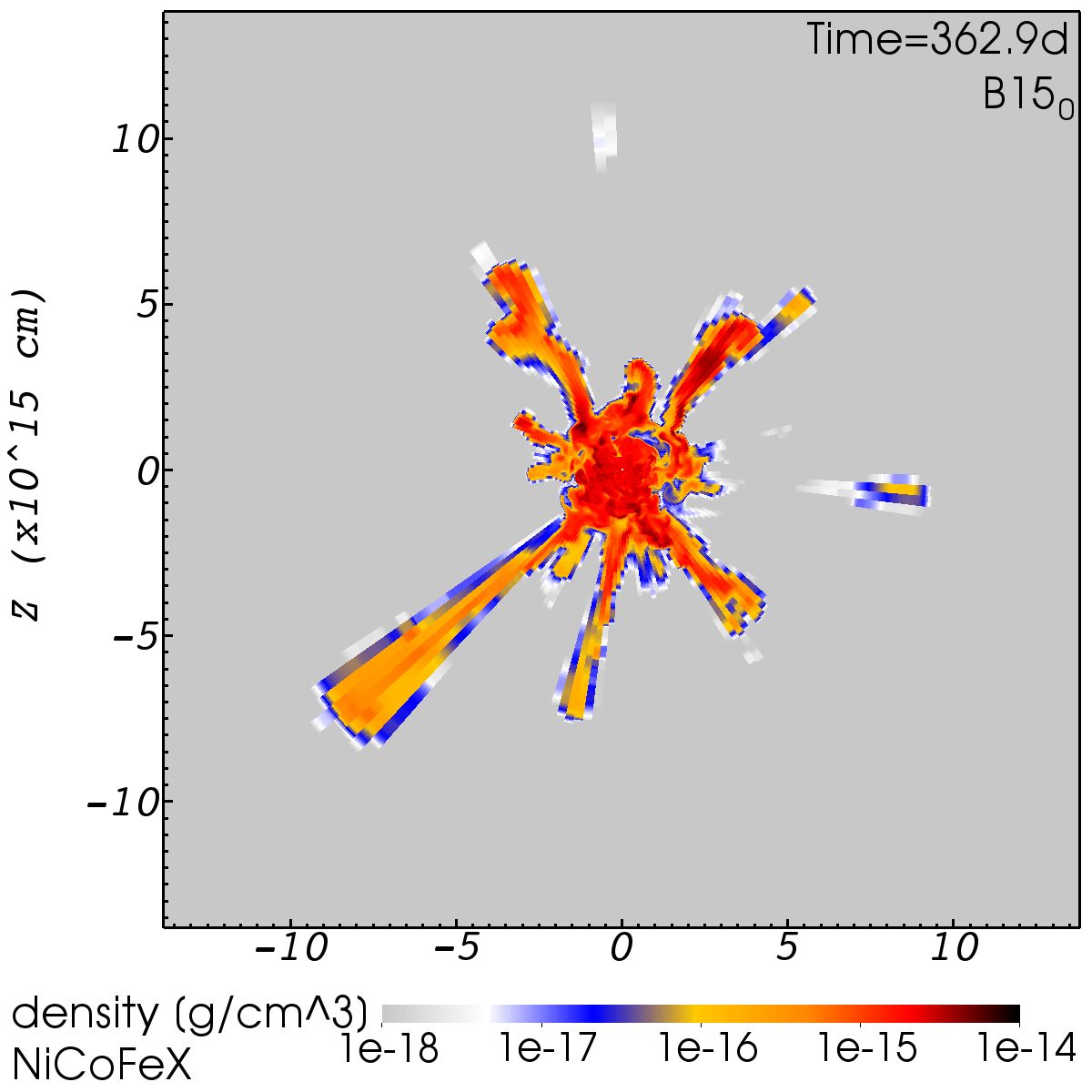}
\includegraphics[width=.325\textwidth]{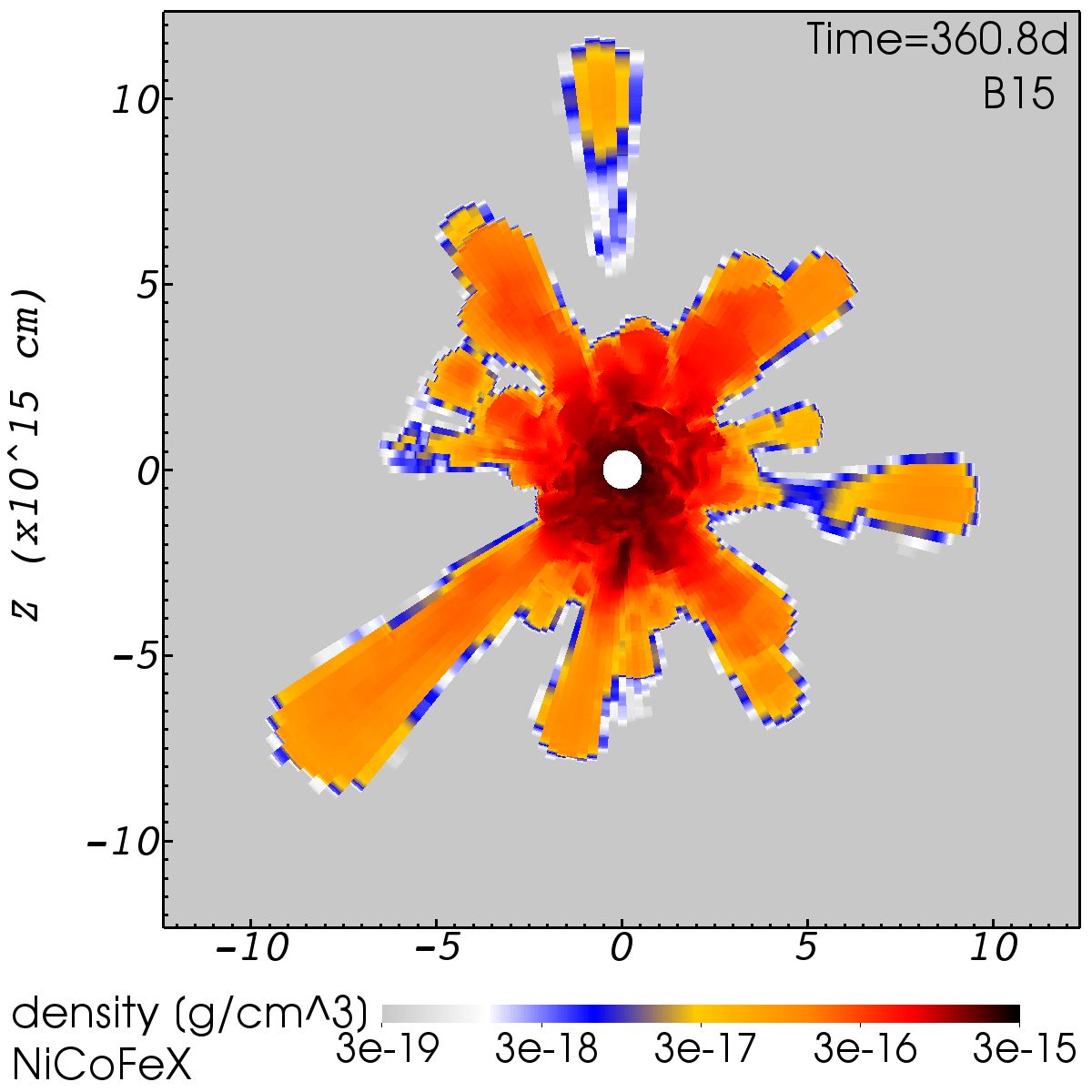}
\includegraphics[width=.325\textwidth]{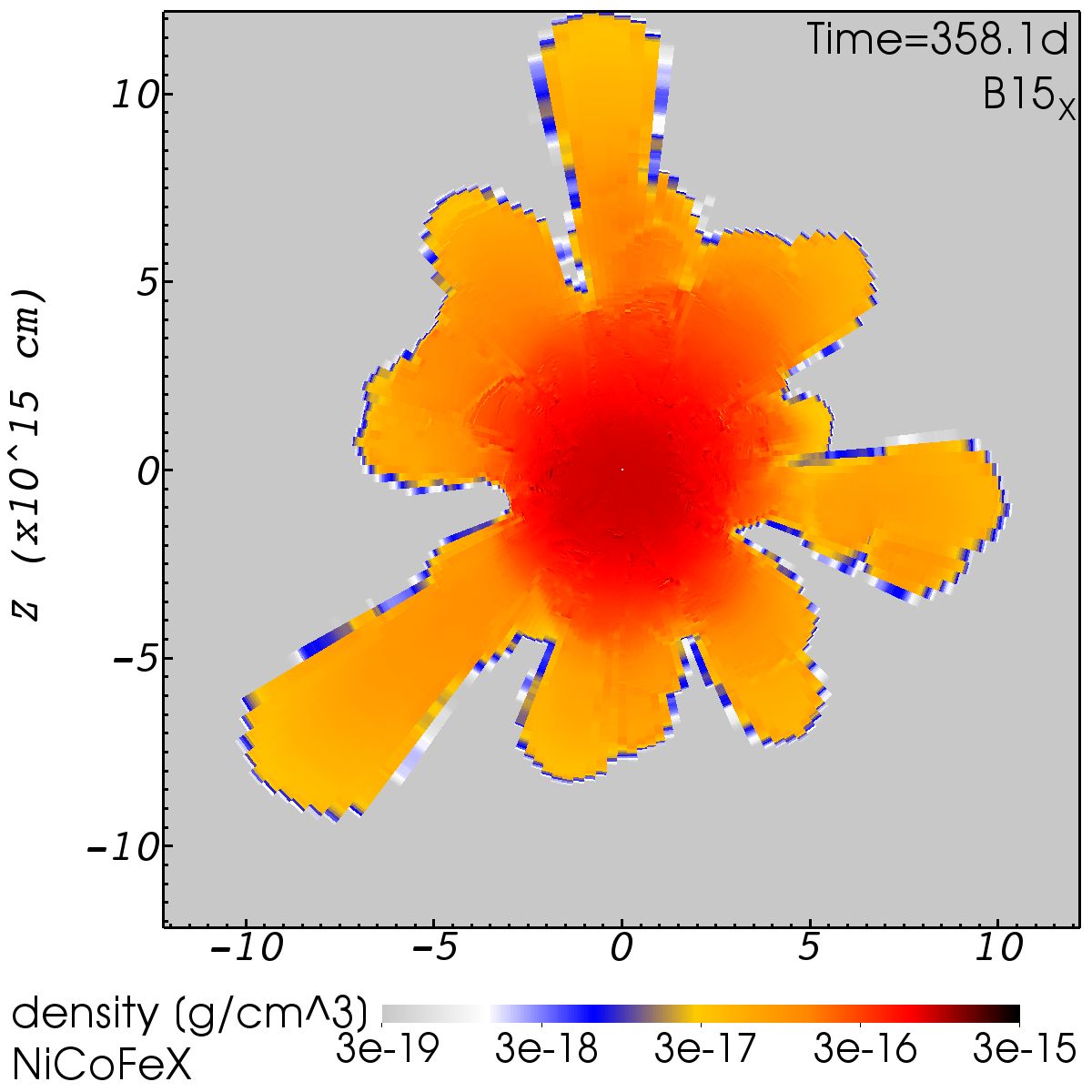}
\caption{{\it Top row}: Density slices of models B15$_0$ (left panel), B15 
(central panel), B15$_\mathrm{X}$ (right panel) in the x-z-plane around 
$t\sim1\,$yr. {\it Bottom row}: Slices of the density of NiCoFeX of the 
corresponding models in the same plane and at the same time. The density scale 
is logarithmic in all panels. Without $\beta$-decay (B15$_0$) the NiCoFeX-rich 
RT 
fingers stay unchanged and are over dense relative to their surroundings. 
When including $\beta$ decay (B15) in the simulation, the fingers inflate and 
become underdense. Due to the larger amount of radioactive material, Model 
B15$_\mathrm{X}$ has a larger inflation of the fingers and the NiCoFeX 
distribution inside the fingers becomes also more uniform. For 
efficiency reasons, as discussed in Section\,\ref{sec_numerics},
we cut out the central white region during the simulation. This is visible in 
particular for model B15 (see also Footnote\,\ref{footnote_problem} in 
Section\,\ref{sec_numerics}).}
\label{fig_slice_B15}
\end{figure*}

\subsubsection{Density distributions}\label{sec_density_dist}
Here, we investigate the effect of the energy input due to the $\beta$ decay on 
the ejecta structures by plotting slices of the total density 
$\rho^\mathrm{tot}$ and the density of NiCoFeX $\rho_\mathrm{NiCoFeX}$. The 
corresponding plots at $t=1\,$yr for models B15$_0$, B15, and 
B15$_\mathrm{X}$ are shown in Fig.\,\ref{fig_slice_B15}. In the top left panel 
for model B15$_0$, there are pronounced elongated ejecta structures, in 
particular in the negative $z$ and negative $x$ direction. These structures 
originate from the growth of RTI during the propagation of the SN shock through 
the progenitor envelope. Note that only in model B15$_0$ the density decreases 
from the centre of these structures towards the exterior, i.e. these RT fingers 
have higher densities than the ambient matter. In the bottom left panel, 
one can see that these overdense RT fingers are very rich in NiCoFeX. 
In the model without $\beta$ decay these structures are completely developed 
already at around $t\sim1\,$d, and do not change morphologically afterwards, 
i.e. in model B15$_0$ the Ni-rich ejecta are already expanding homologously 
after $t\sim1\,$d. In contrast, the models including $\beta$ decay have Ni-rich 
fingers that have lower densities than the matter around them (central and 
right top panels). The decay of radioactive material increases the internal 
energy. This energy increase heats up the NiCoFeX-rich ejecta, and they start 
to expand by doing $pdV$ work against their surroundings. Consequently, the 
densities inside the NiCoFeX-rich clumps decrease (compare bottom left to right 
panels for decreasing densities with increasing $\beta$ decay). This inflation 
of ejecta inside Ni-rich fingers sweeps up ambient matter and compresses this 
material to higher densities. As a result, regions of density enhancements 
build up at the border between decaying and non-decaying ejecta. In model 
B15$_\mathrm{X}$, where the $\beta$-decay energy input is highest (top right 
panel), the fingers or bubbles inflate more than in model B15, and the density 
contrasts between the interior and the walls of the finger borders are also 
more pronounced (compare the bubble borders of the elongated finger in the 
negative z- and x-directions). Regions rich in NiCoFeX expand more than 
NiCoFeX-poor 
regions. Therefore, the density of NiCoFeX smears out and becomes more uniform 
within the finger and the central bubble. Compare central and right bottom 
panels, where the density variations inside the NiCoFeX-rich regions decrease
significantly.

\begin{figure*}  
\includegraphics[width=.325\textwidth]{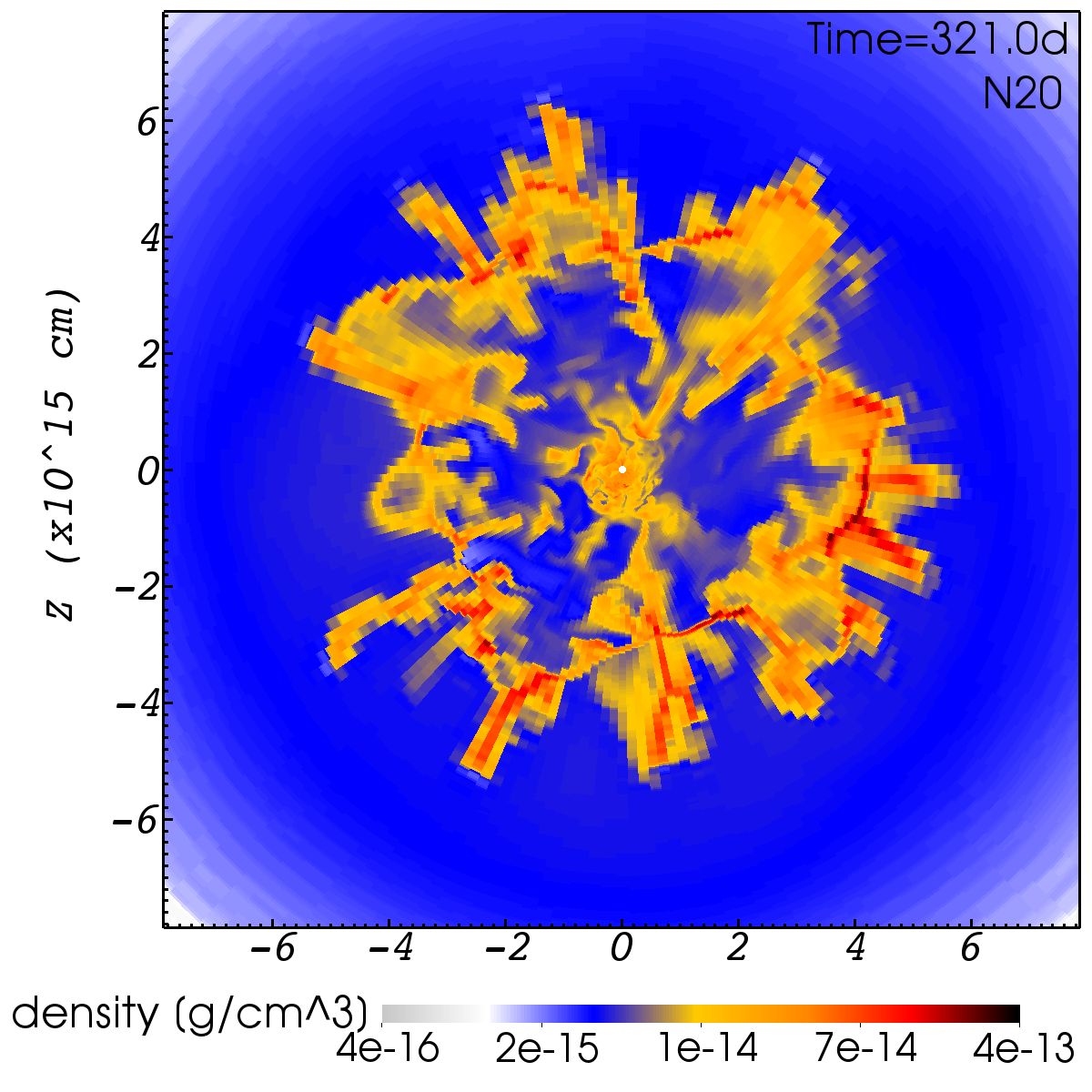}
\includegraphics[width=.325\textwidth]{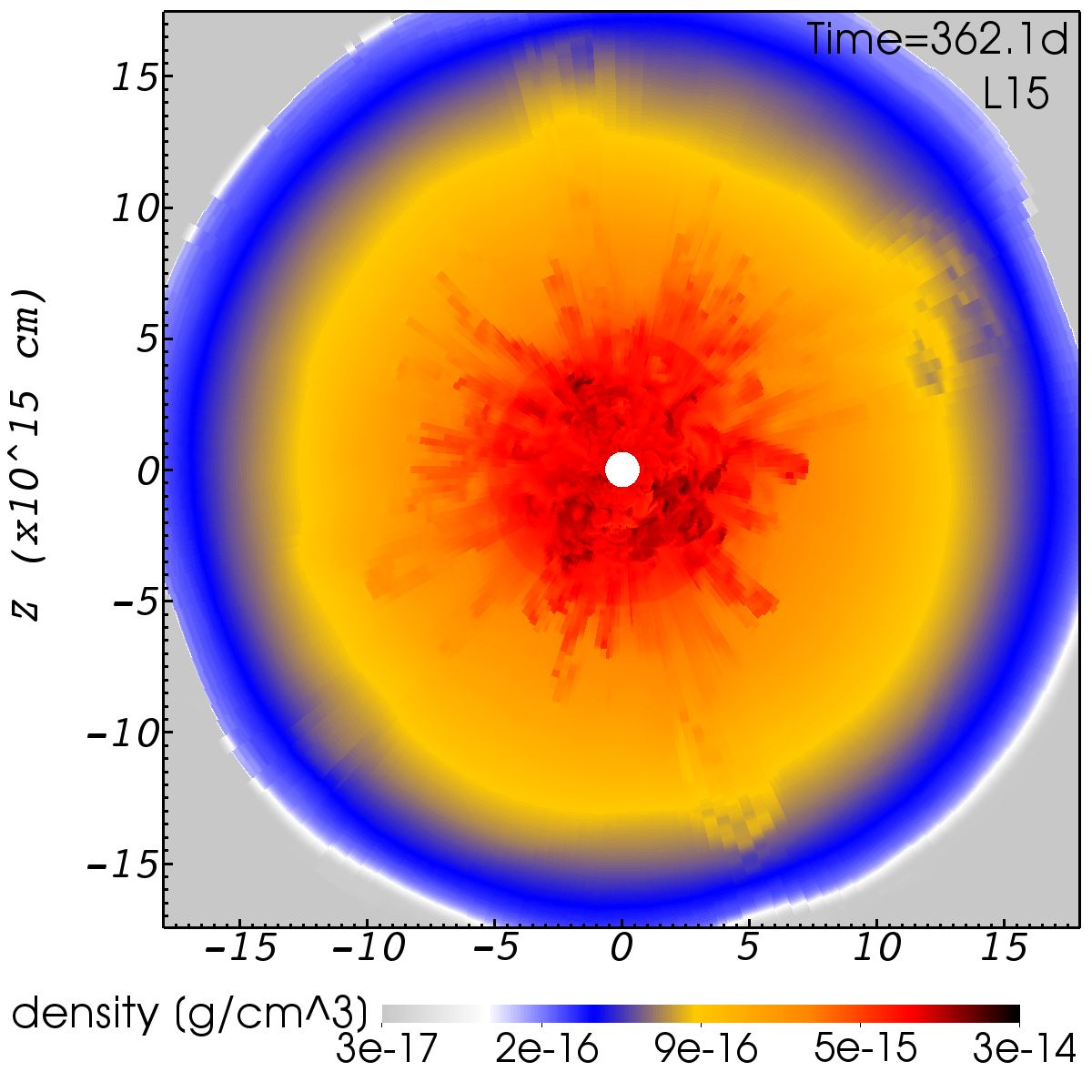}
\includegraphics[width=.325\textwidth]{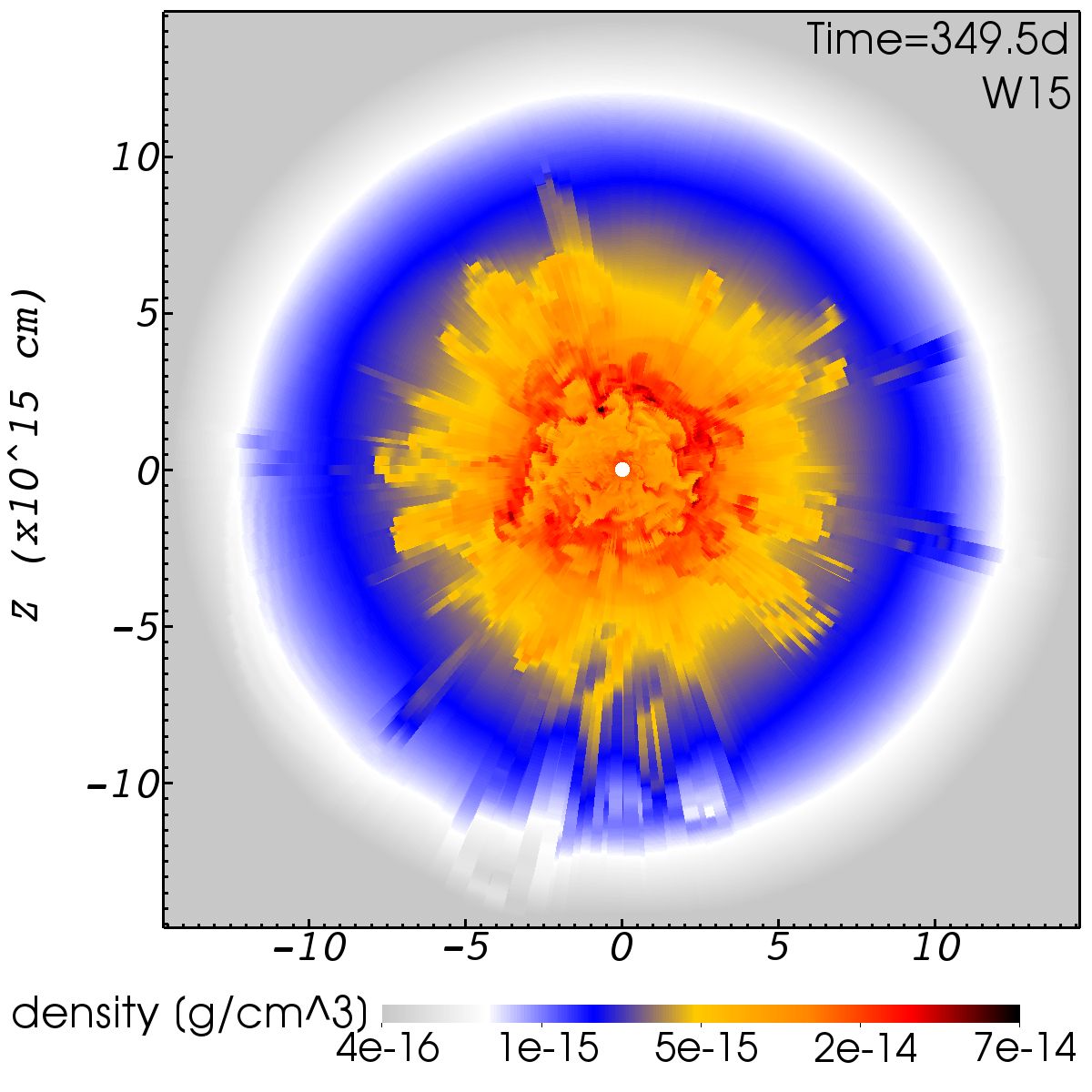}\\
\includegraphics[width=.325\textwidth]{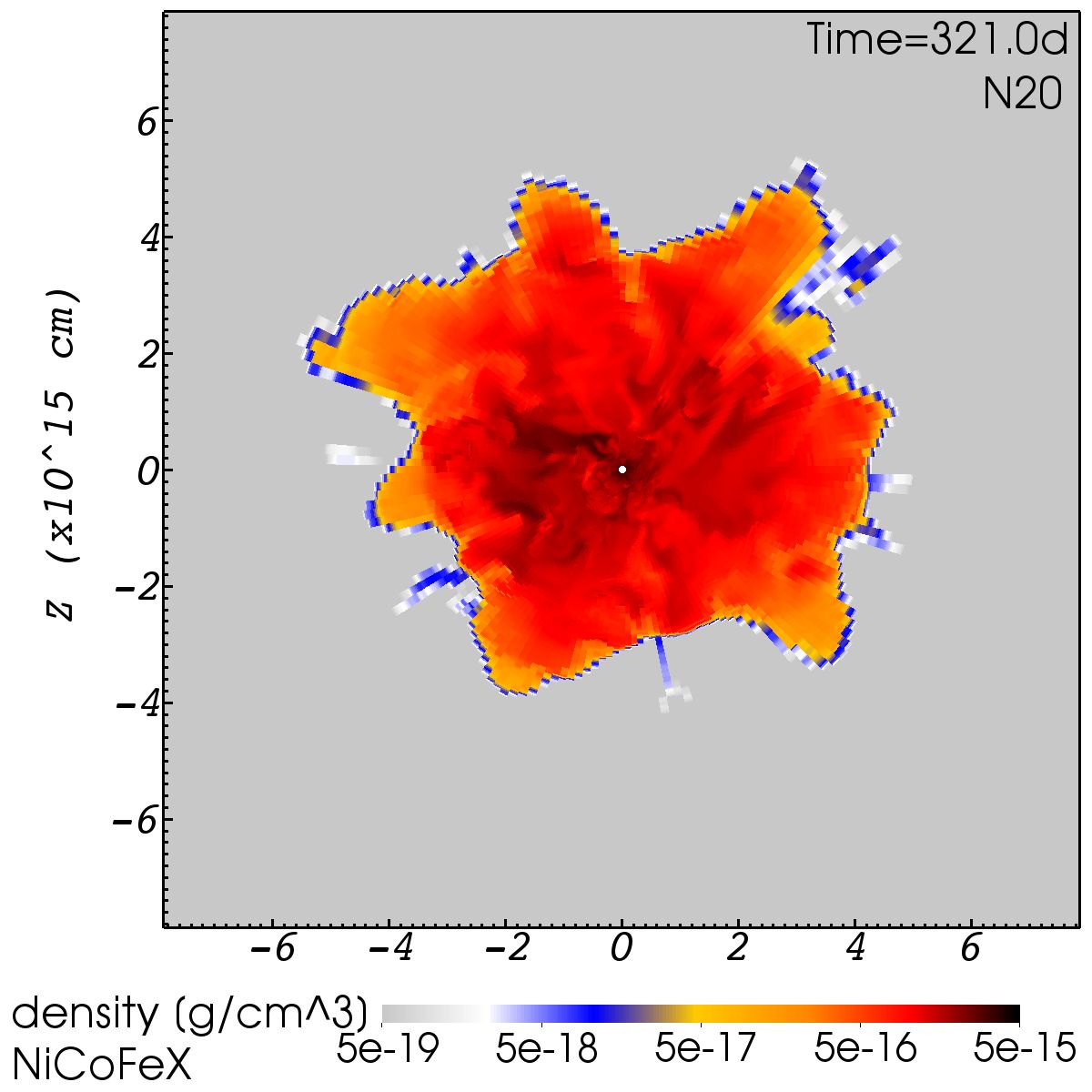}
\includegraphics[width=.325\textwidth]{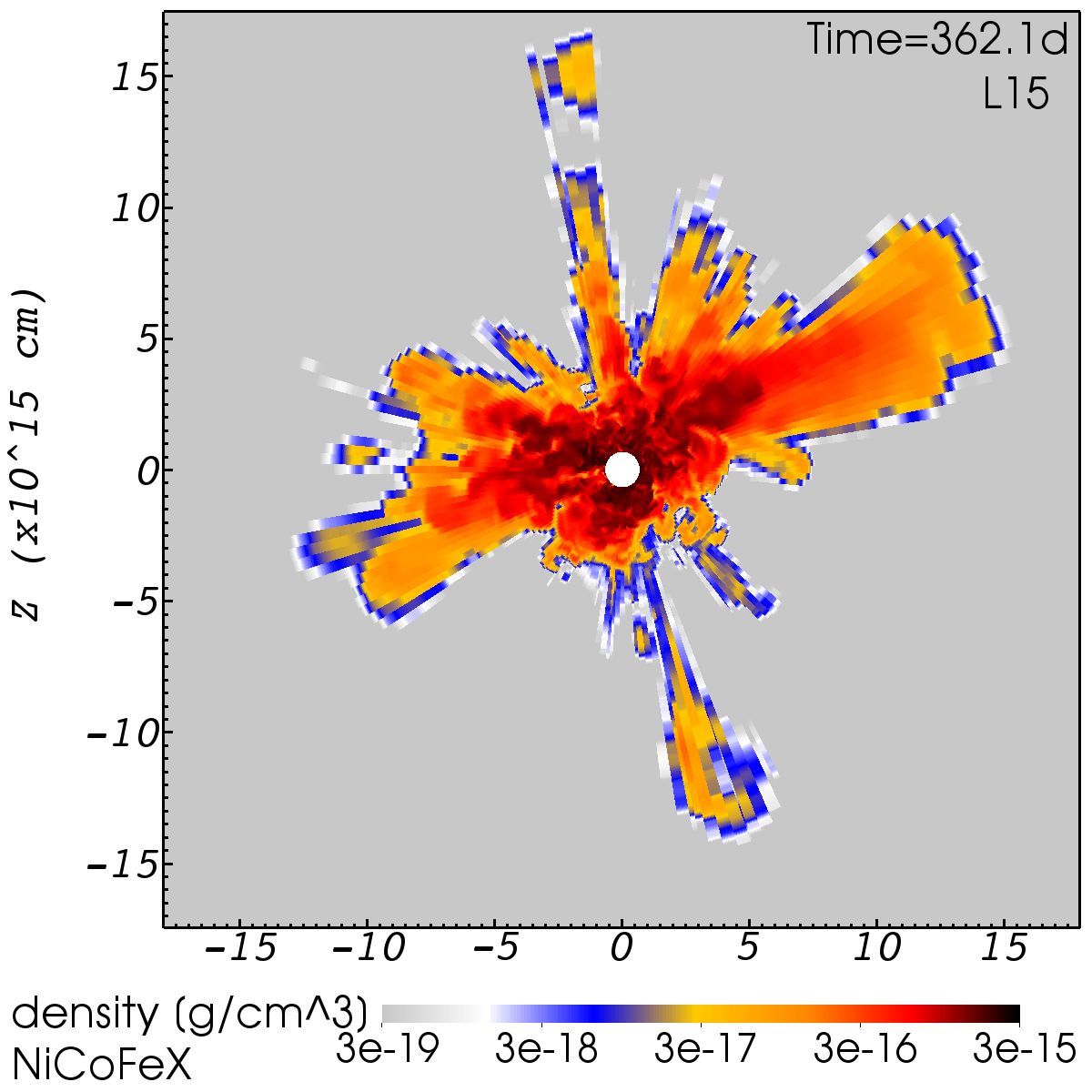}
\includegraphics[width=.325\textwidth]{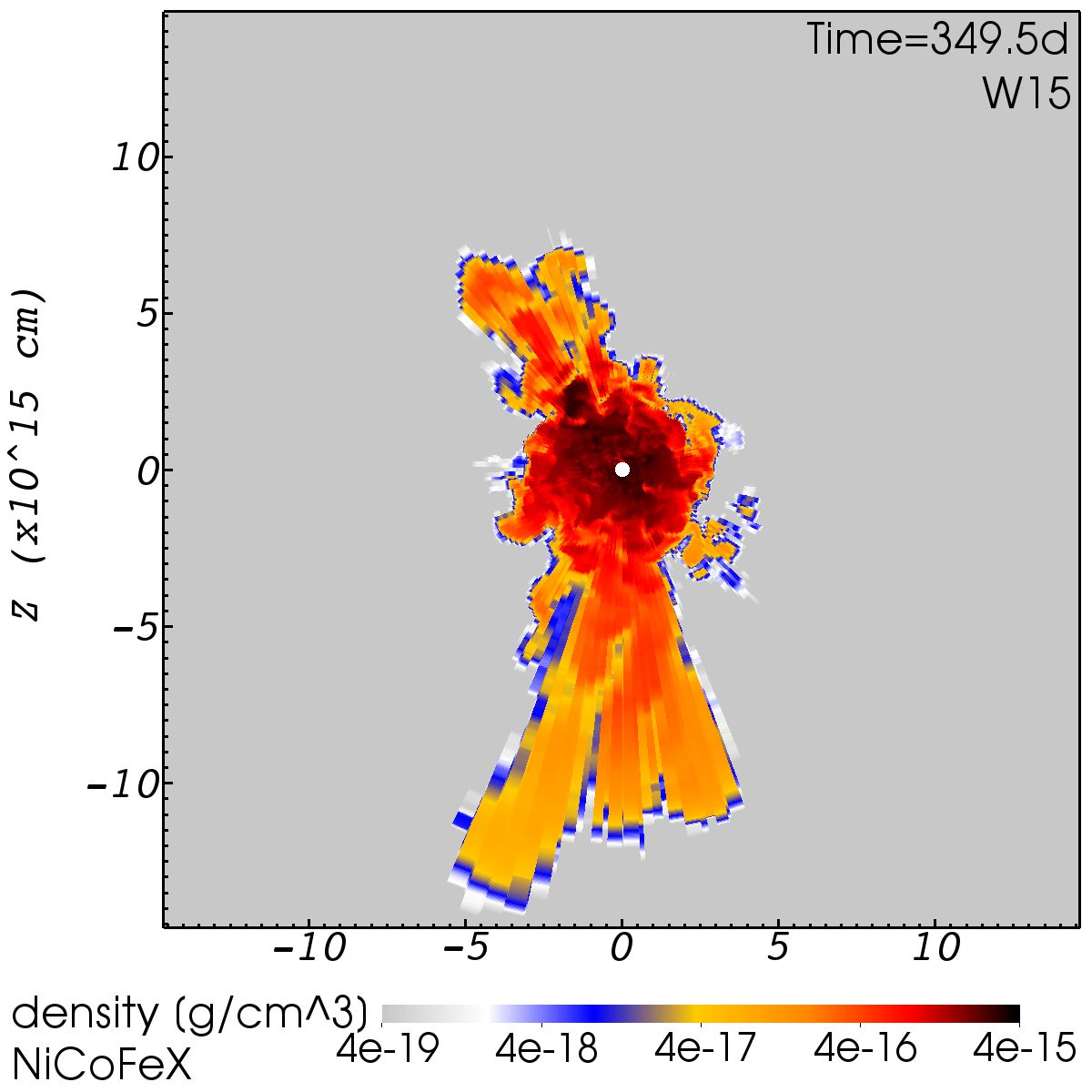}
\caption{{\it Top row}: Density slices of models N20 (left), L15 (centre), W15 
(right) in selected planes containing the most pronounced RT fingers at 
$t\sim1\,$yr. {\it Bottom row}: Slices of the density of NiCoFeX of the 
models in the same planes and at the same time. The density colour 
scale is logarithmic in all cases. In models W15 (top right) and L15 (central 
top panel)  underdense regions can be seen inside of inflated RT fingers 
(compare to the similar to results for model B15 
in Fig.\,\protect\ref{fig_slice_B15}), where the 
corresponding plots of the NiCoFeX densities (bottom right and central panel, 
respectively) display an extended finger. However, in the RSG models, the 
density contrast of these RT fingers compared to their surroundings is less 
pronounced than in model B15. The NiCoFeX-rich ejecta in model N20 are organized 
in a shell-like structure 
with very short RT fingers that do not inflate significantly. During the 
simulation, we cut out the central white region.}
\label{fig_slice_models}
\end{figure*}

Slices of the density distribution of models N20, W15, and L15 are given in 
Fig.\,\ref{fig_slice_models}. Model W15 (top right panel) is the most similar 
model compared to B15. However, the effect of the inflation of the 
NiCoFeX-rich fingers is not as apparent as in B15. See also 
Fig.\,\ref{fig_expansion}, where the rescaled NiCoFeX-rich 
volumes for model B15 roughly double from their minima until the end of the 
simulation, while the corresponding volumes for models W15 and L15 increase by 
at most $50\%$. The absolute volume of the NiCoFeX-rich matter is similar in 
all three models. For model W15, there are three grouped RT fingers in 
negative z-direction (bottom right panel in Fig.\,\ref{fig_slice_models}) that 
have higher density (orange) in the borders between them compared to their 
interiors. In general, the interior volumes are underdense (white) compared to 
the mean density (blue) at the same radius (top right panel). As for model B15 
and B15$_\mathrm{X}$, there is a NiCoFeX-rich 
bubble in the centre, which has lower densities than the surroundings, and 
the walls of this bubble are significantly overdense. The main reason for 
the weaker relative inflation of the RT fingers is that this model is 
already more extended at early times ($t\sim1\,$d) when the $\beta$ decay starts 
to become significant and that the initial structures occupy much larger 
volumes compared to model B15. The region where the $\beta$-decay energy is 
deposited is much less compact and, hence, the internal energy increase 
does not lead to a large growth of the structures in the two RSGs. At 
$t\sim1\,$yr, the sizes of the NiCoFeX-rich structures are similar to 
those in B15, because the fingers and the bubble of model B15 have inflated 
more (see also Fig.\,\ref{fig_expansion}).

The density distribution of model L15 (central panels of 
Fig.\,\ref{fig_slice_models}) is very similar to model W15. As in the 
other models, NiCoFeX-rich regions have slightly lower local densities compared 
to their surroundings as can be seen in particular for the NiCoFeX-fingers in 
all directions, top, bottom, left and right in the bottom central panel. 
However, in model L15, the contrast in $\rho^\mathrm{tot}$ between the 
inner NiCoFeX-rich bubble or the fingers on one side and the corresponding 
borders on the other side is less pronounced than in W15, and much less than in 
B15. The RT fingers in model L15 are almost invisible in the top, central panel 
of Fig.\,\ref{fig_slice_models}, and are hardly visible as the white bubbles in 
the blue shell in model W15 in the top, right panel of the same figure. 
However, the fingers are clearly visible in model B15 in the top central panel 
of Fig.\,\ref{fig_slice_B15}. The weaker density contrast 
between the finger boundaries and their interiors in the RSGs is related to 
 the significantly lower density in the H-envelope of the RSGs compared to 
that in the BSGs \citep[see figures 1 and 2 in][]{Wongwathanarat2015}. 
Therefore, the entropy and pressure in the envelopes of the BSGs after the 
passage of the forward shock is lower than those in the RSGs. This lower 
ambient pressure facilitates the expansion of the NiCoFeX-rich fingers in the 
case of model B15. In addition, the expansion into the denser H-envelope of the 
BSG sweeps up more mass than in the more dilute H-envelopes of the RSGs. 
Consequently, the stronger expansion into a denser environment leads to stronger 
density contrasts for model B15 compared to the RSGs.

Model N20 is very different from all other models. First, the reverse 
shock from the He/H-interface and its self-reflected shock are weaker than in 
the other models (see left panel in Fig.\,\ref{fig_vel_models}). Therefore, 
the central ejecta are almost not decelerated by this reverse shock. In 
addition, they are only mildly reaccelerated almost exclusively by the energy 
input due to $\beta$ decay. This leads to a thin dense, corrugated shell 
of swept-up material with a radius of $r\sim4\times10^{15}\,$cm as shown in the 
top left panel of Fig.\,\ref{fig_slice_models}. This shell is strongly 
fragmented into many RT fingers, which are significantly smaller than the 
extended fingers of the other models. Comparing the top and the bottom left 
panel of Fig.\,\ref{fig_slice_models}, we see that most of the NiCoFeX is 
enclosed by the dense, corrugated shell. Since the shell is not as much affected 
by the RTIs as the other models \citep[see][for a detailed 
discussion]{Wongwathanarat2015}, there is no significant mixing of $^{56}$Ni 
into the small fingers and the latter do not contain sufficient $^{56}$Ni to 
power significant inflation by radioactive decay. Most of 
the $\beta$-decay energy is released in the 
central bubble, leading to a more spherical expansion compared to the other 
models.

\subsubsection{Inflation of NiCoFe-rich fingers}

\begin{figure*}  
\includegraphics[width=.32\textwidth]{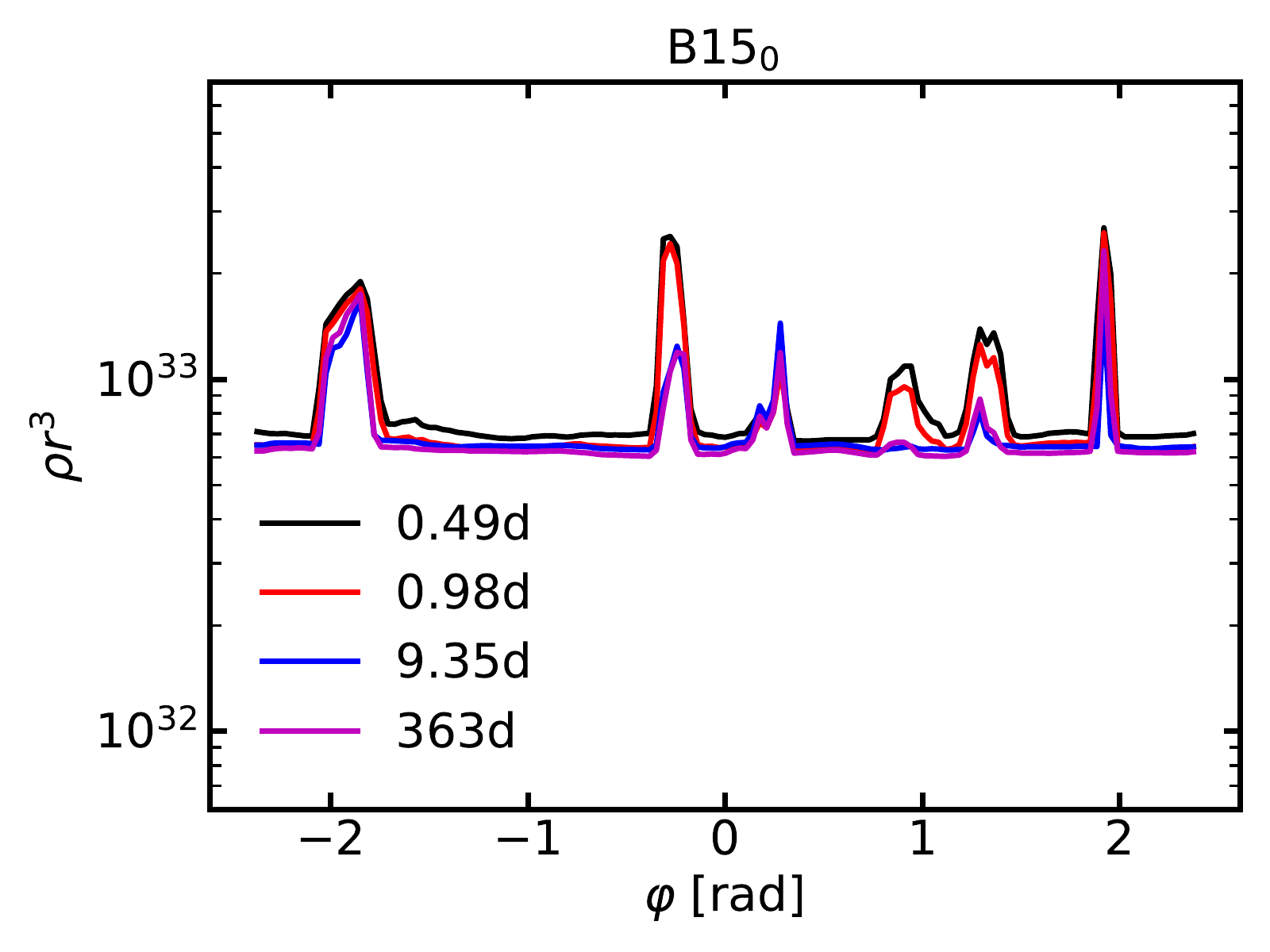}
\includegraphics[width=.32\textwidth]{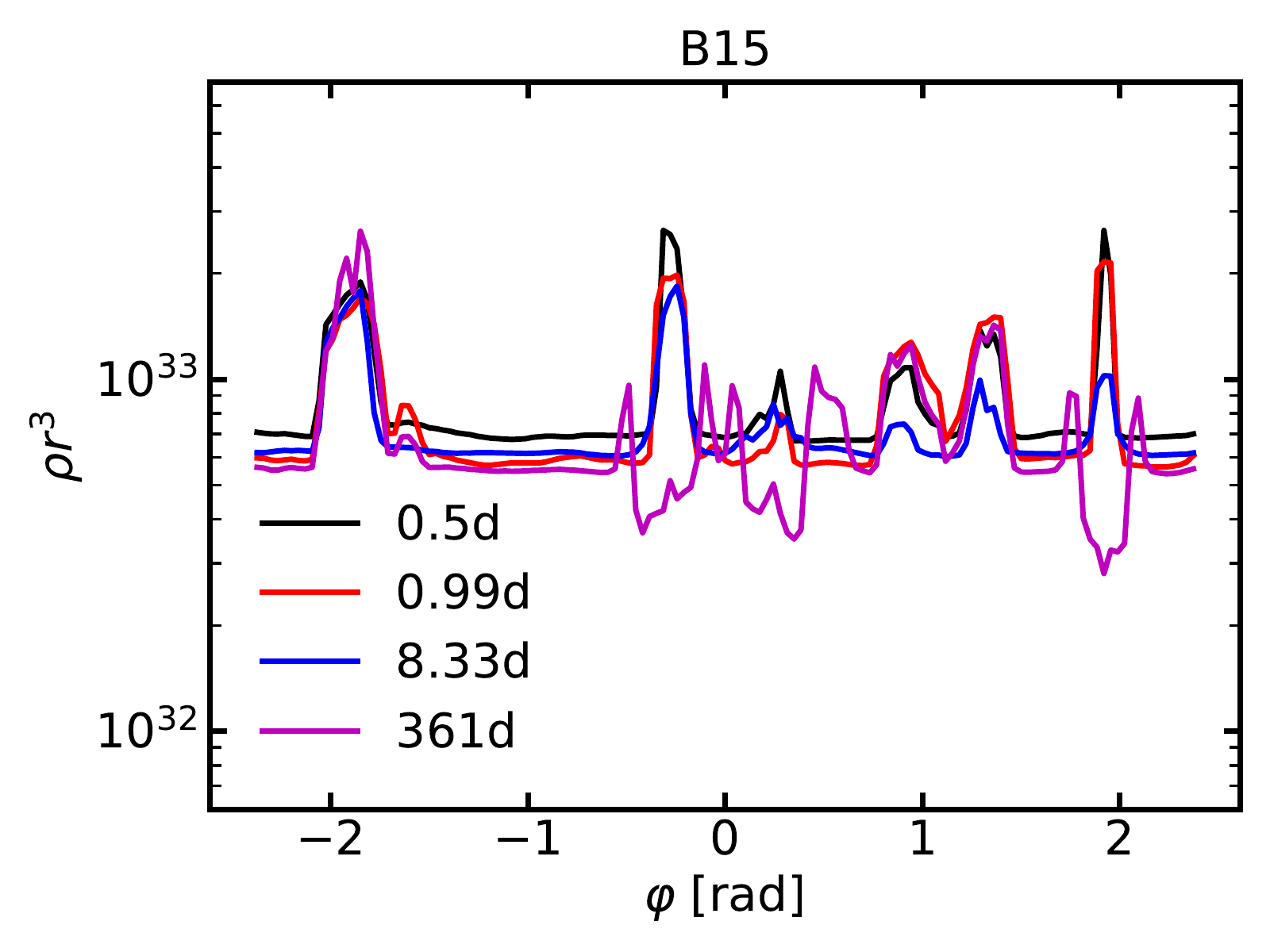}
\includegraphics[width=.32\textwidth]{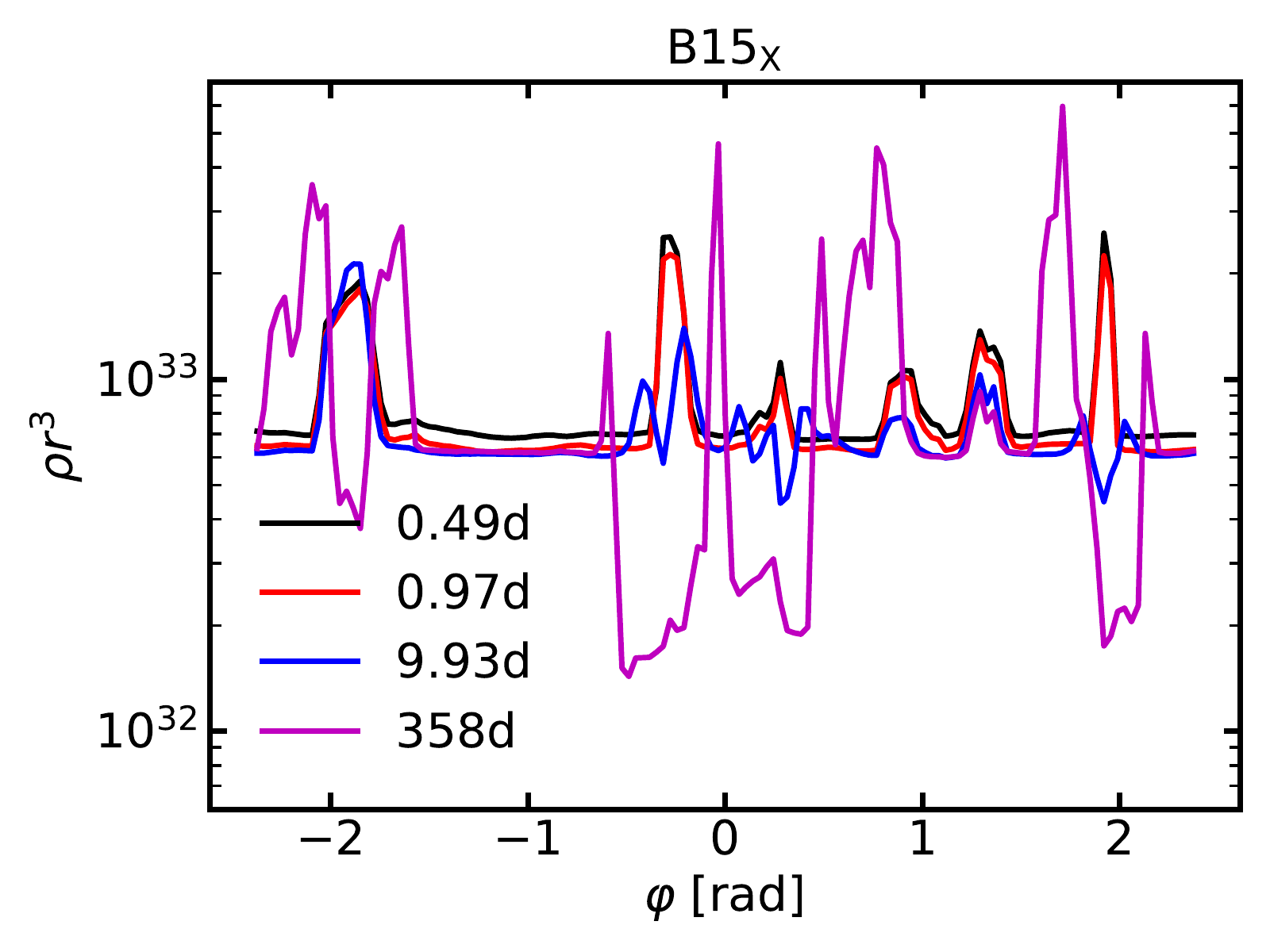}\\
\includegraphics[width=.32\textwidth]{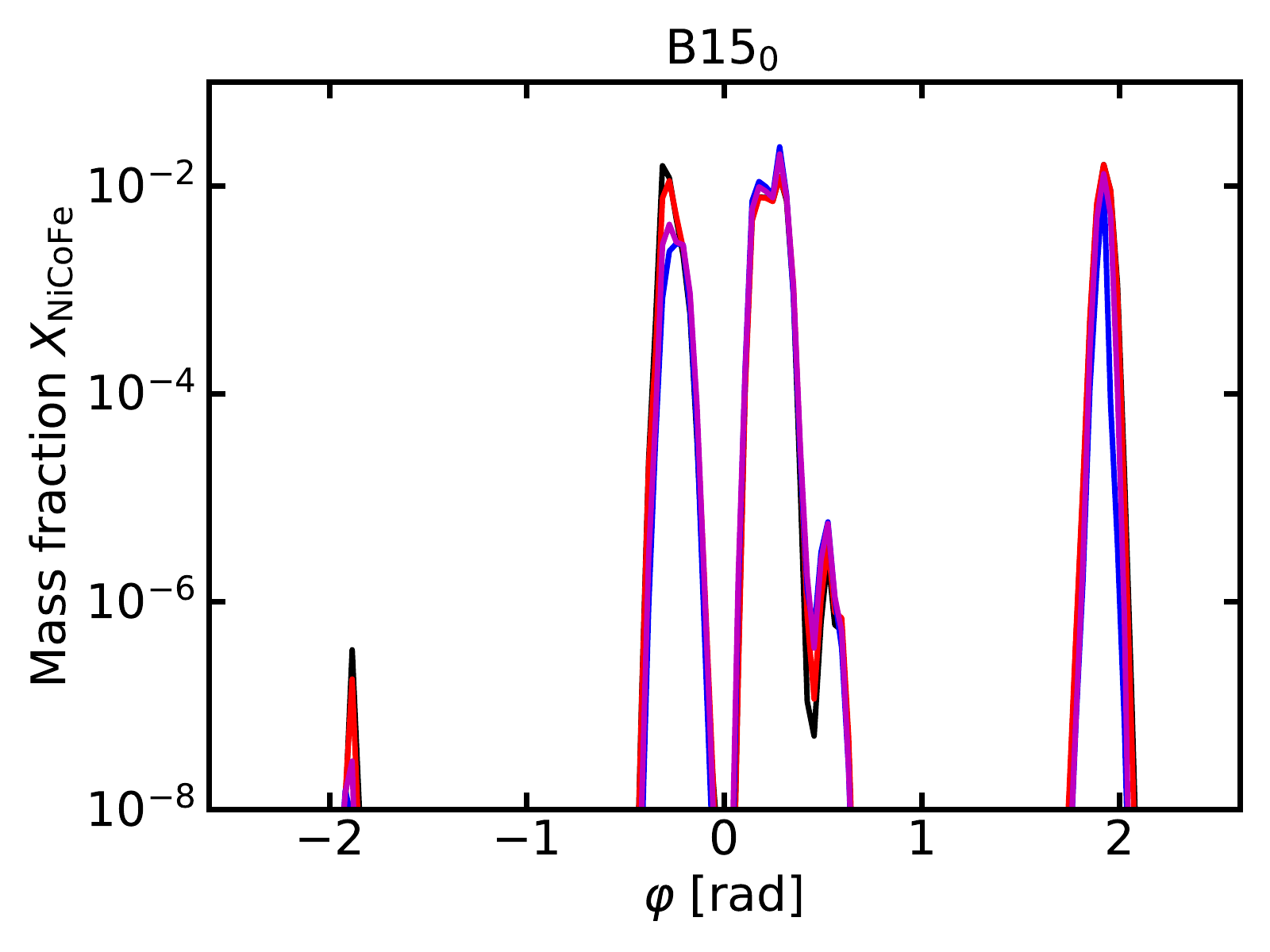}
\includegraphics[width=.32\textwidth]{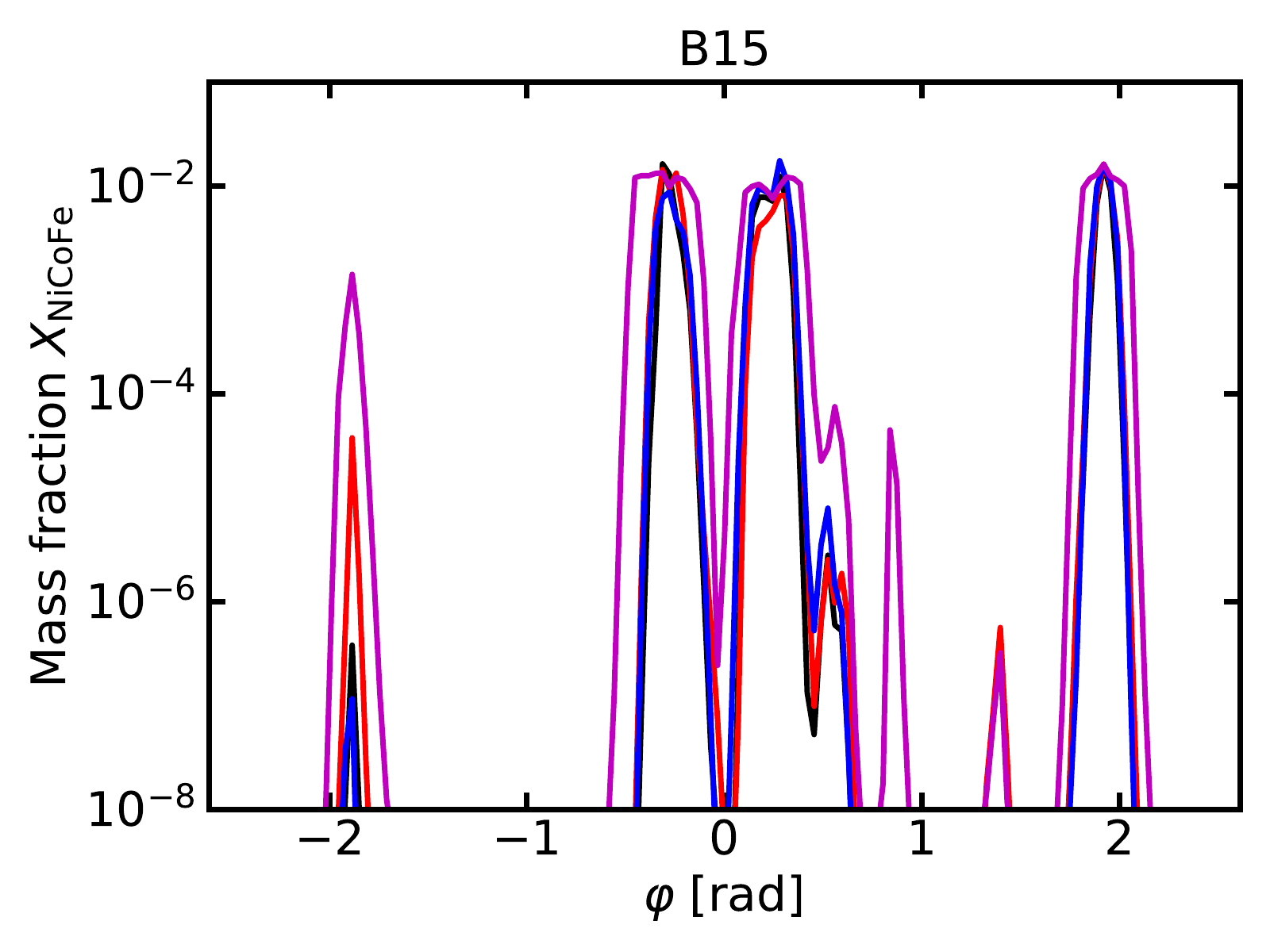}
\includegraphics[width=.32\textwidth]{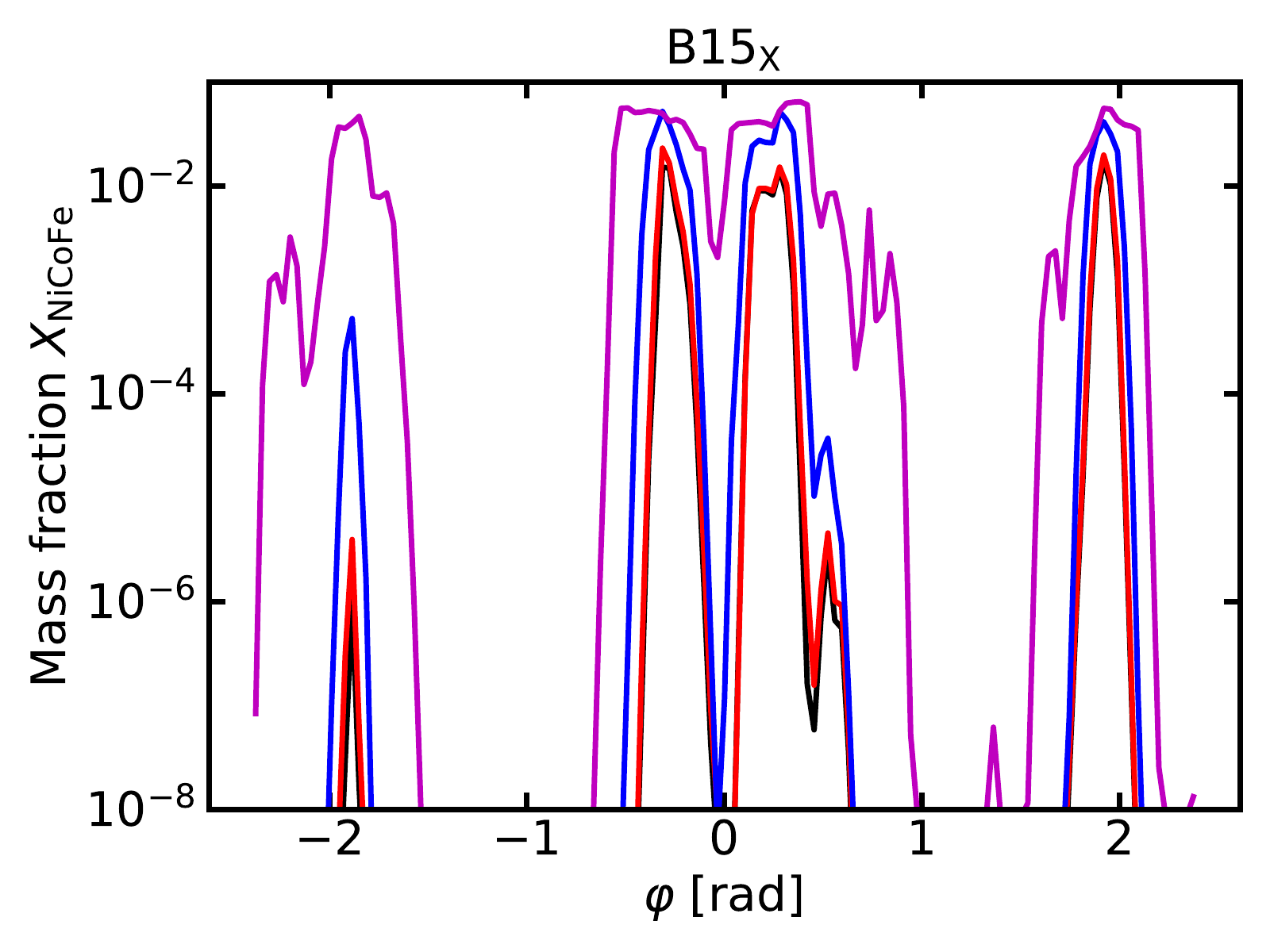}
\caption{{\it Top row}: Density rescaled by the time-evolving $r^3$ of 
initially overdense fingers in models B15$_0$,  
B15, and B15$_\mathrm{X}$ at different times as indicated in the legend. 
Without $\beta$ decay, the density enhancements stay approximately at the same 
angular positions with the same angular extent (left panel), while the $\beta$ 
decay leads to an inflation of the initially overdense regions, which become 
underdense (central panel). If there is more radioactive material (top right 
panel, B15$_\mathrm{X}$), the densities in the underdense regions decrease 
further. At the borders of these regions material gets compressed into thin 
`walls'. {\it Bottom row}: Mass fractions $X_\mathrm{NiCoFe}$. The overdense 
regions (compare with top row) that initially contain significant amounts of 
$^{56}$Ni inflate, while overdense regions initially poor in $^{56}$Ni do not 
change significantly.}
\label{fig_bubbles1D}
\end{figure*}

As discussed in Section\,\ref{sec_beta}, $Vr^{-3}$ or $Vt^{-3}$ should be a 
constant for homologously expanding ejecta. Equivalently, this holds for $\rho 
r^3$. Therefore, we plot the density rescaled by $r^3$ as a function of 
$\varphi$ at given $r$ and $\theta$ for different times and for the three 
models B15$_0$, B15, and B15$_\mathrm{X}$ in the top row of panels in 
Fig.\,\ref{fig_bubbles1D}. At the beginning of the simulation at $t=0.0365\,$d, 
we choose $r=9.7\times10^{11}\,$cm and $\theta=0.578\pi$, which is outside of 
the central bubble. We follow the evolution of this location in the flow by 
integrating the radius in time with the local fluid velocity. We chose this 
initial location because here we see a few of the fastest $^{56}$Ni fingers as 
local density enhancements like at $\varphi\sim[-2.0, -0.2, 0.2, 2.0]$ for 
$t\sim0.5\,$d. These density enhancements are related to the NiCoFe-rich 
fingers. In the bottom panels of the figure, we show the corresponding mass 
fractions of $^{56}$Ni and its decay products. High densities usually appear 
where $X_\mathrm{NiCoFe} \gtrsim5\times10^{-3}$. There are some exceptions 
around $\varphi\sim-2$, $\varphi\sim0.9$, and $\varphi\sim1.1$ where the 
displayed part of the fingers is not $^{56}$Ni enriched initially. In these 
cases, $^{56}$Ni is located further inside the respective finger. The evolution 
of the model without $\beta$ decay is almost homologous from the beginning, 
i.e. $\rho r^3$ is constant and existing structures do not change during the 
evolution (see top and bottom left panels in Fig.\,\ref{fig_bubbles1D}). 
Differences between different time steps mainly arise due to the inaccurate 
time integration of the reference radius. The corresponding integration 
was done as a post processing and, thus, the time stepping was limited to the 
output times of the 3D simulation. In contrast, the models including $\beta$ 
decay show a significant change of their structures. For model B15 (central 
panels in Fig.\,\ref{fig_bubbles1D}), the initial high densities at 
$t=0.5\,$d (black curve)  at $\varphi\sim-0.2$, $\varphi\sim0.2$, and 
$\varphi\sim2.0$ decrease slowly until day $10$ (red and blue curves) and even 
turn into underdense regions at $t\sim1\,$yr.  All this occurs at the highest 
mass fraction $X_\mathrm{NiCoFe}$ (compare bottom central panel of 
Fig.\,\ref{fig_bubbles1D}). The energy input in form of decay energy is used to 
do $pdV$ work on the exterior of the initially $^{56}$Ni-rich fingers which 
inflate. Consequently, the density inside the fingers decreases with respect to 
the surroundings, and the inter-finger, $^{56}$Ni-poor regions get compressed 
into thin filaments with high densities, which we call finger walls or borders. 
During this compression NiCoFe gets also mixed in the border region. Other 
initial overdense regions without significant amounts of $^{56}$Ni 
($\varphi\sim[-2.0,0.9,1.1]$) do not change significantly. However, the mass 
fraction of $^{56}$Ni at $\varphi\sim-2.0$ increases for this model with time. 
This is material, which is accelerated (also radially) due to the $\beta$ decay, 
and, thus, penetrates the previously $^{56}$Ni poor, overdense region. All the 
described effects are even more pronounced in model B15$_\mathrm{X}$ (right 
panels). Here, the maximum density contrast with respect to the mean density 
$\rho r^3\sim 6\times10^{32}\,$g at $t\sim1\,$yr is a factor $~10$ in the 
overdense regions and $1/3$ in the underdense regions, resulting in density 
enhancements of up to more than one order of magnitude between the interior and 
the wall.

\subsubsection{3D structures of NiCoFeX-rich fingers}

\begin{figure*}  
\centering
\includegraphics[width=.4\textwidth]{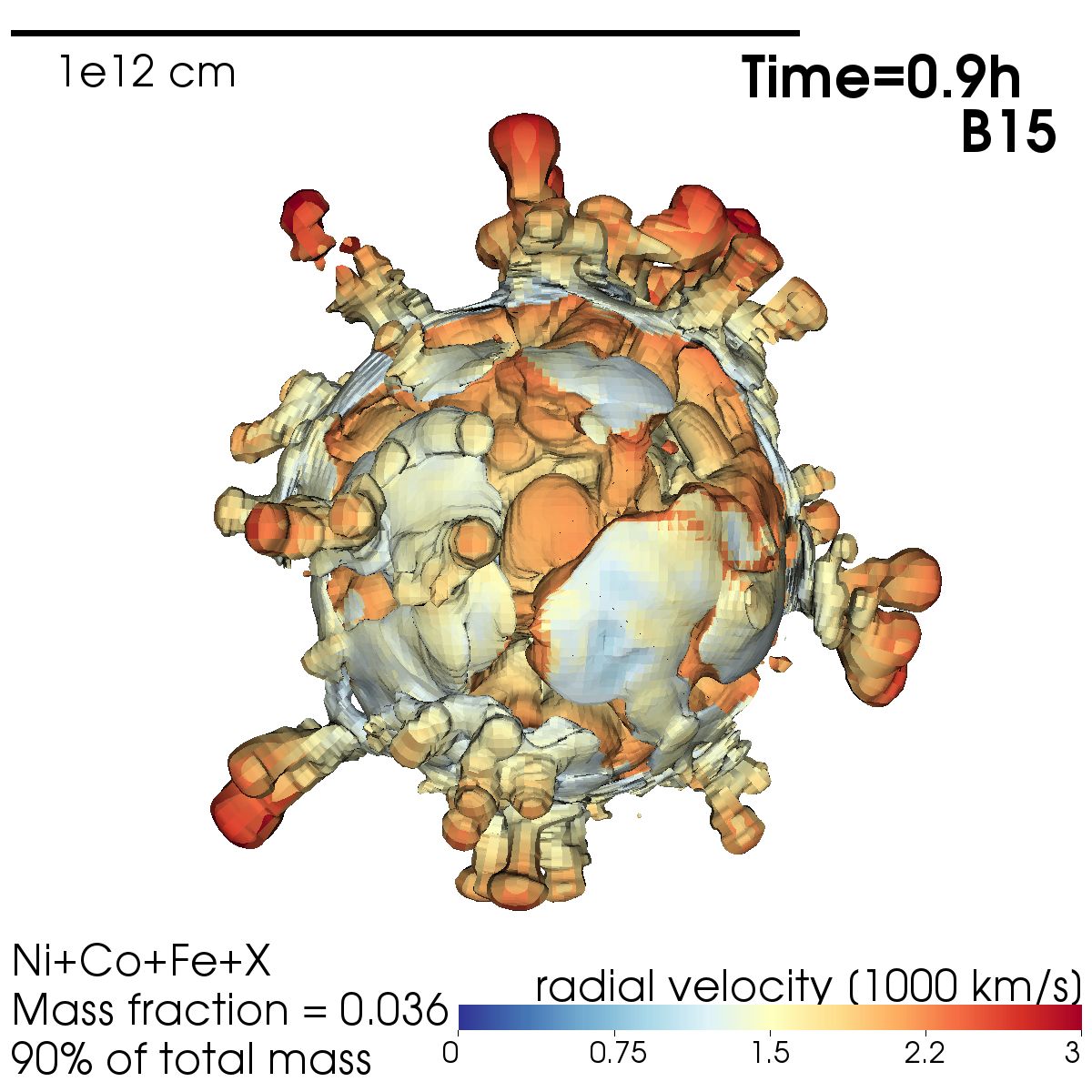}
\includegraphics[width=.4\textwidth]{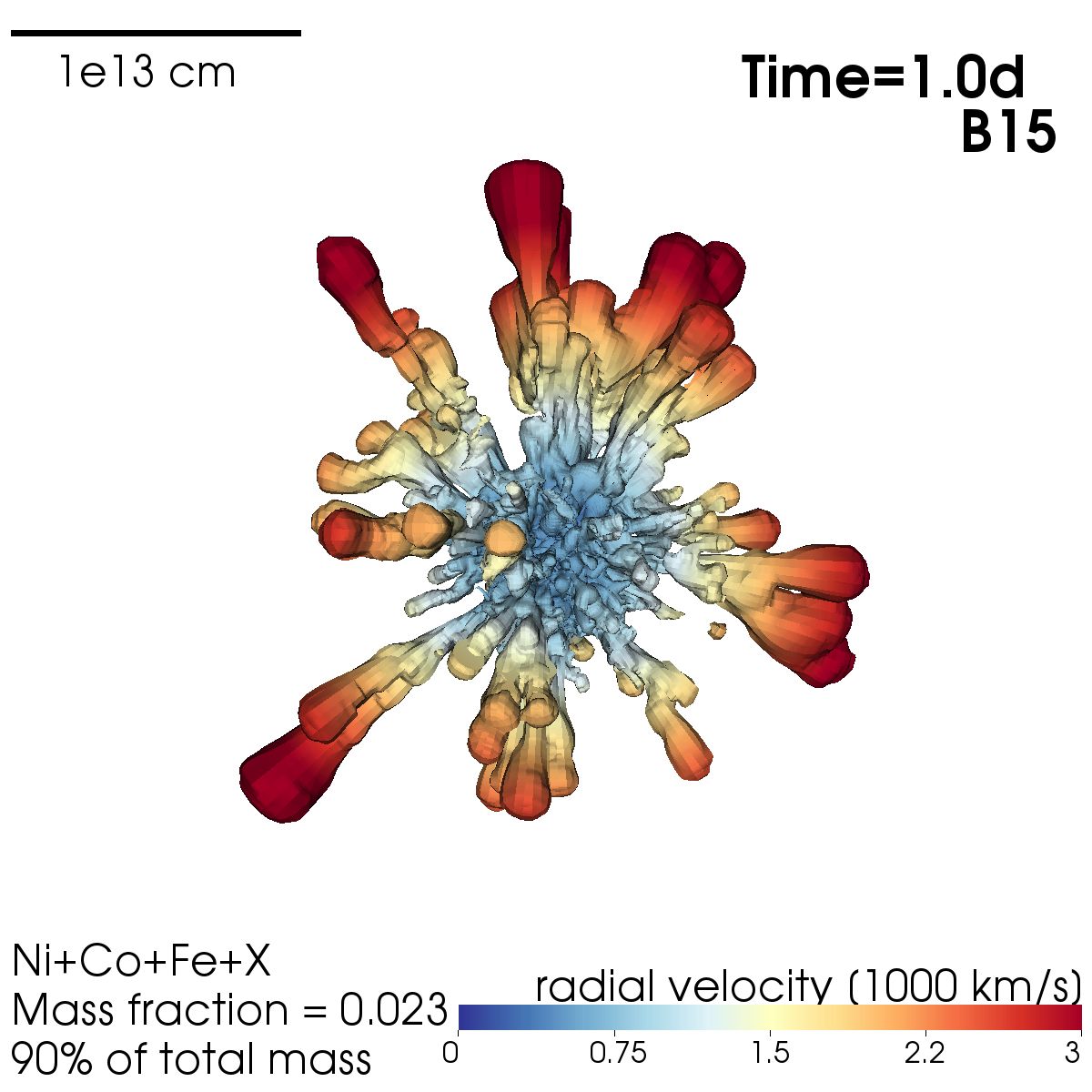}\\
\includegraphics[width=.4\textwidth]{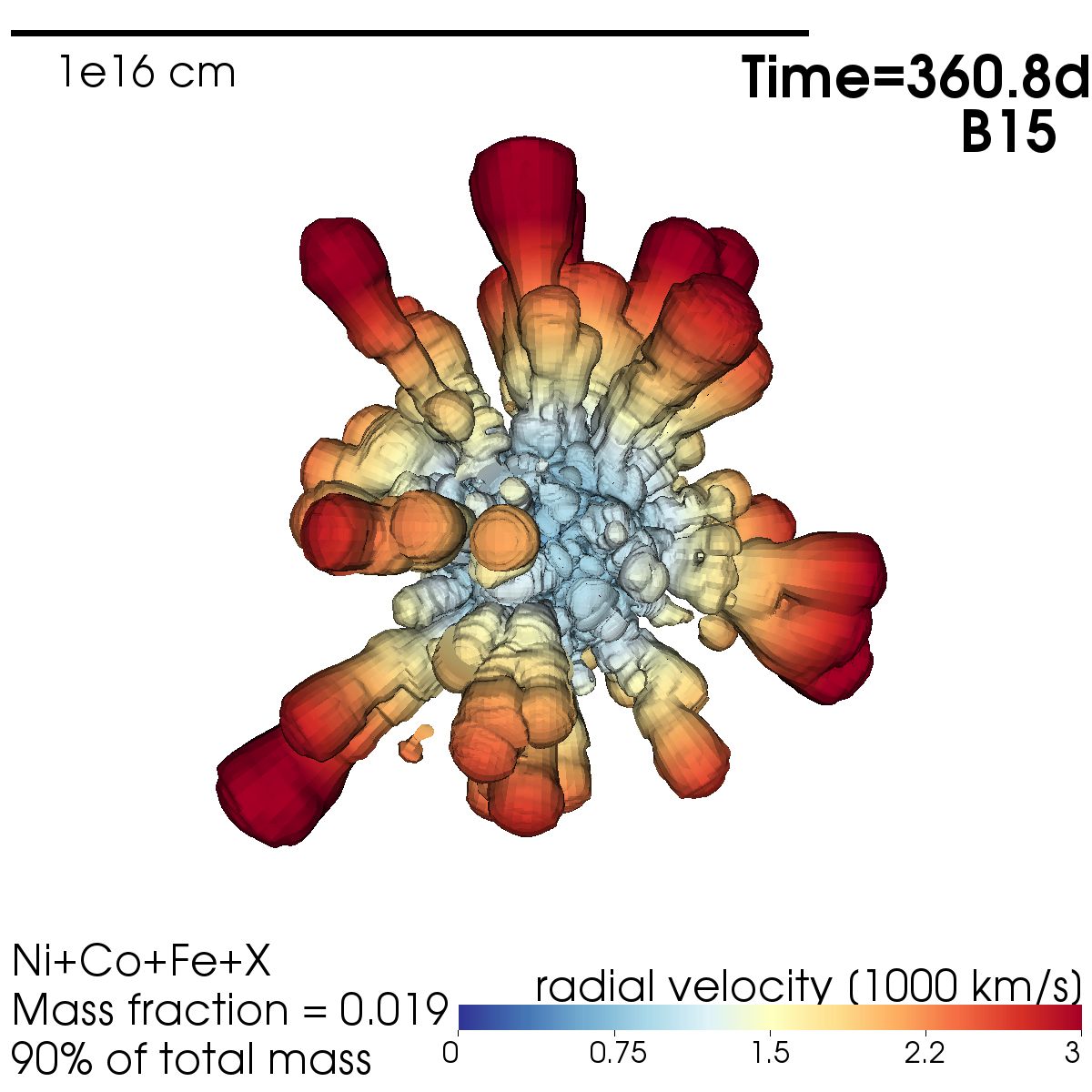}
\includegraphics[width=.4\textwidth]{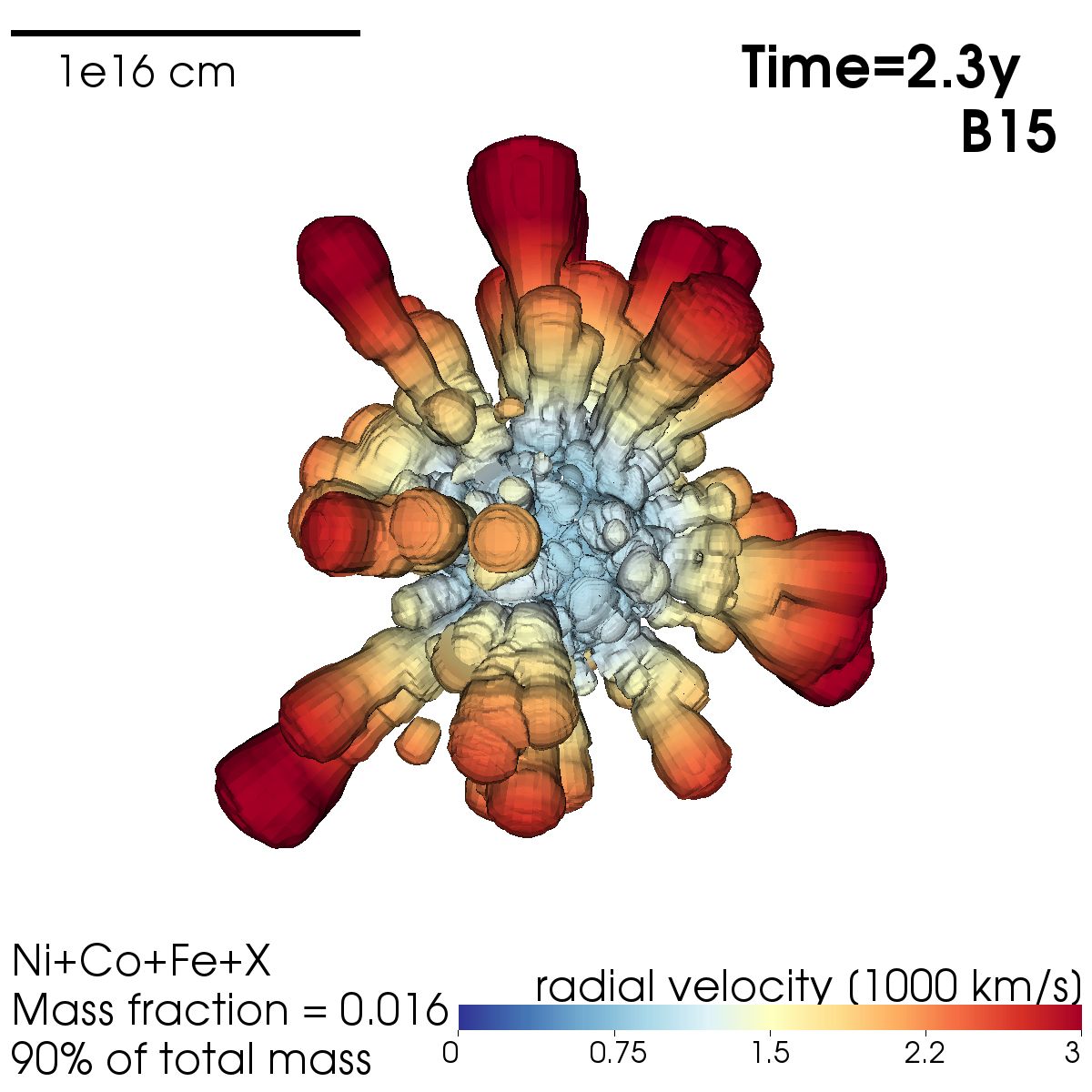}
\caption{Isosurface plots of constant mass fraction containing $90\%$ of the 
mass of NiCoFeX for different times of model B15. After 
the reverse shock (spherical shell in the top, left panel, $t\sim0.9\,$h) 
retreats to the centre and the initial RTIs grow until $t\sim1\,$d 
(top, right panel), the fingers start to inflate due to $\beta$-decay energy 
input (bottom, left panel, $t\sim1\,$yr). After $t\lesssim1\,$yr the inflation 
stalls and the structures do not change significantly. The NiCoFeX mass 
fractions defining the isosurfaces are indicated in each panel.}
\label{fig_iso_B15}
\end{figure*}

In this section we investigate the 3D structures for all of our 
models and how they change with time. Following \cite{Wongwathanarat2015}, 
we use the iron-group elements around $A=56$ like $^{56}$Ni, $^{56}$Co, 
$^{56}$Fe and X (NiCoFeX) to determine the surface of the structures that 
characterize metal mixing from the SN centre into the outer shells of the star. 
To define the isosurfaces, we sum up the mass of all cells with the 
highest mass fractions $X_\mathrm{NiCoFeX}$ until we reach a certain percentage 
of the total mass of NiCoFeX. The mass fraction that encloses all this mass 
defines our isosurface. The corresponding mass fractions are indicated in each 
plot. For example in Fig.\,\ref{fig_iso_B15}, the isosurfaces containing  
$90\%$ of the mass of NiCoFeX in cells that have the highest mass fractions 
are 
plotted. The remaining $10\%$ of the NiCoFeX is contained in ejecta with lower 
NiCoFeX mass fractions. Note that the magnitude of $X_\mathrm{NiCoFeX}$ 
defining the isosurfaces decreases with time. Due to the expansion and the 
related mixing of the matter, $X_\mathrm{NiCoFeX}$ decreases in 
particular in the outermost layers of the fingers.

The top panel of Fig.\,\ref{fig_iso_B15} shows NiCoFeX-rich structures before 
shock breakout. The reverse shock is visible as the spherical shell which is 
penetrated by some faster NiCoFeX-rich fingers. The reverse shock slows down 
the central ejecta compared to the extended RT fingers and leads to an apparent 
contraction of the central part compared to homologous expansion. Note that the 
scale of the plots at different snapshots increases linearly with time and, 
thus, follows homologous expansion. The fingers become more prominent at 
$t\sim1\,$d (second panel). Then the $\beta$-decay energy input becomes 
significant and the thin, 
elongated NiCoFeX-fingers inflate. Some even merge to larger structures 
(third panel, $t\sim1\,$yr). In addition, the central ejecta also inflate, 
leading to a larger central bubble (compare innermost regions in the second and 
third panel). This inflation is caused by the self-reflected reverse shock and 
also by the $\beta$-decay energy input. In the bottom panel, we show our 
model at 
the latest time simulated $t=2.3\,$yr. There is almost no change in the 
structures compared to $t\sim1\,$yr. However, the threshold for the mass 
fraction $X_\mathrm{NiCoFeX}$ is a bit lower than for the earlier time, because 
we cut some material with higher densities in the centre and because there is 
still some inflation of the NiCoFeX-rich ejecta.

\begin{figure*}  
\centering
\includegraphics[width=.33\textwidth]{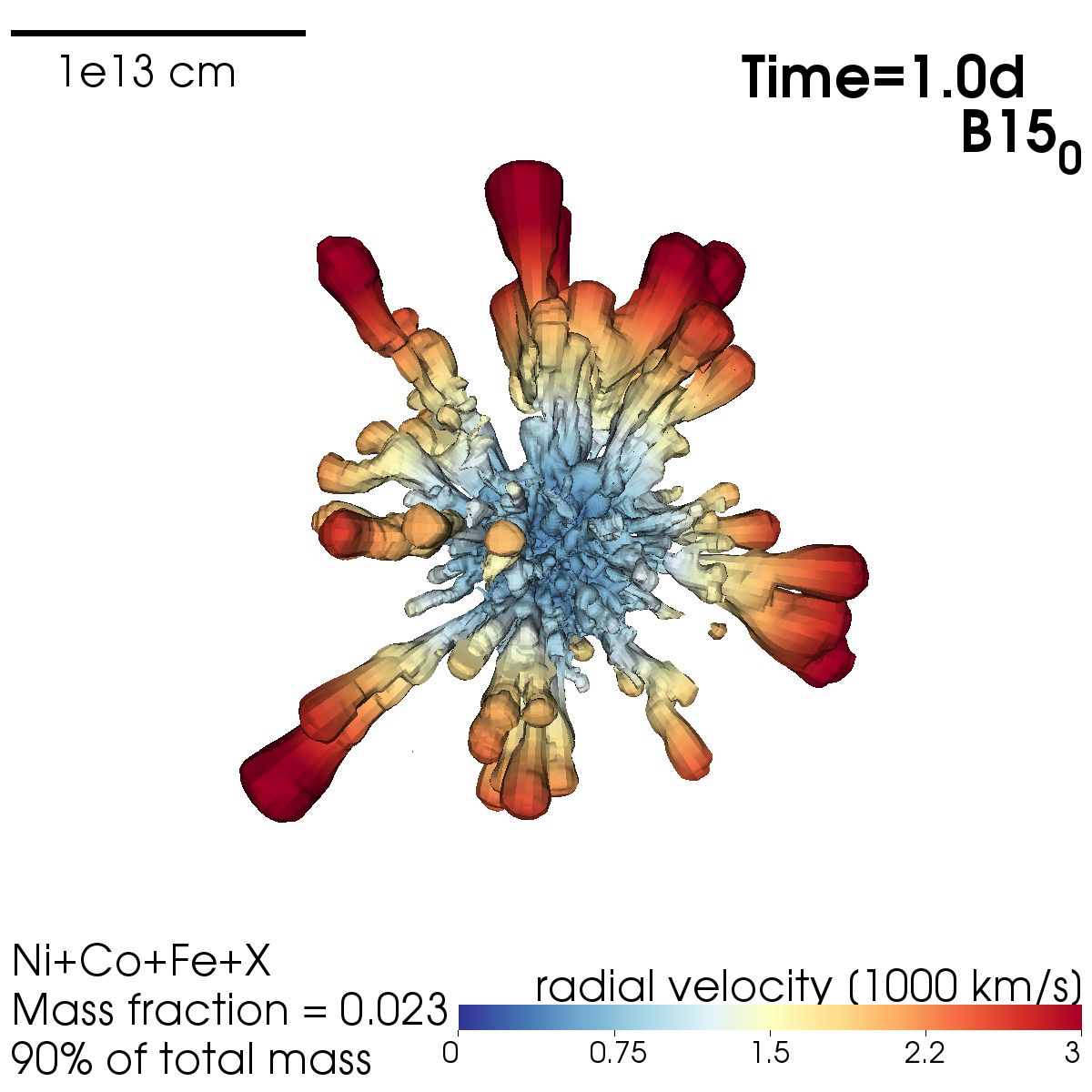}
\includegraphics[width=.33\textwidth]{iso/B15_Ni56_09_1.jpeg}
\includegraphics[width=.33\textwidth]{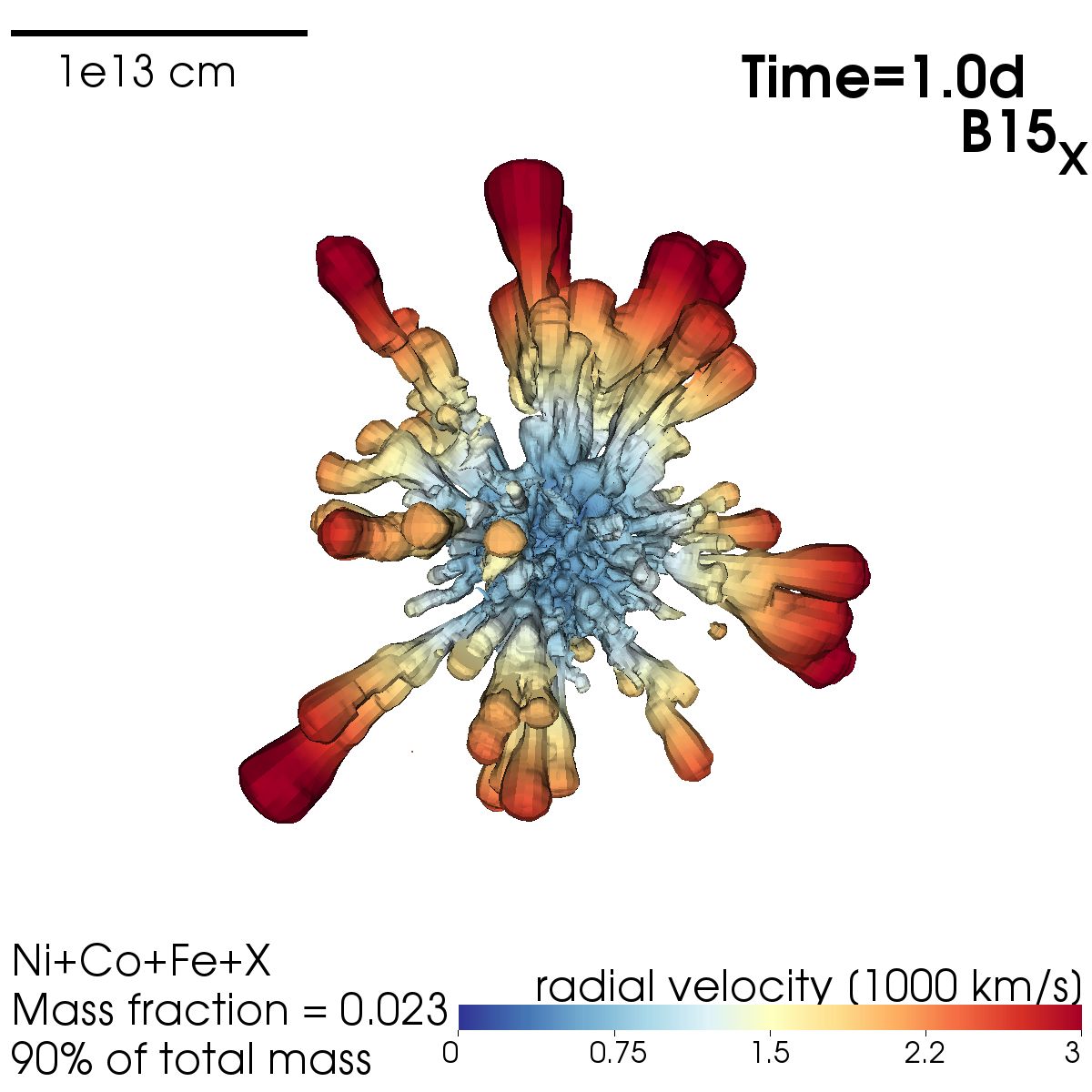}\\
\includegraphics[width=.33\textwidth]{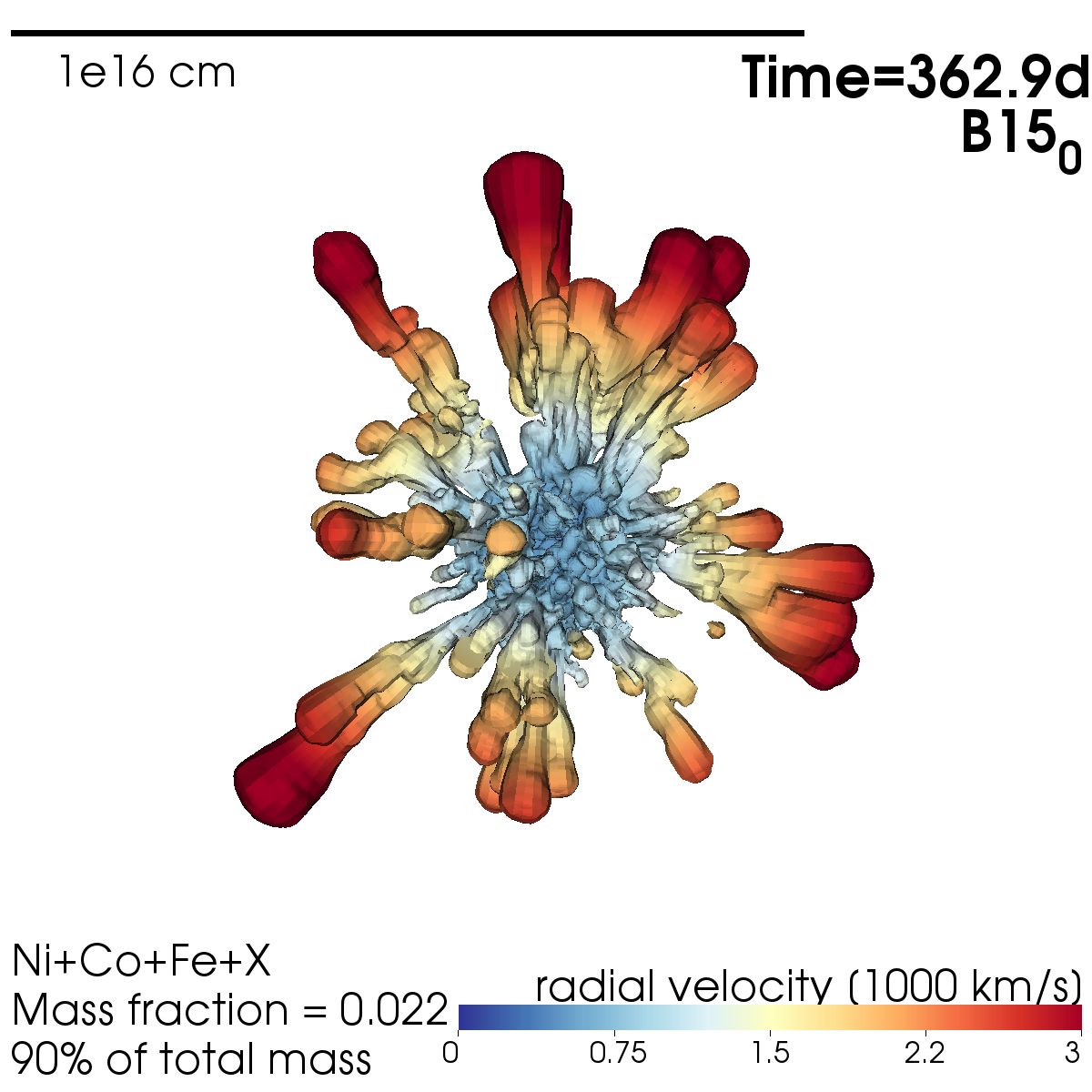}
\includegraphics[width=.33\textwidth]{iso/B15_Ni56_09_360.jpeg}
\includegraphics[width=.33\textwidth]{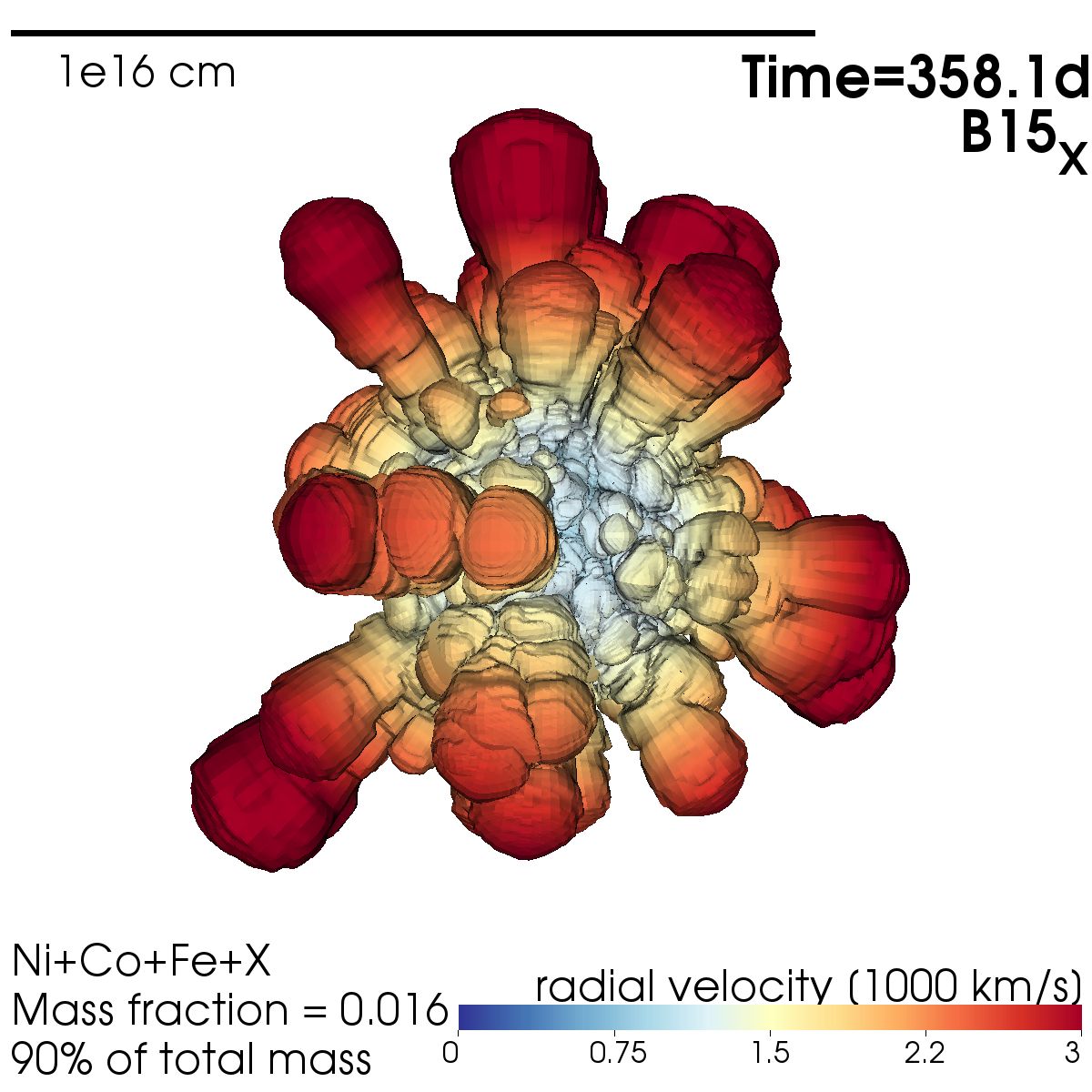}
\caption{Isosurface plots of constant mass fraction of NiCoFeX for models 
B15$_0$, B15, and B15$_\mathrm{X}$ at 
$t\sim1\,$d and $t\sim1\,$yr, respectively. At $t\sim1\,$d, all three models, 
B15$_0$ (top, left panel), B15 (top middle panel), and 
B15$_\mathrm{X}$ (top right panel), are almost indistinguishable because the 
$\beta$ decay did not significantly influence the evolution. After 
$t\sim1\,$yr, the structures of B15$_0$ are almost the same as at $t\sim1\,$d 
(compare upper and lower left panels), while the NiCoFeX-rich fingers of 
the other two models still inflate significantly after $t\sim1\,$d. The final 
structures of model B15$_\mathrm{X}$ are more extended than those of model B15. 
The NiCoFeX mass fractions defining the isosurfaces are indicated in each panel.
}
\label{fig_iso_B15_compare}
\end{figure*}

In Fig.\,\ref{fig_iso_B15_compare}, we plot the models B15$_0$ (left), B15 
(central) and B15$_\mathrm{X}$ (right) at two different times $t\sim1\,$d (top 
row) and $t\sim1\,$yr (bottom row). At $t\sim1\,$d (top row) all three models 
have almost identical structures and the mass fraction thresholds are the same 
for all. This is expected since the only difference between the models is the 
treatment of the $\beta$ decay, which should not have any significant influence 
at this early time. At $t\sim1\,$yr, the structures of the models B15 (bottom, 
central panel) and in particular B15$_\mathrm{X}$ (bottom, right panel) are 
significantly inflated. Model B15$_0$ is almost unchanged compared to 
$t\sim1\,$d (left column). It also still has almost the same mass fraction 
threshold as in the beginning. The threshold $X_\mathrm{NiCoFeX}$ containing 
$90\%$ of the total mass of NiCoFeX of model B15 decreases from $0.023$ to 
$0.019$, and, due to the stronger inflation and the correspondingly stronger 
mixing, the one of B15$_\mathrm{X}$ decreases to $X_\mathrm{NiCoFeX}=0.016$.

\begin{figure*}  
\centering
\includegraphics[width=.33\textwidth]{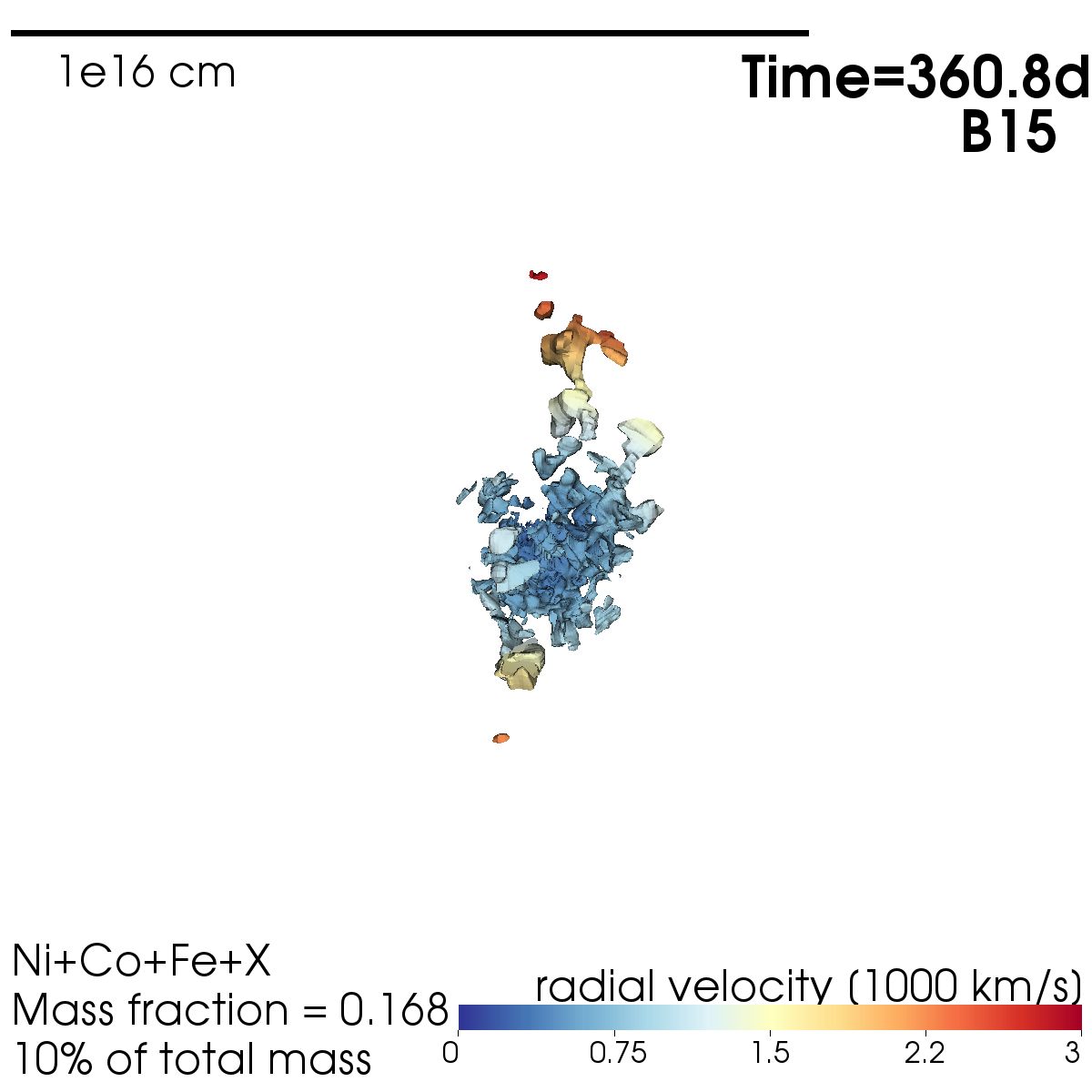}
\includegraphics[width=.33\textwidth]{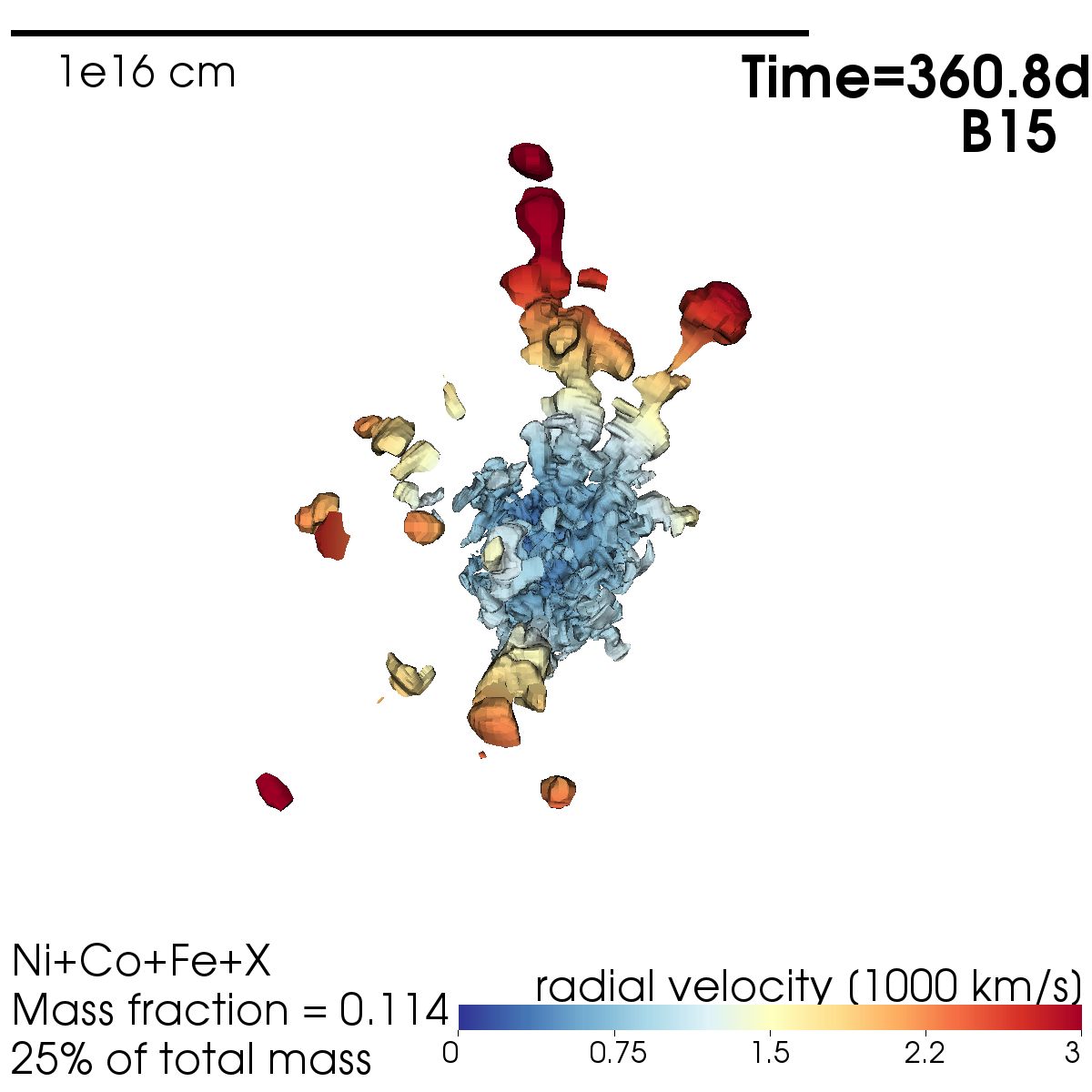}
\includegraphics[width=.33\textwidth]{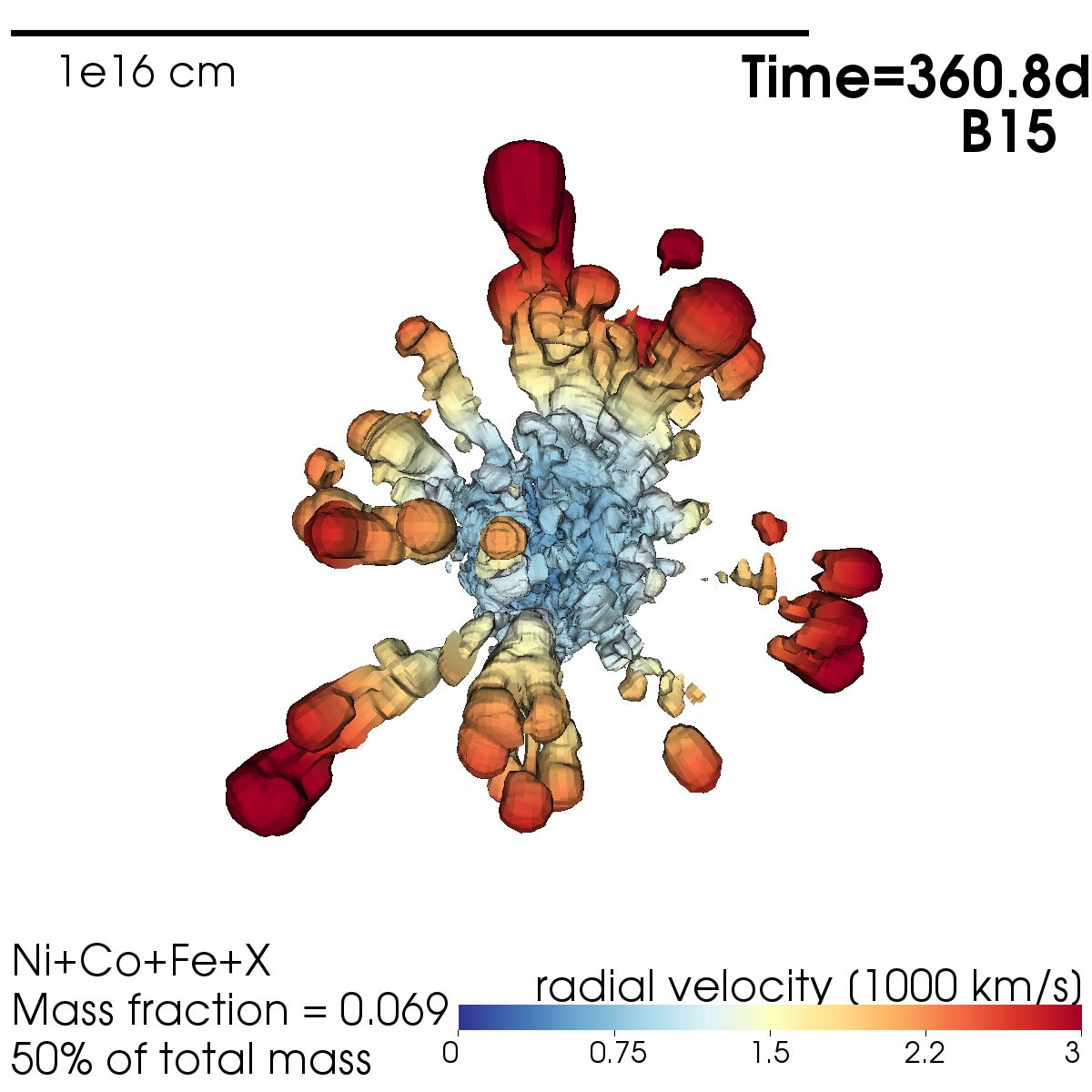}
\caption{Isosurface plots of of constant mass fraction of NiCoFeX-rich 
ejecta containing different percentages of the total mass of these nuclei in 
model B15, from left to right $10\%$, $25\%$, and $50\%$, respectively. For 
lower percentages of the mass, more of the fingers disappear and mainly a 
central region with the highest mass fractions remains visible. The NiCoFeX 
mass fractions defining the isosurfaces are indicated in each panel.
}
\label{fig_iso_B15_fractions}
\end{figure*}
In Fig.\,\ref{fig_iso_B15_fractions}, we show the isosurfaces 
containing different mass percentages 
of NiCoFeX in the ejecta: $10\%$, $25\%$, and $50\%$, respectively. The 
morphologies are significantly different from each other depending on the 
mass fractions corresponding to the isosurfaces. In the left panel for $10\%$ 
of the NiCoFeX mass, the ejecta seem elongated preferentially along a 
particular axis. For 
the $25\%$ limit (central panel), the structures look similar, but there are 
additional small clumps distributed also on the left side of the image, while 
the 
right side is almost empty. Increasing to $50\%$ (right panel) more 
NiCoFeX-rich fingers and clumps appear. For a very low mass fraction threshold 
and, thus, for a plot that shows most of the NiCoFeX-rich ejecta like the third 
panel in Fig.\ref{fig_iso_B15} for $90\%$, the fingers are more isotropically 
distributed. When comparing to observations, these significant differences 
should be kept in mind.

\begin{figure*}  
\centering
\includegraphics[width=.295\textwidth]{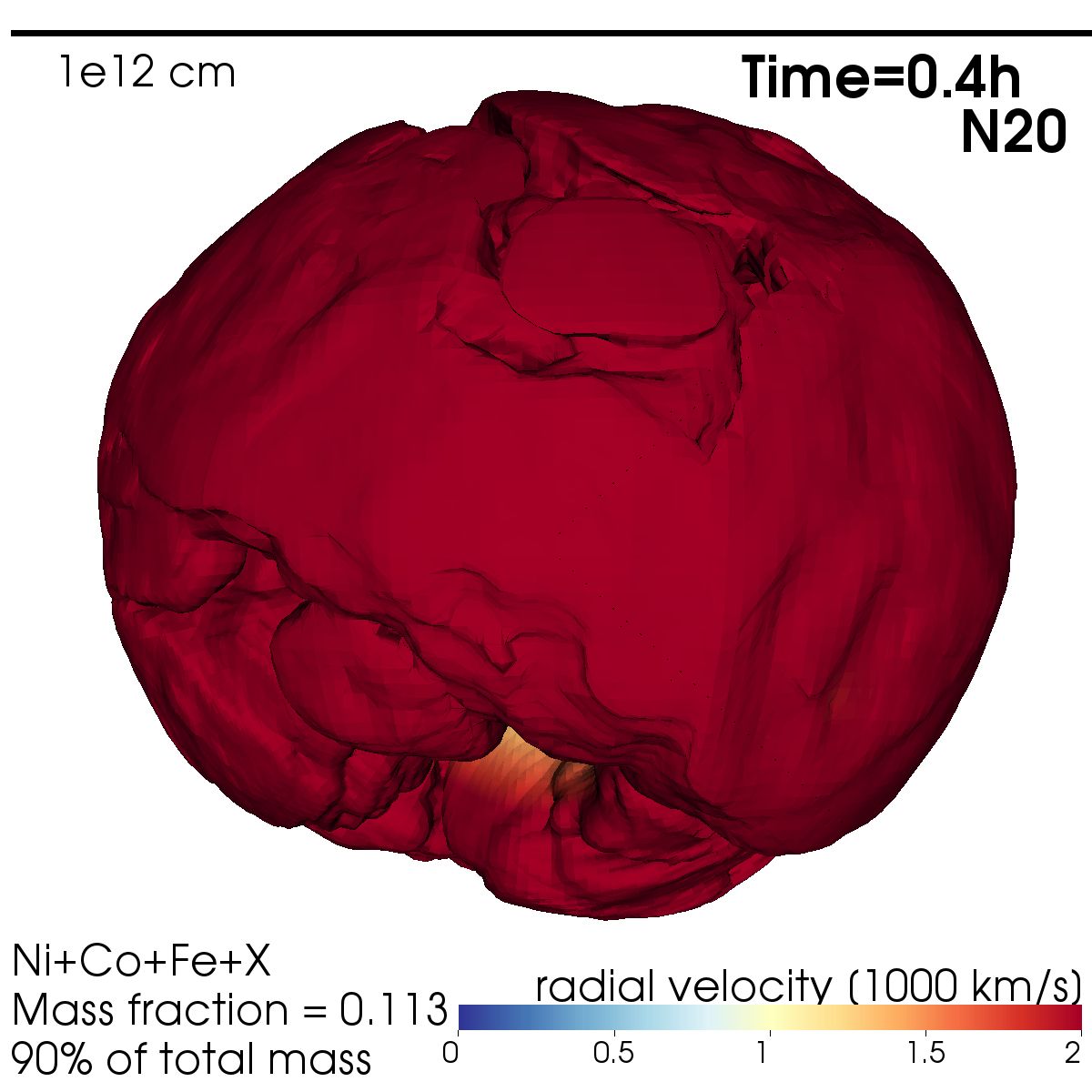}
\includegraphics[width=.295\textwidth]{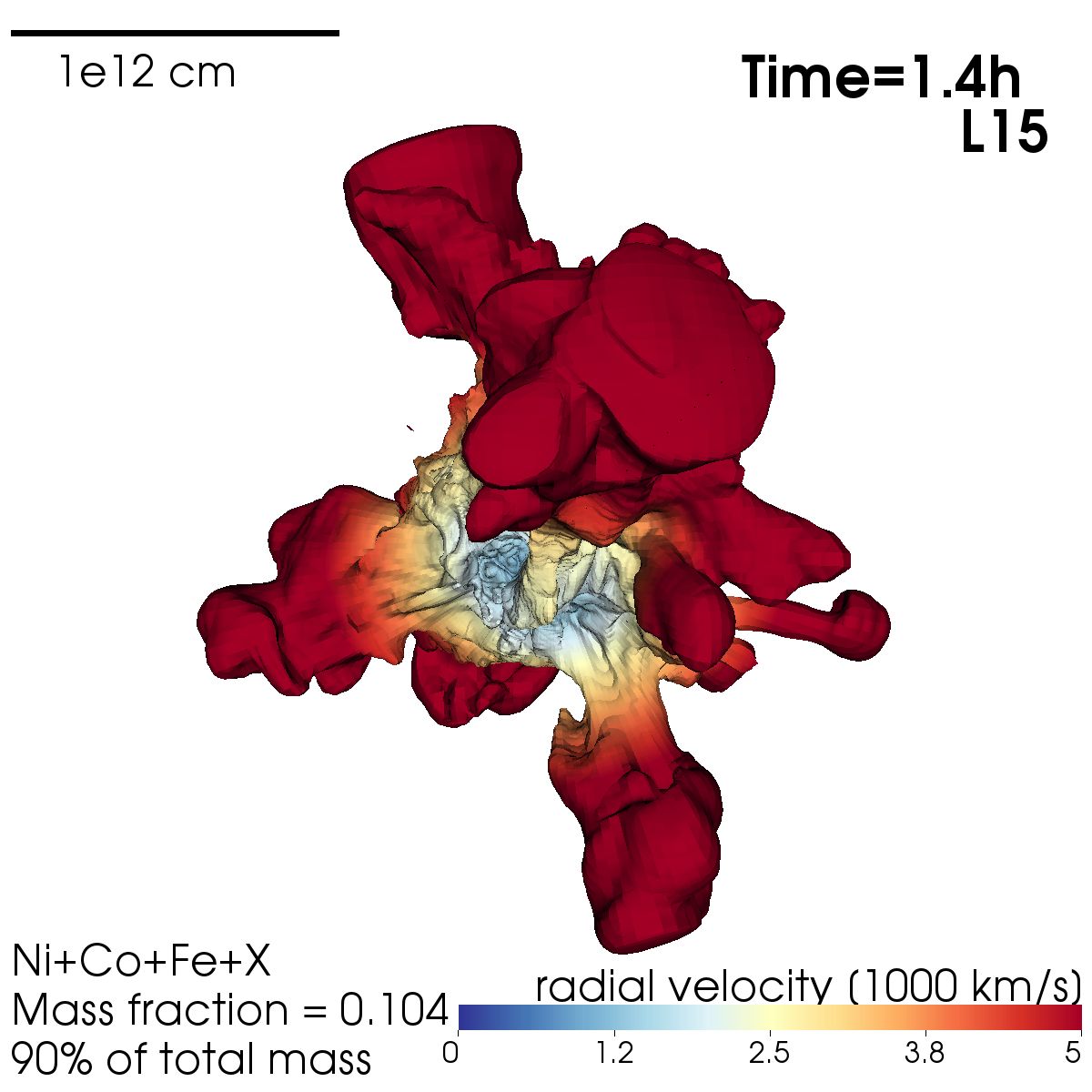}
\includegraphics[width=.295\textwidth]{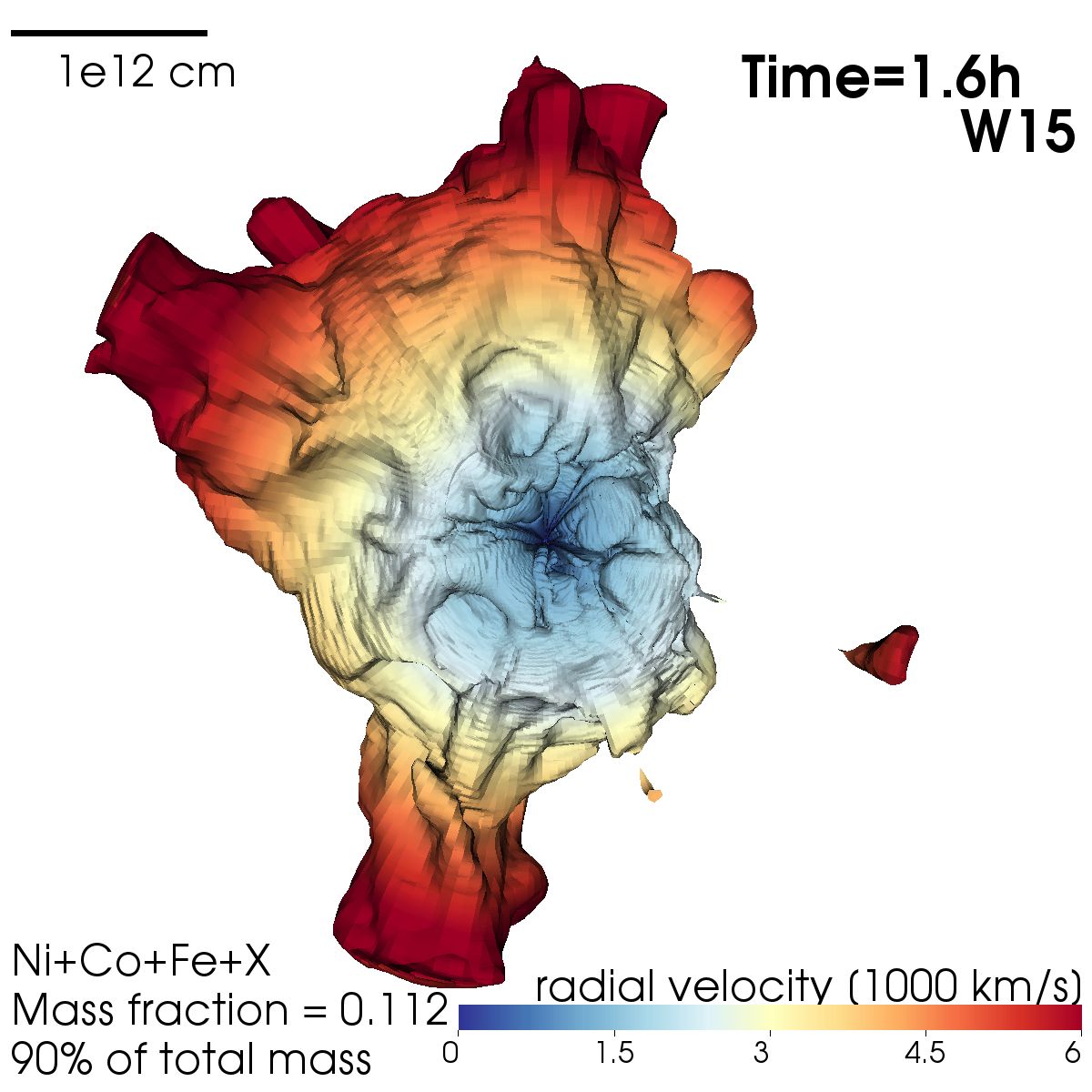}\\
\includegraphics[width=.295\textwidth]{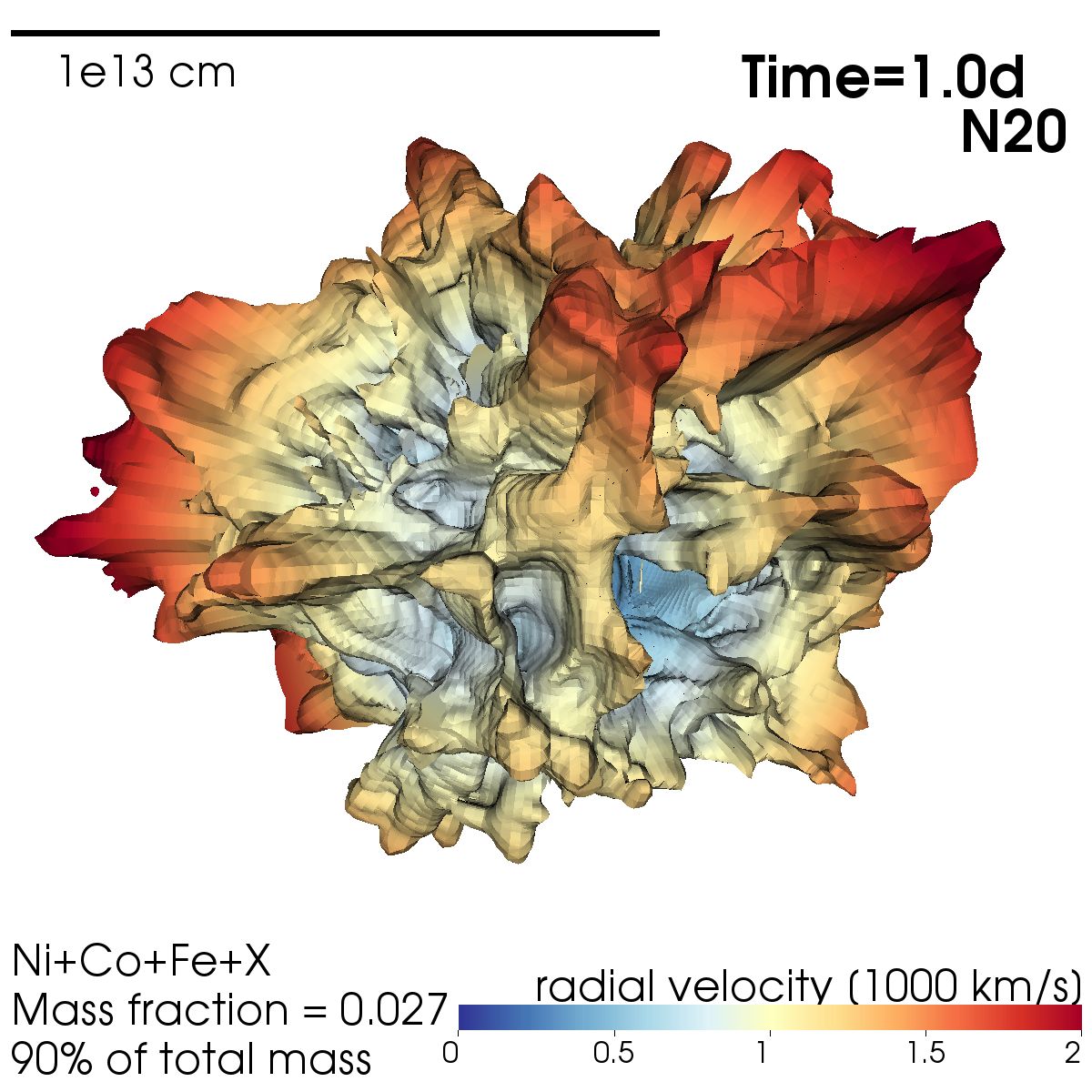}
\includegraphics[width=.295\textwidth]{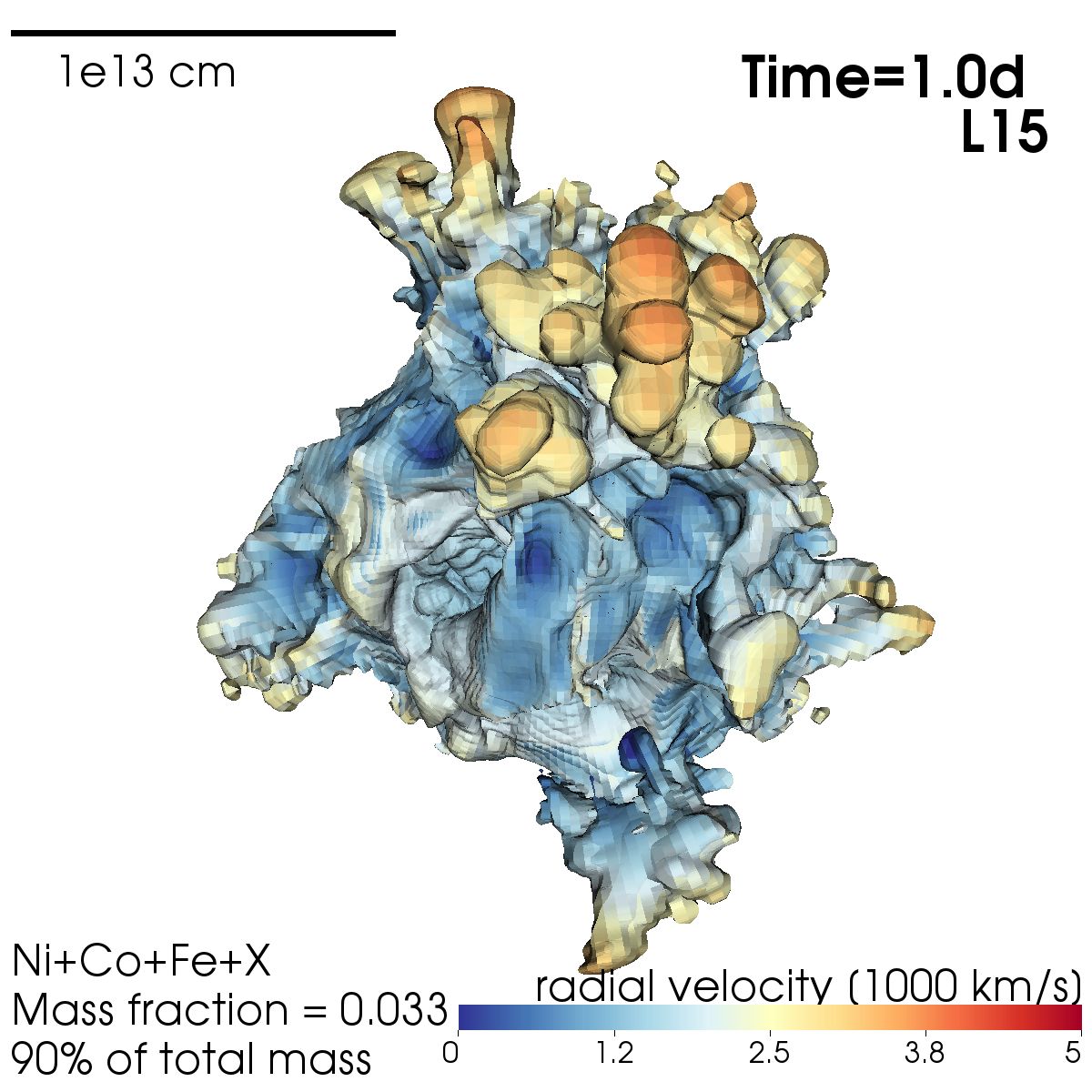}
\includegraphics[width=.295\textwidth]{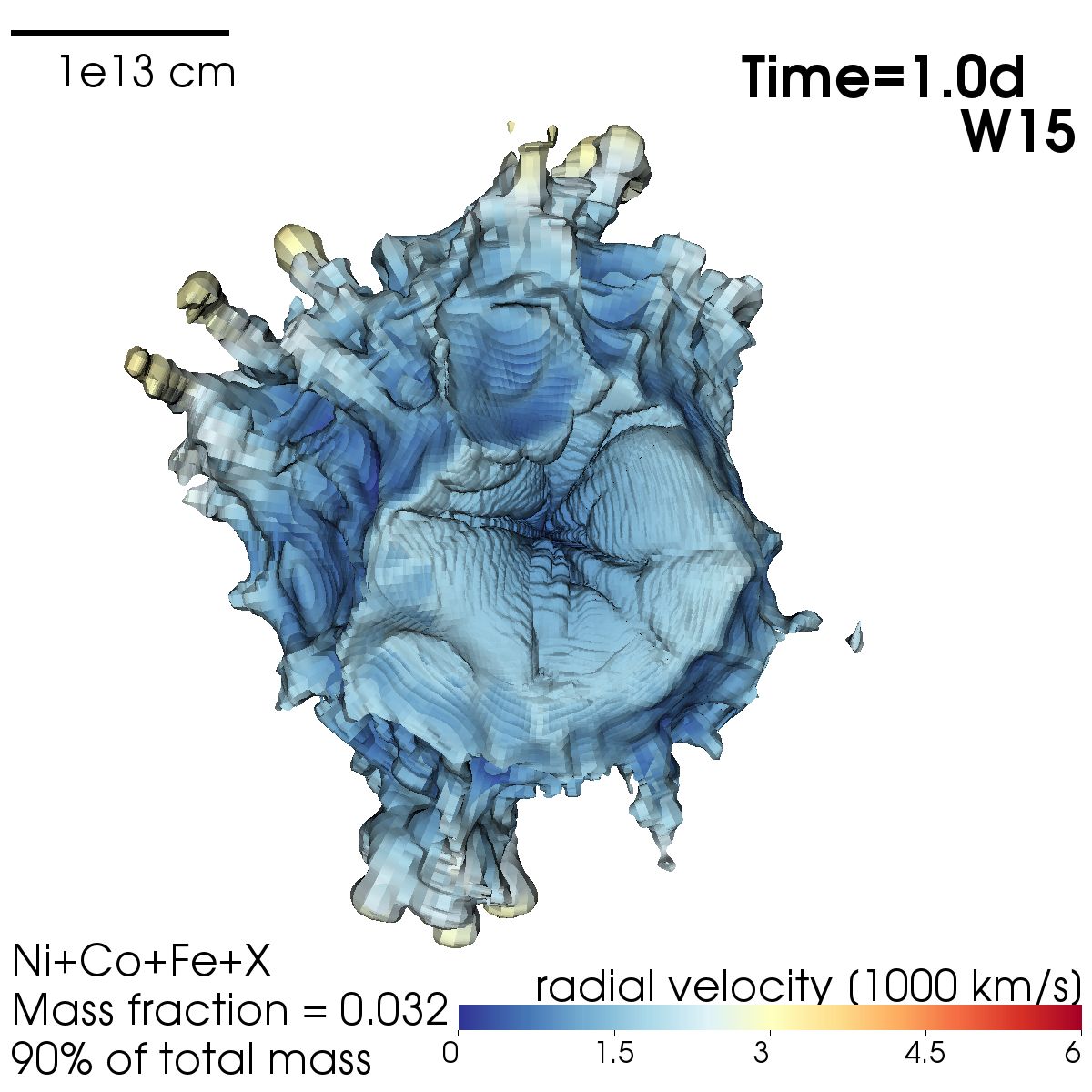}\\
\includegraphics[width=.295\textwidth]{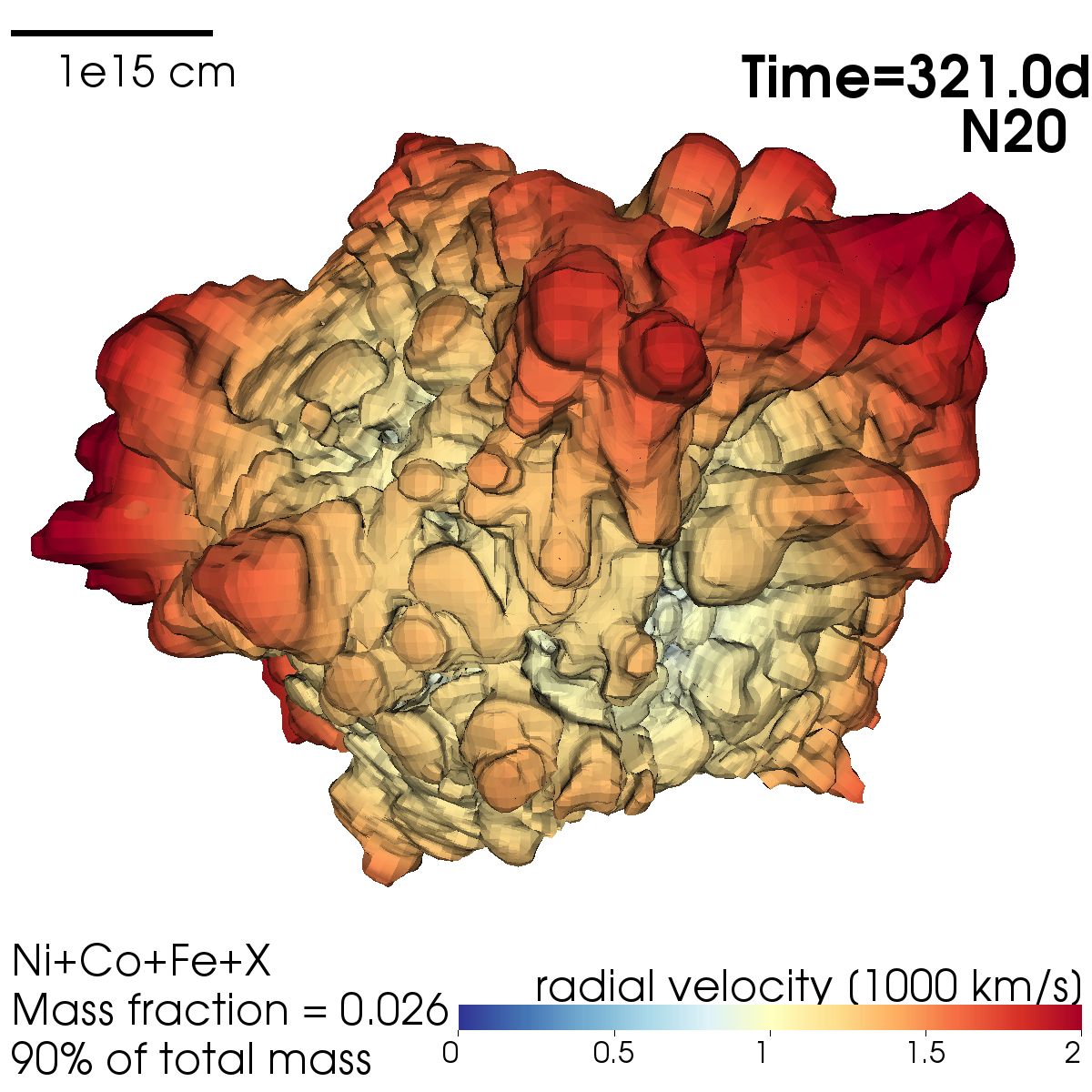}
\includegraphics[width=.295\textwidth]{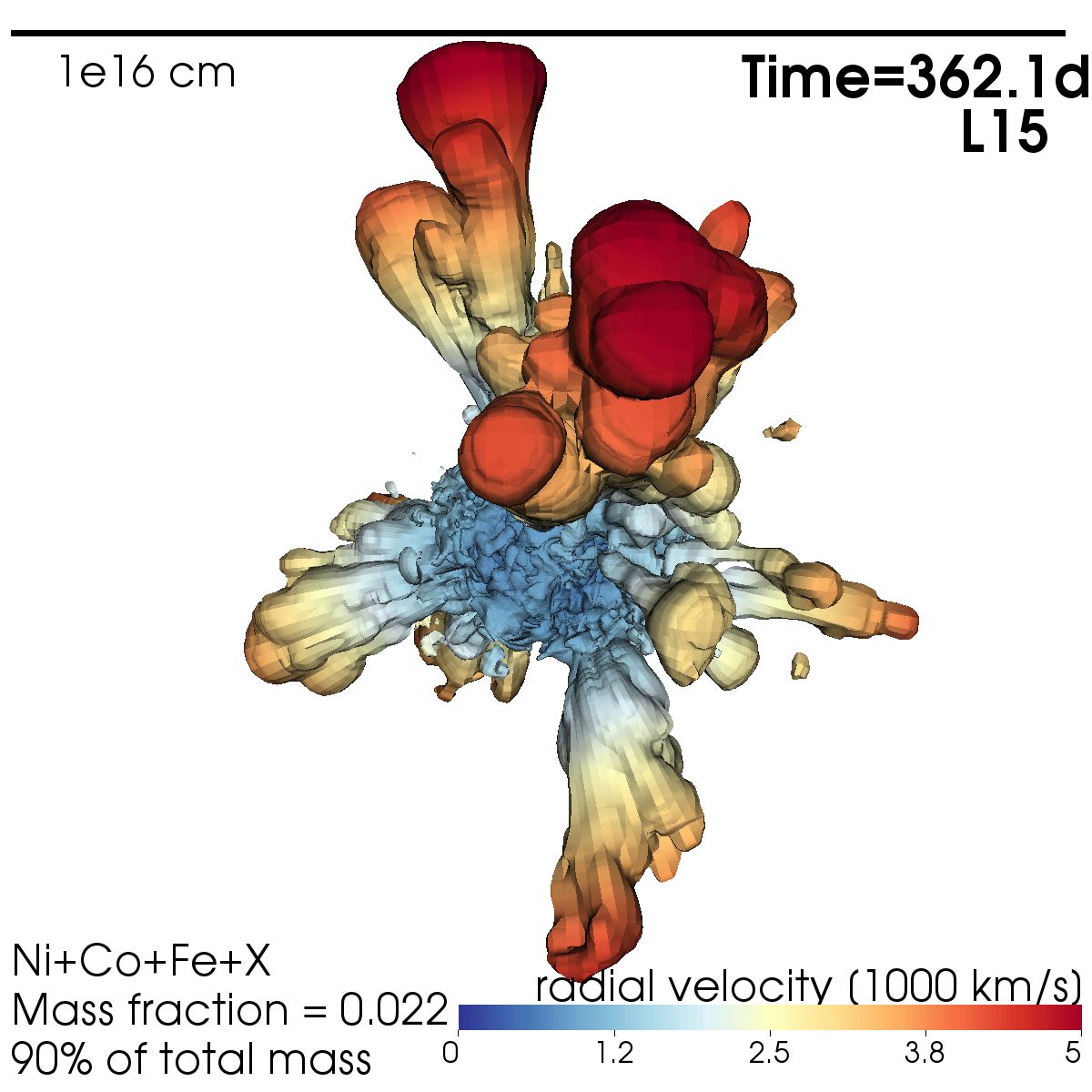}
\includegraphics[width=.295\textwidth]{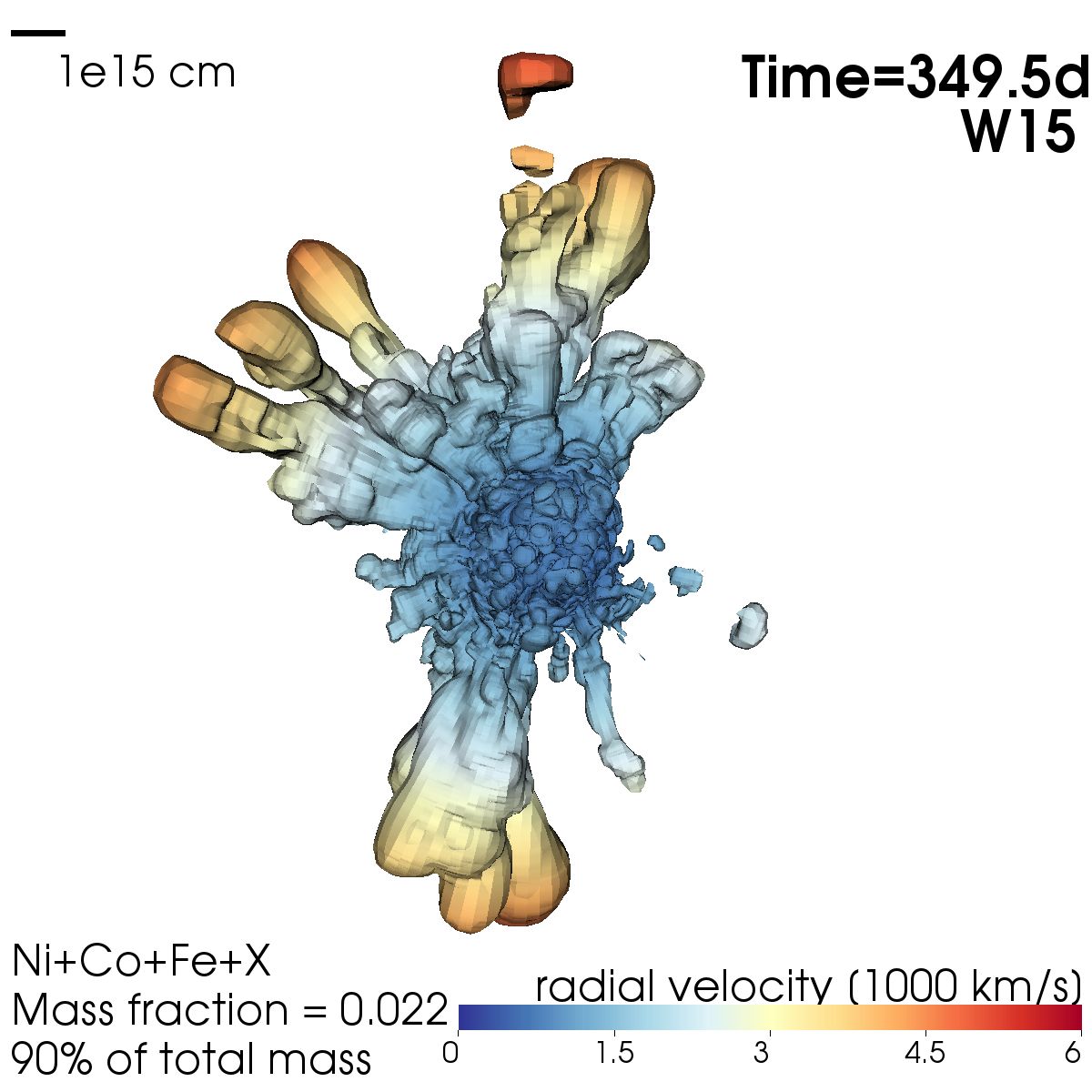}\\
\includegraphics[width=.295\textwidth]{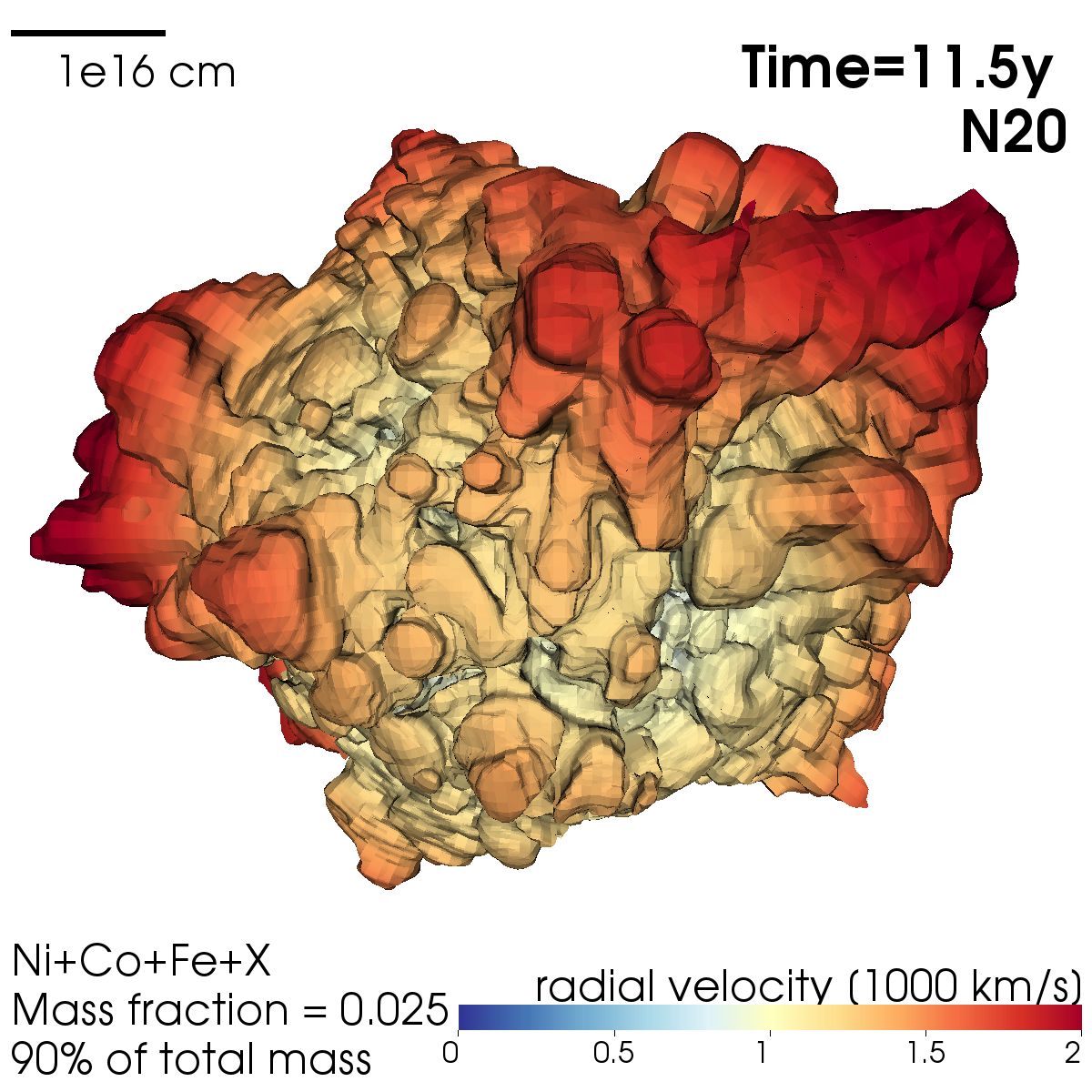}
\includegraphics[width=.295\textwidth]{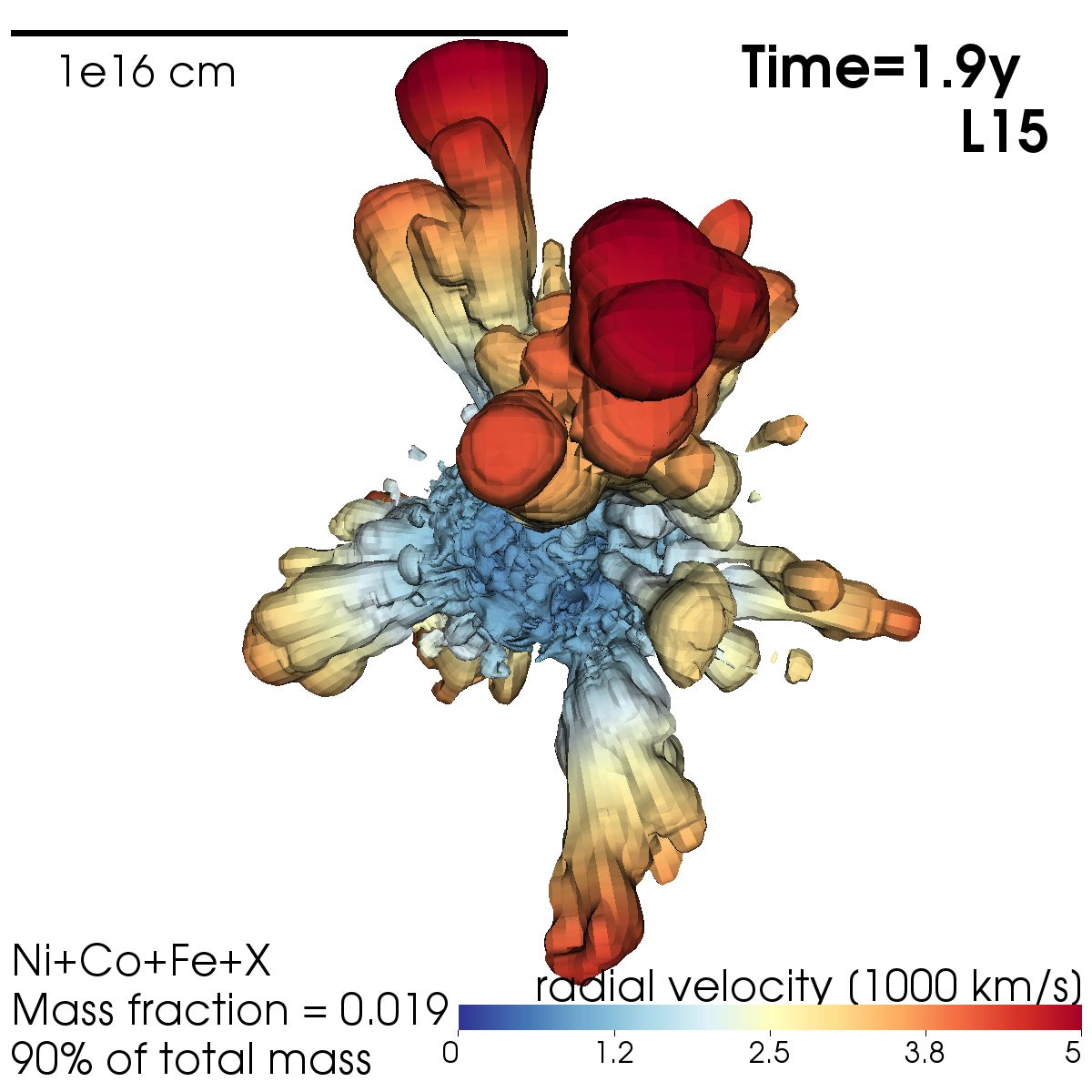}
\includegraphics[width=.295\textwidth]{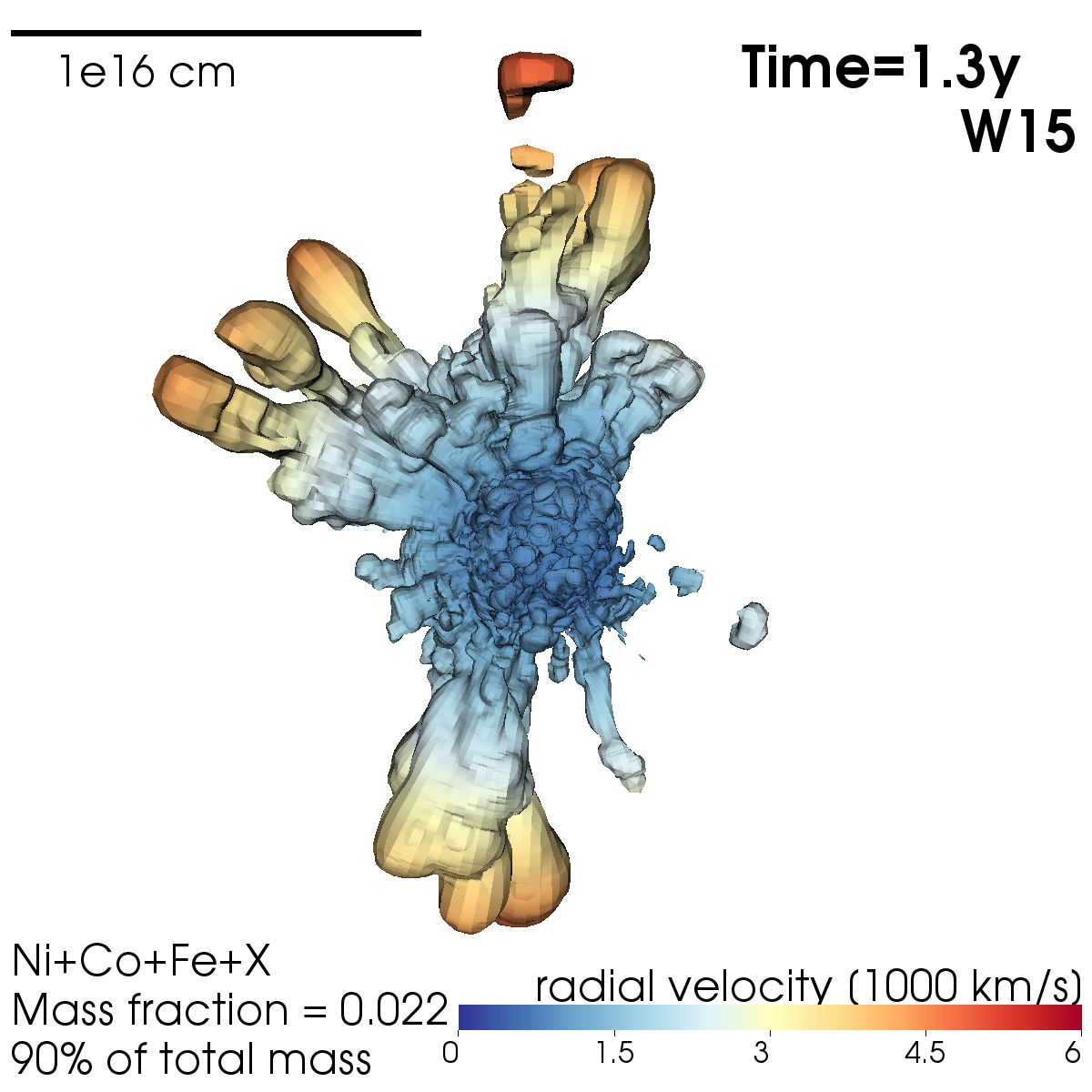}
\caption{Isosurface plots of constant mass fraction containing 
$90\%$ of the mass of NiCoFeX in regions with highest mass fractions for 
different times and different models. Model N20 (left column) is almost 
spherical initially (top, left panel) 
and becomes slightly asymmetric after the shock breakout $t\sim1\,$d (second 
row, left panel). These small asymmetries are then partially erased due to the 
inflation of the NiCoFeX-rich material. The two RSGs L15 (central column) and 
W15 
(right column) are similar to each other. Initially at $t\gtrsim1\,$h (top 
panels), they have large-scale plumes that fragment into smaller fingers 
(second and third row) and the reverse shock slows down the central ejecta 
compared to homologous expansion. A few strong NiCoFeX-rich clumps extend much 
farther out than the central bulk material. The decay of $^{56}$Ni leads to an 
inflation of the central bubble and of the RT fingers. In all 
models, the structures no longer change significantly after $t\lesssim1\,$yr 
(compare third and bottom row for the respective models).
The NiCoFeX mass fractions defining the isosurfaces are indicated in each panel.
}
\label{fig_iso_all}
\end{figure*}

The NiCoFeX-rich structures of the other models are plotted in 
Fig.\,\ref{fig_iso_all}. The left column is for model N20, the central for L15 
and the right for W15. The two RSG models L15 and W15 are qualitatively 
similar to each other, i.e. the initially large plumes (top central and right 
panels) fragment into smaller fingers due to RTIs, which occur during the SN 
shock propagation through the progenitor \citep[see also][for a detailed 
discussion]{Wongwathanarat2015}. The reverse shock begins to slow down the 
central ejecta compared to homologous expansion at about $t\sim1\,$d (second 
row, central and right panels). Consequently, the central NiCoFeX-rich bubble 
shrinks relative to the extended fingers. Then, the reverse shock 
self-reflects and accelerates the innermost, central ejecta, supported by the 
input from the $\beta$-decay energy. Also the initially big, but later 
fragmented plumes inflate due to $\beta$ decay. After the fragmentation is 
finished, and the inflation due to $\beta$ decay becomes significant, these 
transiently fine-structured fingers merge to large-scaled clumps again, which 
have a similar shape compared to the initial plumes. They are even more 
prominent 
at this late time, because the innermost ejecta were decelerated by the reverse 
shock for some time and, thus, the velocity difference between the outermost RT 
fingers and the central ejecta is larger. The corresponding final structures 
after $t\sim1\,$yr are shown in the third row (central and right panels) and at 
the end of our simulations in the bottom row.

The NiCoFeX-rich structures we find in model N20 (left column in 
Fig.~\ref{fig_iso_all}) are qualitatively very different from the other models. 
Shortly before the shock 
breakout from the progenitor (top left panel), the NiCoFeX-rich ejecta are 
almost 
spherically symmetrically distributed. 
At about $t\sim 1\,$d the model becomes slightly more asymmetric (second row, 
left panel), but then the expansion of the NiCoFeX-rich ejecta leads to a more 
spherical configuration again (third row, left panel). No significant 
asymmetries or RT fingers can be found. As there is no significant difference 
between the corresponding plots of all models between the third and the bottom 
row of Fig.\,\ref{fig_iso_all}, which shows the last times simulated, we 
conclude that the evolution of the structures becomes homologous after about 
$t\lesssim1\,$yr in all models.

\begin{figure*}  
\centering
\includegraphics[width=.48\textwidth]{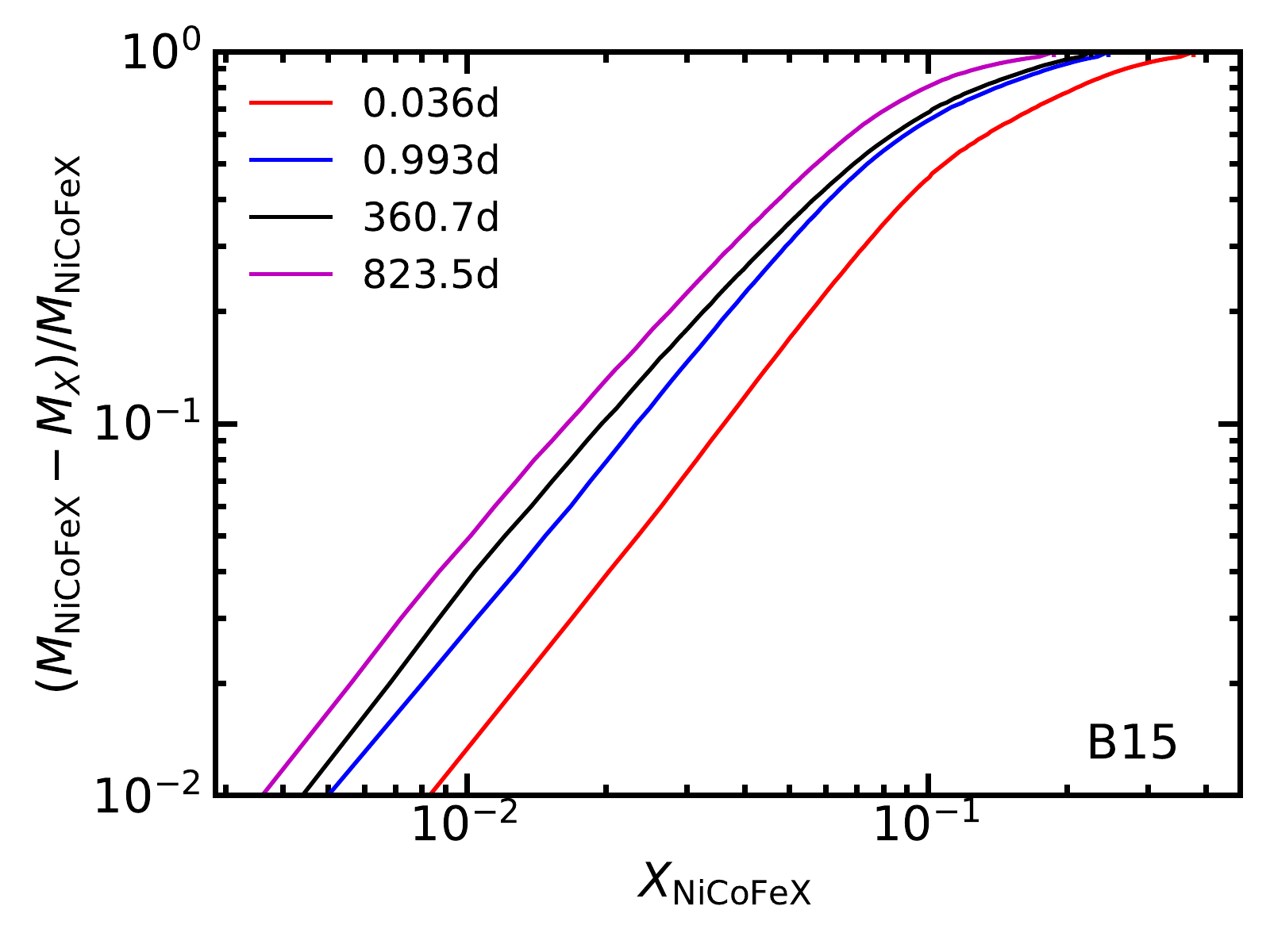}
\includegraphics[width=.485\textwidth]{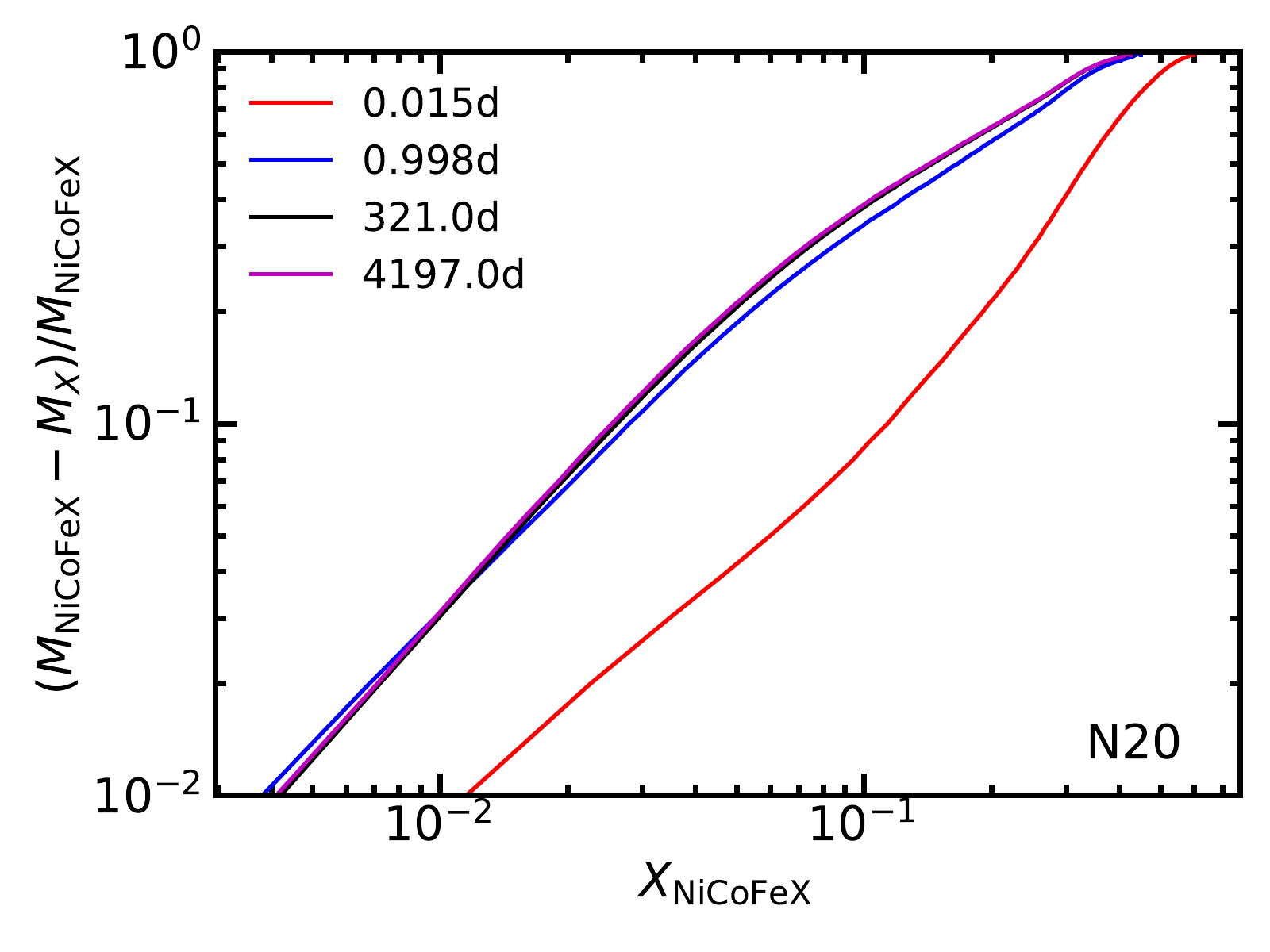}
\includegraphics[width=.48\textwidth]{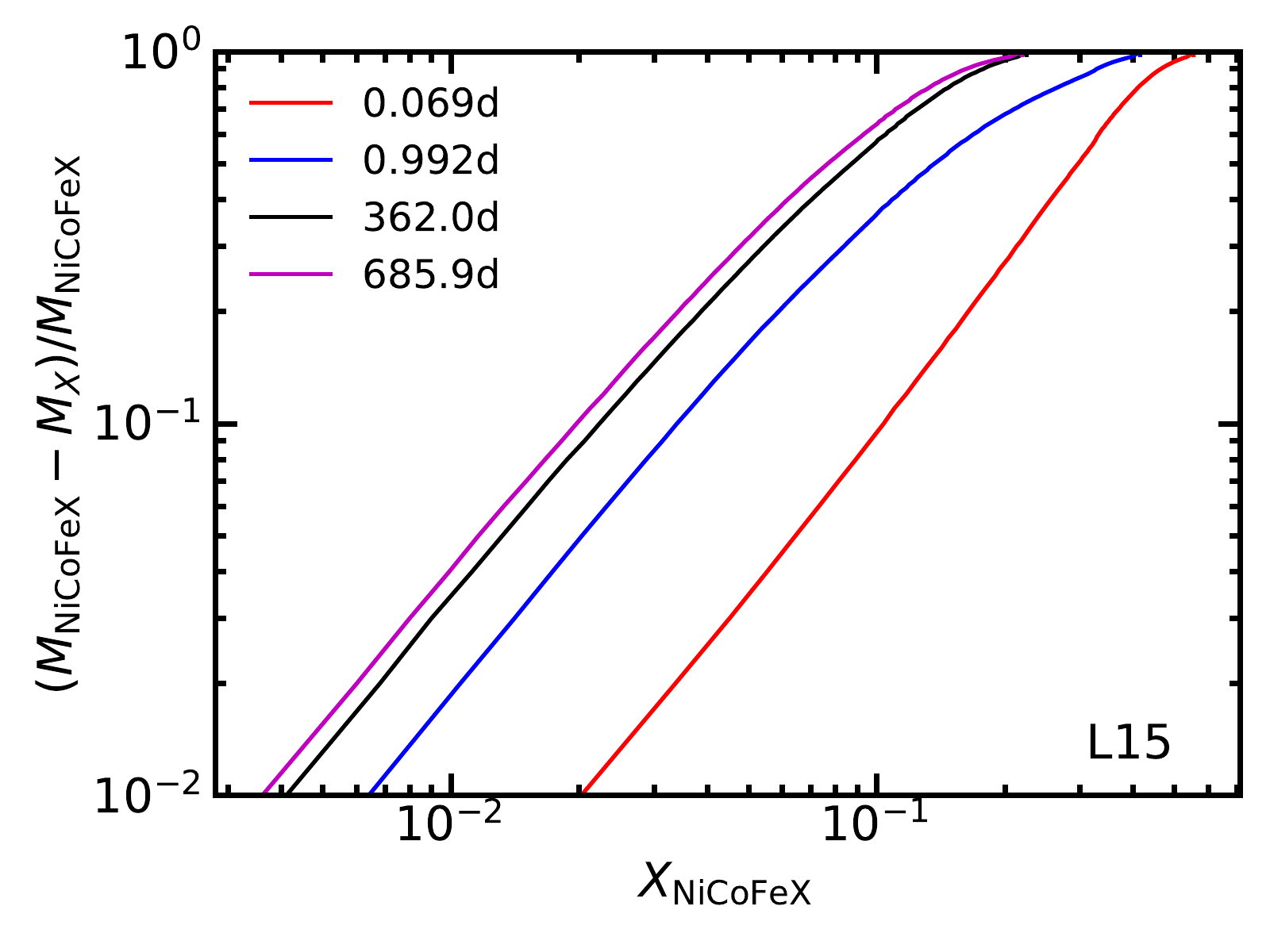}
\includegraphics[width=.48\textwidth]{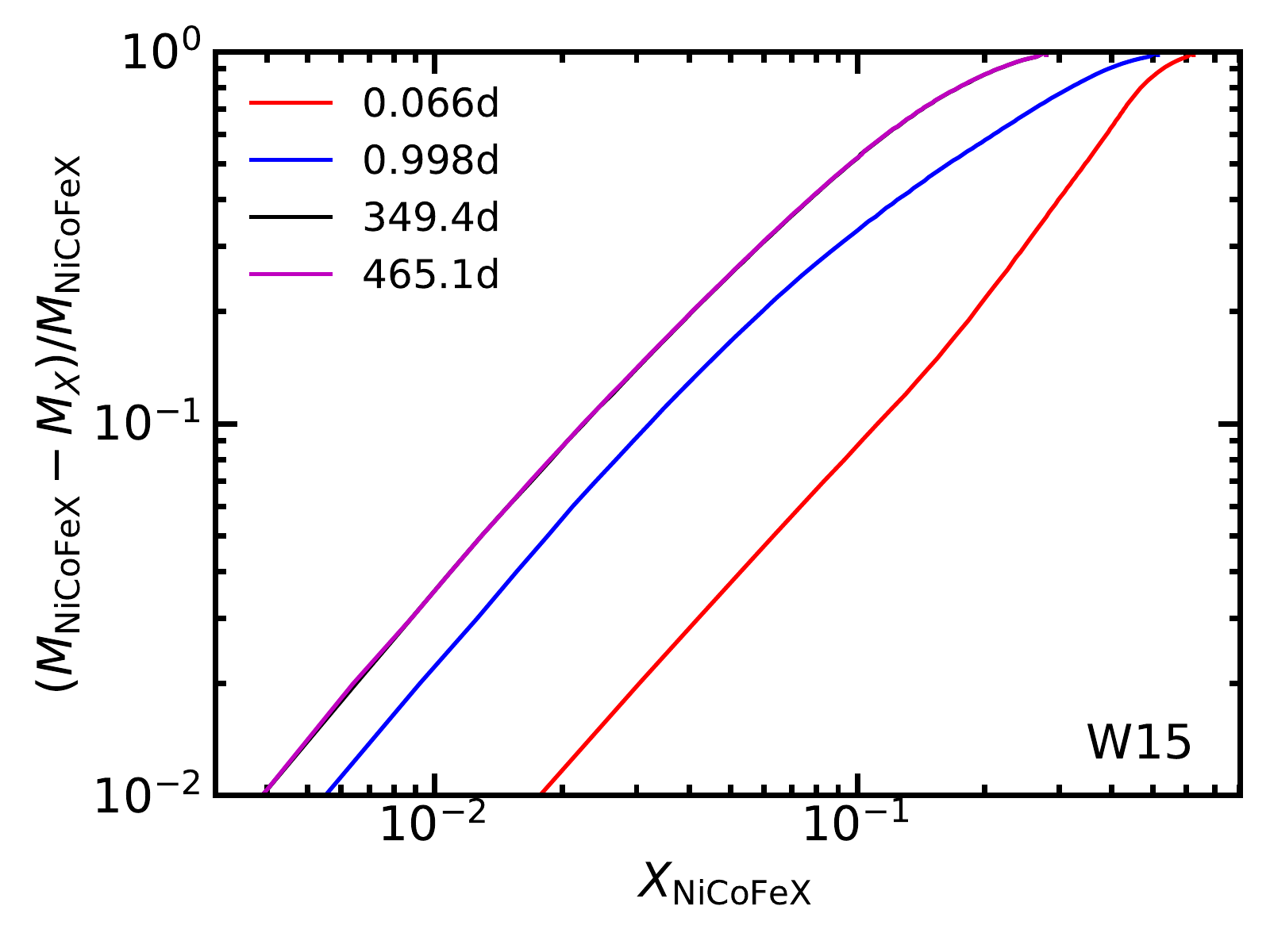}
\caption{Fraction of mass of NiCoFeX
$(M_\mathrm{NiCoFeX}-M_X) / M_\mathrm{NiCoFeX}$ which has a smaller 
$X_{\mathrm{NiCoFeX}}$ than given at the corresponding x-coordinate for 
the different models at different times. Note that the black curves for models 
N20 and W15 are almost exactly covered by the magenta curves, indicating a 
perfectly homologous expansion between the corresponding last time steps.}
\label{fig_mass_Xc}
\end{figure*}

To get a better feeling for the mass cut-offs applied to reach a certain 
fraction of the total mass $M_X/ M_\mathrm{NiCoFeX}$ contained inside the 
isosurfaces defined by the minimal mass fractions $X_\mathrm{NiCoFeX}$ 
indicated in the plots, we provide the plots of the fraction of the mass 
$(M_\mathrm{NiCoFeX} - M_X)/ M_\mathrm{NiCoFeX}$ outside the corresponding 
isosurface as a function of $X_{\mathrm{NiCoFeX}}$ in 
Fig.\,\ref{fig_mass_Xc} for the different models at different times. We 
choose to plot the complement to $M_X/ M_\mathrm{NiCoFeX}$ for a better 
visualization of the fractions of the mass for small $X_\mathrm{NiCoFeX}$. The 
relatively large change at late times between the black and magenta curves for 
model B15 (top left 
panel) is related to the faster cut out of the central volume during the 
evolution. This did not influence the main results of the current work and was 
done for numerical efficiency as briefly described in 
Footnote\,\ref{footnote_problem}.

\subsubsection{Quantitative analysis}\label{sec_clumps}

In the preceding sections, we described the structures obtained in the 
long-time evolution qualitatively. Here, we provide quantitative 
characteristics of the NiCoFeX-rich clumps for the different models. 
Since the particular choice of $F_\rho$ (or equivalently 
$\rho_\mathrm{NiCoFeX}^\mathrm{min}$) is somewhat arbitrary, we provide 
the characteristics of the clumps for different choices of 
$F_\rho$ for our models at $t\sim1\,$yr in 
Table\,\ref{tab_clumps}. The data in the table contain the minimal 
density $\rho^\mathrm{min}_\mathrm{NiCoFeX}$ above which we define the clump, 
the total number of clumps, the number of clumps with masses larger than 
$10^{-6}M_\odot$, and the volume of the clumps  
$V^\mathrm{x}_\mathrm{NiCoFeX}\equiv{V_\mathrm{NiCoFeX}}/{V_\mathrm{x} }$ 
compared to the volume of the sphere defined by the mean radius where the 
ejecta move with $\bar v_{1500}=1500\,$km/s, or $\bar v_{2500}=2500\,$km/s, 
respectively. The super- and subscripts $\mathrm{x}$ of 
$V^\mathrm{x}_\mathrm{NiCoFeX}$ 
or $V_\mathrm{x}$ represent the respective velocities.
We also give the ratio of clump volume to the volume of the sphere defined by 
the radius of the fastest moving NiCoFeX. These fastest blobs are the 
outermost NiCoFeX-rich ejecta which have a mass fraction 
$X_\mathrm{NiCoFeX}>10^{-3}$. The corresponding maximal velocities 
$v_\mathrm{fastest}$ are given in Table\,\ref{tab_clump_max_vel}. 
To have a measure to describe the clumpiness of the ejecta when 3D 
information in observations is not available, we provide the surface filling 
factors $A^\mathrm{x}_\mathrm{NiCoFeX}\equiv{A_\mathrm{NiCoFeX}}/{A_\mathrm{x}}$ 
of the corresponding clumps in the last three columns of 
Table\,\ref{tab_clumps}. 
The reference line of sight to obtain the $A^\mathrm{x}_\mathrm{NiCoFeX}$ is in 
the y-direction such that we are looking at the x-z plane. This is the same 
viewing direction used in the Figs.\,\ref{fig_iso_B15} - \ref{fig_iso_all}.
\begin{table*}
 \begin{tabular}{c c c c c c c c c c c c}
\multirow{2}{*}{Model}& 
\multirow{2}{*}{$F_\rho$}
&$\rho^\mathrm{min}_\mathrm{NiCoFeX}$&number&clumps with
&\multirow{2}{*}{$V_\mathrm{NiCoFeX}^\mathrm{1500}$}
&\multirow{2}{*}{$V_\mathrm{NiCoFeX}^\mathrm{2500}$}
&\multirow{2}{*}{$V_\mathrm{NiCoFeX}^\mathrm{fastest}$}
&\multirow{2}{*}{$A_\mathrm{NiCoFeX}^\mathrm{1500}$}
&\multirow{2}{*}{$A_\mathrm{NiCoFeX}^\mathrm{2500}$}
&\multirow{2}{*}{$A_\mathrm{NiCoFeX}^\mathrm{fastest}$}\\ 
&&[g/cm$^3$]&of clumps&$M>10^{-6}M_\odot$\\\hline
\multirow{9}{*}{B15$_0$}
&0.9& 0.021& 21 & 8& 0.591 &0.127 &0.0349& 0.799&0.562&0.232\\
&0.8& 0.034& 28 &11& 0.406 &0.088 &0.0240& 0.693&0.483&0.198\\
&0.7& 0.046& 41 &20& 0.288 &0.062 &0.0170& 0.592&0.407&0.166\\
&0.6& 0.057& 54 &39& 0.200 &0.043 &0.0118& 0.514&0.338&0.136\\
&0.5& 0.070& 58 &36& 0.134 &0.029 &0.0079& 0.441&0.264&0.105\\
&0.4& 0.085& 60 &37& 0.085 &0.018 &0.0050& 0.372&0.193&0.075\\
&0.3& 0.105& 51 &30& 0.051 &0.011 &0.0030& 0.294&0.129&0.049\\
&0.2& 0.134& 70 &34& 0.028 &0.006 &0.0017& 0.202&0.081&0.029\\
&0.1& 0.172& 129&36& 0.012 &0.003 &0.0007& 0.125&0.046&0.017\\\hline
\multirow{9}{*}{B15}
&0.9&  0.018&  9 &6  &1.509& 0.324&0.0866&0.952&0.721&0.305\\
&0.8&  0.029&  13&11 &1.210& 0.259&0.0694&0.912&0.674&0.283\\
&0.7&  0.041&  25&13 &0.971& 0.208&0.0557&0.864&0.624&0.259\\
&0.6&  0.052&  36&23 &0.759& 0.163&0.0436&0.790&0.565&0.231\\
&0.5&  0.065&  50&40 &0.565& 0.121&0.0324&0.715&0.490&0.199\\
&0.4&  0.079&  61&44 &0.388& 0.083&0.0223&0.643&0.402&0.160\\
&0.3&  0.098&  51&32 &0.237& 0.051&0.0136&0.554&0.300&0.116\\
&0.2&  0.123&  53&36 &0.121& 0.026&0.0070&0.409&0.181&0.066\\
&0.1&  0.163&  64&28 &0.047& 0.010&0.0027&0.268&0.100&0.036\\\hline
\multirow{9}{*}{B15$_\mathrm{X}$}
&0.9& 0.016& 5 &3 &3.154 &0.643 &0.1372 &0.998&0.849&0.387\\
&0.8& 0.025& 9 &5 &2.755 &0.562 &0.1198 &0.986&0.815&0.368\\
&0.7& 0.039& 18&10&2.386 &0.487 &0.1038 &0.978&0.780&0.349\\
&0.6& 0.044& 23&11&2.011 &0.410 &0.0875 &0.948&0.742&0.324\\
&0.5& 0.055& 39&24&1.616 &0.330 &0.0703 &0.892&0.668&0.291\\
&0.4& 0.068& 63&39&1.200 &0.245 &0.0522 &0.824&0.567&0.242\\
&0.3& 0.085& 77&43&0.792 &0.161 &0.0344 &0.740&0.442&0.182\\
&0.2& 0.107& 71&35&0.423 &0.086 &0.0184 &0.574&0.277&0.105\\
&0.1& 0.145& 91&28&0.151 &0.031 &0.0066 &0.399&0.155&0.056\\\hline
\multirow{9}{*}{N20}
& 0.9&0.026& 5 &2 &0.766&0.163&0.1381&0.782&0.316&0.282\\
& 0.8&0.047& 7 &2 &0.607&0.129&0.1095&0.735&0.279&0.249\\
& 0.7&0.071& 6 &3 &0.488&0.104&0.0881&0.671&0.243&0.217\\
& 0.6&0.100& 23&4 &0.396&0.084&0.0715&0.608&0.215&0.192\\
& 0.5&0.138& 25&1 &0.315&0.067&0.0568&0.558&0.195&0.174\\
& 0.4&0.180& 32&3 &0.243&0.052&0.0438&0.510&0.178&0.159\\
& 0.3&0.228& 36&8 &0.175&0.037&0.0316&0.443&0.155&0.138\\
& 0.2&0.280& 51&13&0.110&0.024&0.0199&0.361&0.128&0.114\\
& 0.1&0.333& 37&17&0.051&0.011&0.0092&0.274&0.097&0.087\\\hline
\multirow{9}{*}{L15}
& 0.9&0.021& 31 &13&2.534&0.527&0.0495&0.867&0.664&0.214\\
& 0.8&0.037& 35 &21&1.822&0.379&0.0356&0.798&0.597&0.187\\
& 0.7&0.052& 62 &37&1.378&0.287&0.0269&0.732&0.538&0.166\\
& 0.6&0.068& 51 &28&1.056&0.220&0.0206&0.644&0.476&0.147\\
& 0.5&0.085& 58 &30&0.798&0.166&0.0156&0.554&0.412&0.127\\
& 0.4&0.102& 54 &27&0.588&0.122&0.0115&0.479&0.353&0.108\\
& 0.3&0.121& 72 &32&0.398&0.083&0.0078&0.400&0.274&0.085\\
& 0.2&0.143& 116&46&0.230&0.048&0.0045&0.333&0.199&0.059\\
& 0.1&0.175& 125&59&0.084&0.018&0.0016&0.260&0.132&0.028\\\hline
\multirow{9}{*}{W15}
& 0.9& 0.022& 20 &8 &1.451&0.307&0.0279&0.750&0.523&0.167 \\
& 0.8& 0.039& 25 &17&0.974&0.206&0.0187&0.688&0.443&0.125\\
& 0.7& 0.057& 20 &12&0.698&0.148&0.0134&0.646&0.388&0.103\\
& 0.6& 0.074& 42 &22&0.500&0.106&0.0096&0.601&0.341&0.085\\
& 0.5& 0.093& 61 &34&0.354&0.075&0.0068&0.555&0.294&0.069\\
& 0.4& 0.113& 60 &28&0.238&0.050&0.0046&0.501&0.238&0.053\\
& 0.3& 0.137& 78 &34&0.152&0.032&0.0029&0.421&0.184&0.039\\
& 0.2& 0.167& 95 &29&0.088&0.019&0.0017&0.328&0.128&0.026\\
& 0.1& 0.208& 130&30&0.040&0.008&0.0007&0.230&0.082&0.017\\\hline
\end{tabular}
\caption{Characteristics of the clumps of NiCoFeX after $t\sim1\,$yr for 
models B15$_0$, B15, B15$_\mathrm{X}$, N20, L15, and W15, respectively. In the 
different columns we give the model name, the fraction of mass of the 
clumps compared to the total mass of NiCoFeX, $F_\rho$ , 
the threshold density above which we define the clumps, the number of clumps, 
the number of clumps with NiCoFeX mass larger than $10^{-6}M_\odot$, 
the volume of the clumps compared to the volumes inside a sphere with the 
radius where the mean velocities of the material are $\bar 
v_{1500}=1500\,$km/s, $\bar v_{2500}=2500\,$km/s, and 
$v_\mathrm{fastest}^\mathrm{NiCoFeX}$, and finally the 
surface area in the x-z plane covered by the NiCoFeX clumps compared to a 
square with side length of twice the radius where the ejecta move with $\bar 
v_{1500}$, $\bar v_{2500}$, and $v_\mathrm{fastest}^\mathrm{NiCoFeX}$, 
respectively.
}
\label{tab_clumps}
\end{table*}
\begin{table}
 \begin{tabular}{c c c c c c c c}
{Model}&B15$_0$ & B15 & B15$_\mathrm{X}$ & N20 & L15 & W15 \\\hline
$v^\mathrm{NiCoFeX}_\mathrm{fastest}$ [km/s] & 
3813&3899&4199&2646&5484&5544\\[3pt]
$r^\mathrm{NiCoFeX}_\mathrm{fastest}$ 
&\multirow{2}{*}{12.0}&\multirow{2}{*}{12.1}& \multirow{2}{*}{12.8} 
&\multirow{2}{*}{7.3}&\multirow{2}{*}{17.1}&\multirow{2}{*}{16.7}\\

[$10^{15}$cm]
\end{tabular}
\caption{Velocity $v_\mathrm{fastest}^\mathrm{NiCoFeX}$ and radius of the 
fastest NiCoFeX  at $t\sim1\,$yr. The minimum mass fraction to 
define the fastest NiCoFeX is $10^{-3}$.}
\label{tab_clump_max_vel}
\end{table}

As expected, there are more clumps when the density threshold is increased. For 
low densities $\rho^\mathrm{min}_\mathrm{NiCoFeX}$, large volumes are connected 
and form big clumps. If the threshold for the definition of the clump is 
increased, different high-density `islands' get disconnected from each other and 
form separate clumps. However, as a secondary effect some clumps disappear 
completely because their highest density of NiCoFeX elements falls below the 
selected threshold. For example see model B15$_0$ or B15, where the number 
of clumps decreases despite an increase of $\rho^\mathrm{min}_\mathrm{NiCoFeX}$ 
from $F_\rho=0.4$ to $F_\rho=0.3$. For the volume and surface filling factors,
we see a monotonic trend of decreasing values with increasing density threshold 
for all models. Note that we allow for volume filling factors larger than one, 
which states that the NiCoFeX-rich ejecta fill a larger volume than that given 
by a sphere of a particular radius. For model B15, $\bar v_{1500}$ and 
$F_\rho=0.9$, we find $V_\mathrm{NiCoFeX}^\mathrm{1500}=1.509$, which means that 
significant parts of the NiCoFeX-rich ejecta move faster than $\bar 
v=1500\,$km/s. 

Let us compare the different prescriptions for the $\beta$ decay in model B15. 
The density threshold of the clumps containing $90\%$ of the NiCoFeX mass 
decreases from $\rho^\mathrm{min}_\mathrm{NiCoFeX}=0.021\,$g/cm$^3$ for B15$_0$ 
to $\rho^\mathrm{min}_\mathrm{NiCoFeX}=0.018\,$g/cm$^3$ for B15 and finally to 
$\rho^\mathrm{min}_\mathrm{NiCoFeX}=0.016\,$g/cm$^3$ for B15$_\mathrm{X}$. This 
decrease has two main reasons: the extra mixing in particular at the finger 
borders caused by instabilities due to the inflation \citep[see 
also][]{Basko1994,Blondin2001,Chevalier2005}, and the reduction of the 
densities inside the NiCoFeX-rich ejecta due to the inflation. The same trend of 
decreasing densities with stronger $\beta$ decay holds for all fractions of the 
total NiCoFeX mass, B15$_0$ has always the highest and B15$_\mathrm{X}$ the 
lowest $\rho^\mathrm{min}_\mathrm{NiCoFeX}$.
The number of clumps is also related to the inflation of the clumps. The more 
the initially separated clumps inflate, the more of the clumps merge. For 
$F_\rho=0.9$, there are 21 clumps for B15$_0$, 9 
clumps for B15, and only 5 clumps for B15$_\mathrm{X}$. The opposite trend 
holds 
for the respective volume and area filling factors. The stronger the $\beta$ 
decay is, the larger are the filling factors $V_\mathrm{NiCoFeX}^\mathrm{x}$ 
and $A_\mathrm{NiCoFeX}^\mathrm{x}$. This is expected because the 
inflation leads to an increase of volume and area. 

Model N20 has the smallest number of individual clumps. This can already be 
seen in Fig.\,\ref{fig_iso_all}, where this model is the most spherically 
symmetric. It is also the only model without significantly extended 
NiCoFeX-rich fingers. Therefore, all ejecta are at comparable radii and one big 
central bubble dominates. When increasing the density threshold only a small 
number of clumps show up. The two RSG models L15 and W15 have comparable 
numbers of clumps, which are significantly larger than the one for model N20. 
The NiCoFeX-rich 
fingers in Fig.\,\ref{fig_iso_all} extend to larger radii than the bulk of the 
material. These fast ejecta form many separated clumps when the density 
threshold is increased, and the structures get disconnected from the central 
bubble. Model B15 has an intermediate number of clumps.

Among the models with standard $\beta$ decay, model L15 has volume filling 
factors for $\bar v_{1500}$ and $\bar v_{2500}$ that are at least $50\%$ 
higher than those of all other models (B15, N20, W15). 
$V_\mathrm{NiCoFeX}^\mathrm{fastest}$ 
seems not to follow the same trend, however, remember that each model has a 
different value of $v_\mathrm{fastest}$, see Table\,\ref{tab_clump_max_vel}. 
Compared to models B15 and N20, which have larger 
$V_\mathrm{NiCoFeX}^\mathrm{fastest}$, model L15 has the fastest moving 
NiCoFeX, and, hence, the volume of the sphere with the corresponding radius is 
the largest among these models. So despite of having the largest 
$V_\mathrm{NiCoFeX}^\mathrm{1500}$ and $V_\mathrm{NiCoFeX}^\mathrm{2500}$, model 
L15 does not have the largest $V_\mathrm{NiCoFeX}^\mathrm{fastest}$. 
In Table\,\ref{tab_clump_max_vel}, we note that the velocity of 
the fastest moving ejecta of model W15 is even slightly faster than that of 
model L15, which seems to contradict our discussions related to 
Tables\,\ref{tab_vel_final} and \ref{tab_vel_final_1percent}. However, in those 
tables we considered the mean velocities of the bulk and of the fastest one 
percent of the ejecta. Here, we take the absolute value of the velocity of the 
very fastest ejecta having $X_\mathrm{NiCoFeX}>10^{-3}$, which make up less than
$10^{-6}$ of the total NiCoFeX-mass only.

The surface filling factors of all the models are more similar to each 
other. Model B15 has the largest values for 
$A_\mathrm{NiCoFeX}^\mathrm{1500}$ and $A_\mathrm{NiCoFeX}^\mathrm{2500}$. The 
different behaviour of the 2D projections and 3D analysis can be explained by the 
different morphologies of the models: The NiCoFeX-rich clumps and fingers of 
model B15 are distributed more isotropically than in the other models (see 
Figs.\,\ref{fig_iso_B15} and \ref{fig_iso_all}). In a surface projection, this 
leads to almost complete coverage of the entire surface within a square of side 
length $2\times r_{1500}$. Models L15 and W15, which only have a few 
NiCoFeX-rich fingers in distinct directions, only have 
$A_\mathrm{NiCoFeX}^\mathrm{1500}\simeq0.87$ and $0.75$ for 
$F_\rho=0.9$, respectively.  The large 
volume filling factor of L15 can be explained by the few large fingers that 
extend to very large radii compared to the central bubble of the NiCoFeX-rich 
ejecta. The fingers of model B15 are less extended and, hence, the volume 
filled by these structures is comparably smaller than those in model L15.
Without any extended NiCoFeX-rich fingers, model N20 has the
smallest $V_\mathrm{NiCoFeX}^\mathrm{1500}$ and 
$V_\mathrm{NiCoFeX}^\mathrm{2500}$ of all models with standard $\beta$ decay. 
Again, $V_\mathrm{NiCoFeX}^\mathrm{fastest}$ does not follow this trend 
because the $v_\mathrm{fastest}$ are different for all models.
As for the volume filling factors, model N20 has also the smallest surface 
filling factor $A_\mathrm{NiCoFeX}^\mathrm{2500}$. 
However, the surface filling factor for $\bar v_\mathrm{1500}$ is not following 
this trend. $A_\mathrm{NiCoFeX}^\mathrm{1500}$ is slightly larger for model N20 
than that for model W15 for all $F_\rho$. The larger 
occupied surface area shows that the fastest ejecta of the central spherical 
bubble of model N20 is moving faster than that of model W15 (see also 
Fig.\,\ref{fig_vel_1y}), and that this difference cannot be cured by the few 
extended and fast-moving NiCoFeX-rich fingers.
Note that the volume and surface filling factors  
$V_\mathrm{NiCoFeX}^\mathrm{1500}, V_\mathrm{NiCoFeX}^\mathrm{2500}, 
A_\mathrm{NiCoFeX}^\mathrm{1500}$, and $A_\mathrm{NiCoFeX}^\mathrm{2500}$ 
depend sensitively on the explosion energy of the model. The higher the 
explosion energy is, the faster the ejecta should propagate and, hence, the 
larger the filling factor for the volumes and surfaces determined by fixed 
velocities should be after $1\,$yr. Only the entries compared to the fastest 
moving NiCoFeX-rich ejecta, $V_\mathrm{NiCoFeX}^\mathrm{fastest}$ and 
$A_\mathrm{NiCoFeX}^\mathrm{fastest}$, should be less sensitive to the 
particular value of the explosion energy, since in this case 
$v_\mathrm{fastest}$ also scales with the explosion energy. 

\begin{figure*}  
\includegraphics[width=.49\textwidth]{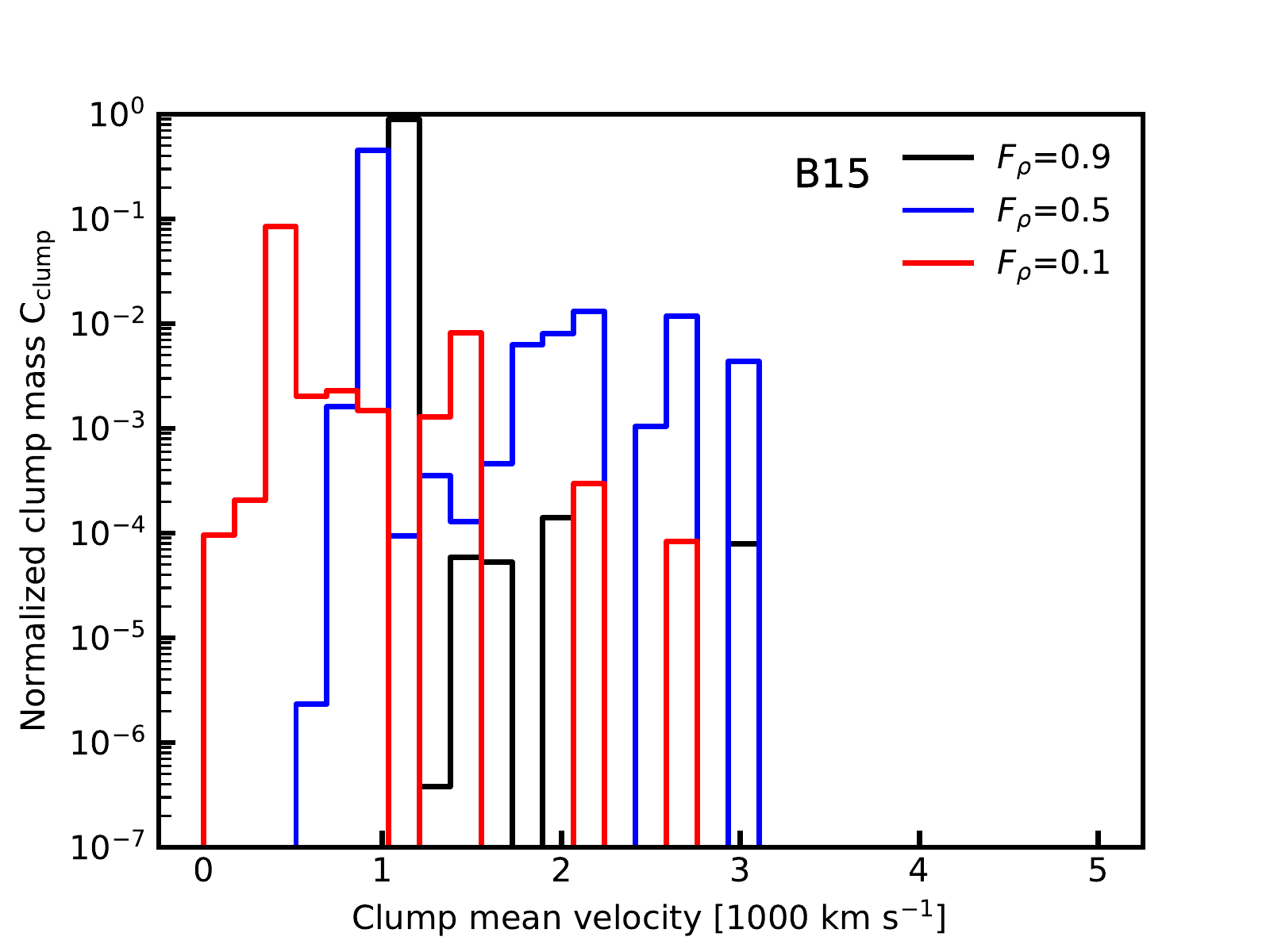}
\includegraphics[width=.49\textwidth]{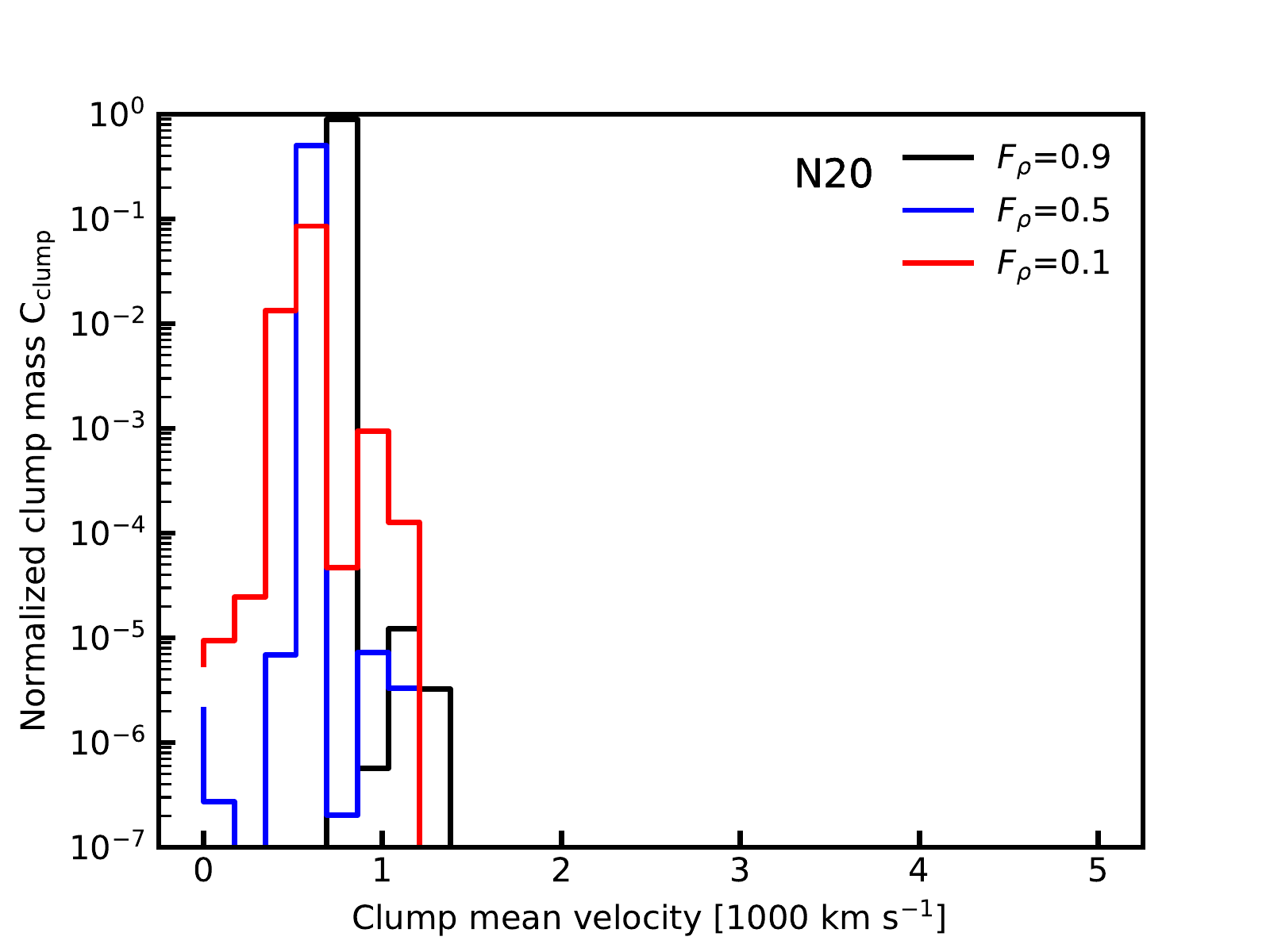}\\
\includegraphics[width=.49\textwidth]{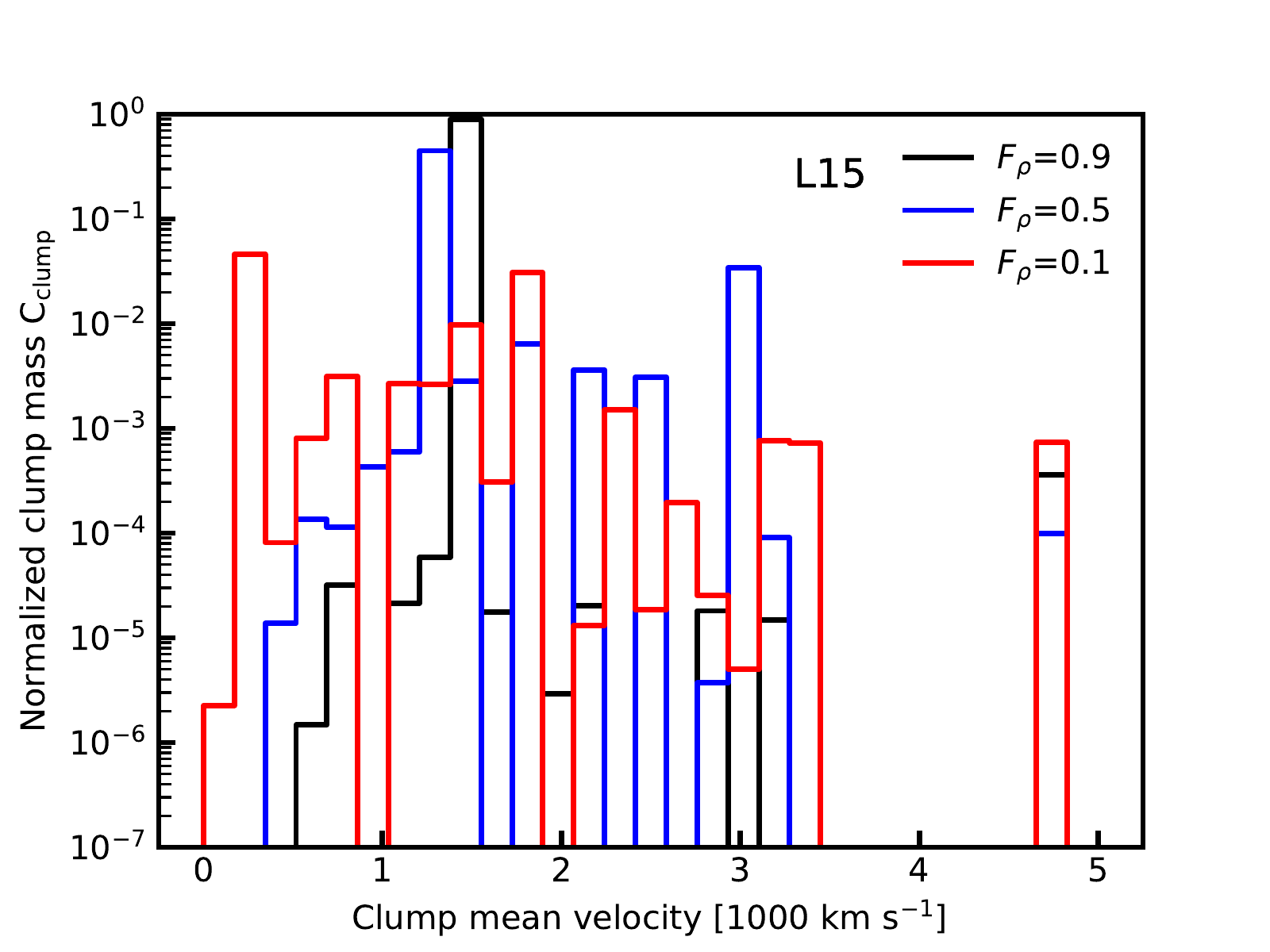} 
\includegraphics[width=.49\textwidth]{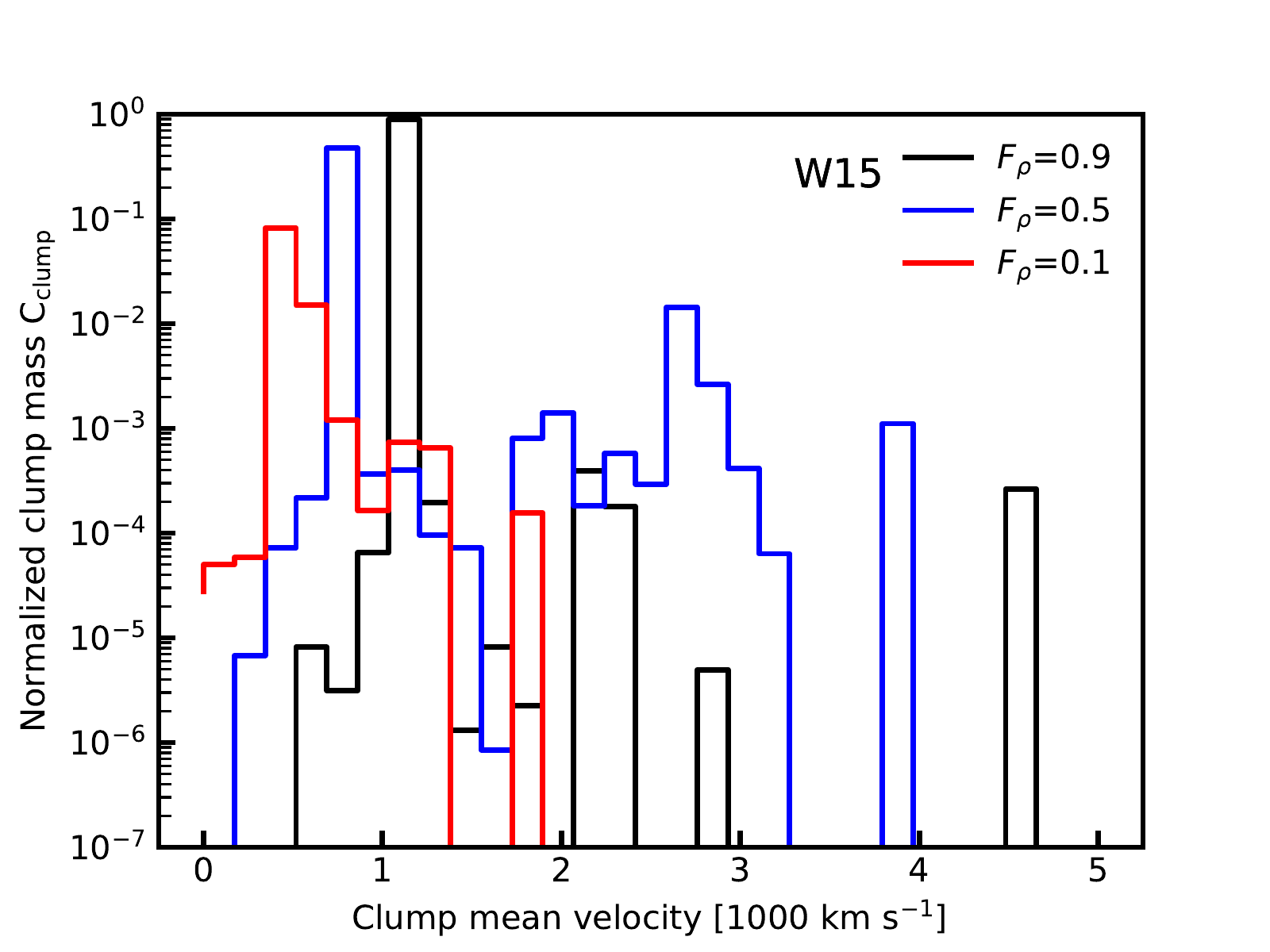}
\caption{ Distributions $C_\mathrm{clump}$ of the 
normalized mass in clumps with a given mean 
velocity $\bar v_\mathrm{clumps}$ containing $90\%$, $50\%$, and
$10\%$ of the ejecta with the highest densities of NiCoFeX for models B15 ({\it 
top left panel}), N20 ({\it top right 
panel}), L15 ({\it bottom left panel}), and W15 ({\it bottom right panel}), 
respectively. The normalization is relative to the total NiCoFeX mass.
When considering $90\%$ of the NiCoFeX mass ($F_\rho=0.9$), almost all material 
is concentrated in one central bubble with mean clump velocity of 
approximately $\bar v\sim1000\,$km/s for all models. When reducing 
$F_\rho$ and, consequently, increasing the density threshold, the main bubble 
shifts to lower velocities, and clumps at different mean clump velocities 
disconnect from the central bubble. Due to the little mixing and weak 
self-reflected shock, model N20 has the slowest clumps. The bins of the 
mean velocity have a width of $167\,$km/s.
}
\label{fig_mass_vel}
\end{figure*}

When considering the clumps as completely disconnected, we can define a mean 
velocity for each individual clump $\bar v_\mathrm{clump}$:
\begin{equation}\label{eq_mean_clump_vel}
\bar v_\mathrm{clump}=\frac{\int\limits_{\mathrm{clump}} \rho v dV} 
{M_\mathrm{clump}}\,.
\end{equation}
To see how many clumps and also how much mass propagate with
a certain velocity, we plot the mass inside the clumps normalized to 
the total NiCoFeX mass 
\begin{equation}
C_\mathrm{clump}(\bar v_\mathrm{clump})
\equiv\frac{M_\mathrm{clump}(\bar 
v_\mathrm{clump})}{M^\mathrm{tot}_\mathrm{NiCoFeX} } \, ,
\end{equation}
as a function of $\bar v_\mathrm{clump}$ in Fig.\,\ref{fig_mass_vel}.
Here, $M_\mathrm{clump}(\bar v_\mathrm{clump})$ is the NiCoFeX mass of the 
clumps with mean velocity $\bar v_\mathrm{clump}$, and 
$M^\mathrm{tot}_\mathrm{clumps}$ is the total NiCoFeX mass (including the 
NiCoFeX not contained in the clumps).  If more than one clump falls within the 
same velocity bin, we add up the normalized masses of these clumps.
For the central bubble, i.e. the `clump' with 
the largest $C_\mathrm{clump}$, the latter is not a 
useful measure. For large $F_\rho$, the central ejecta 
can be very extended and be connected to very elongated NiCoFeX-rich fingers. 
Consequently, the integral in Eq.\,(\ref{eq_mean_clump_vel}) gives essentially 
the bulk velocity of the NiCoFeX elements. 

Comparing $C_\mathrm{clump}$ for $F_\rho=0.9$ (black curve, top left panel 
of Fig.\,\ref{fig_mass_vel}) and 
$F_\rho=0.5$ (red) of model B15, we see that there are more clumps with higher 
velocities for $F_\rho=0.5$. The big central ejecta bubble at $\bar 
v_\mathrm{clump}\gtrsim1000\,$km/s for $F_\rho=0.9$ splits into many smaller 
clumps for $F_\rho=0.5$, some with higher velocities, but there are also more 
clumps with lower velocities $\bar v_\mathrm{clump}<1000\,$km/s. The 
mean velocity of the central bubble reduces from 
$\bar v_\mathrm{clump}\gtrsim1000\,$km/s for $F_\rho=0.9$ 
(black line) to $\bar v_\mathrm{clump}\lesssim1000\,$km/s for 
$F_\rho=0.5$ (red line) to $\bar v_\mathrm{clump}\lesssim500\,$km/s for 
$F_\rho=0.1$ (blue line). 
This decrease of the mean velocity is a consequence of the fragmentation 
of the big central clump. The still connected part of the central 
ejecta shrinks and has higher densities. Since this means considering denser 
material farther inside, the mean velocity of these ejecta decreases compared 
to those 
of a more extended central bubble for $F_\rho=0.9$.
Similar trends are found for all models shown in the other panels of 
Fig.\,\ref{fig_mass_vel}. 

\begin{figure*}  
\includegraphics[width=.92\textwidth]{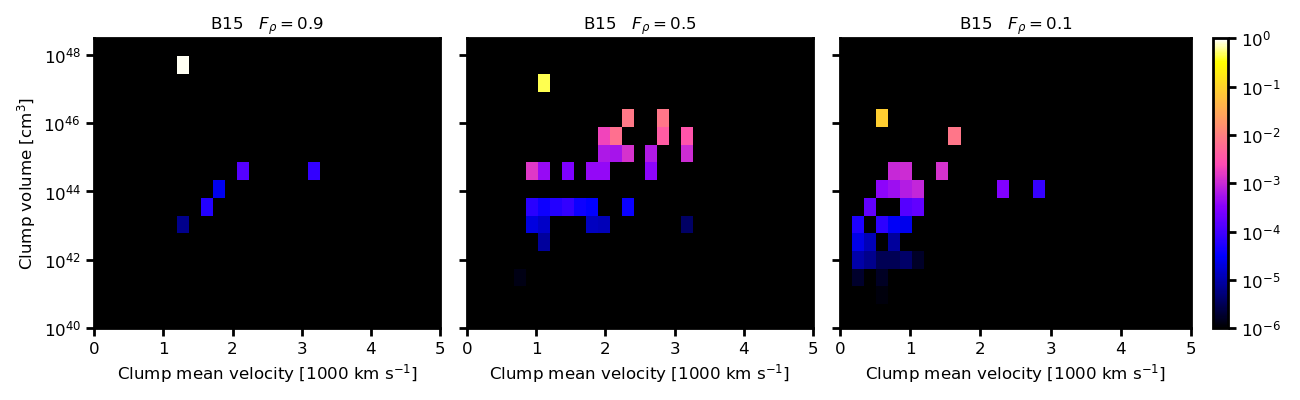}\\
\includegraphics[width=.92\textwidth]{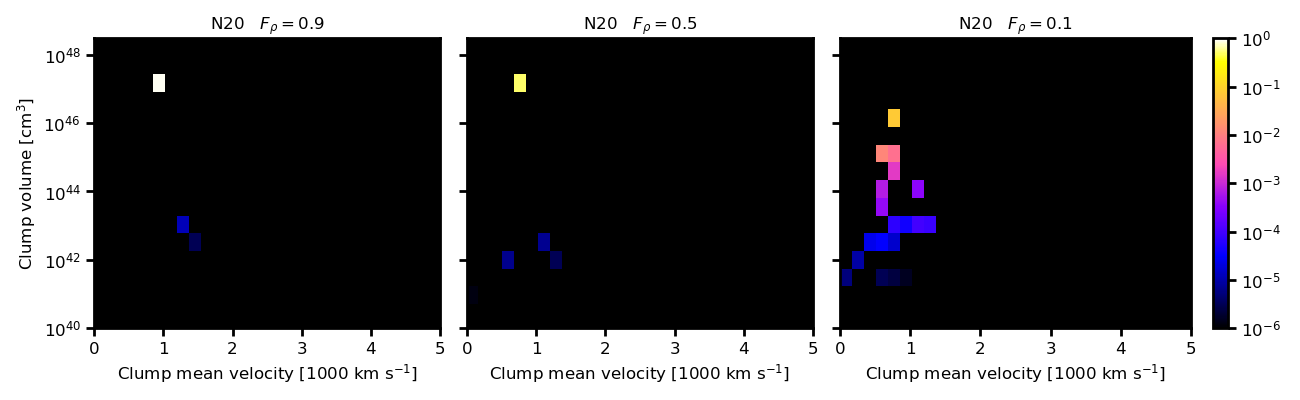}\\
\includegraphics[width=.92\textwidth]{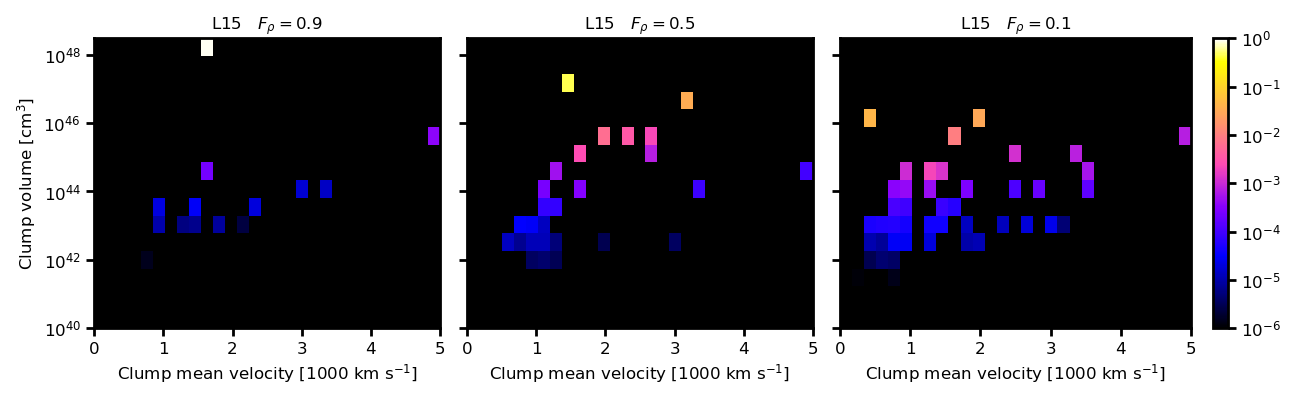}\\
\includegraphics[width=.92\textwidth]{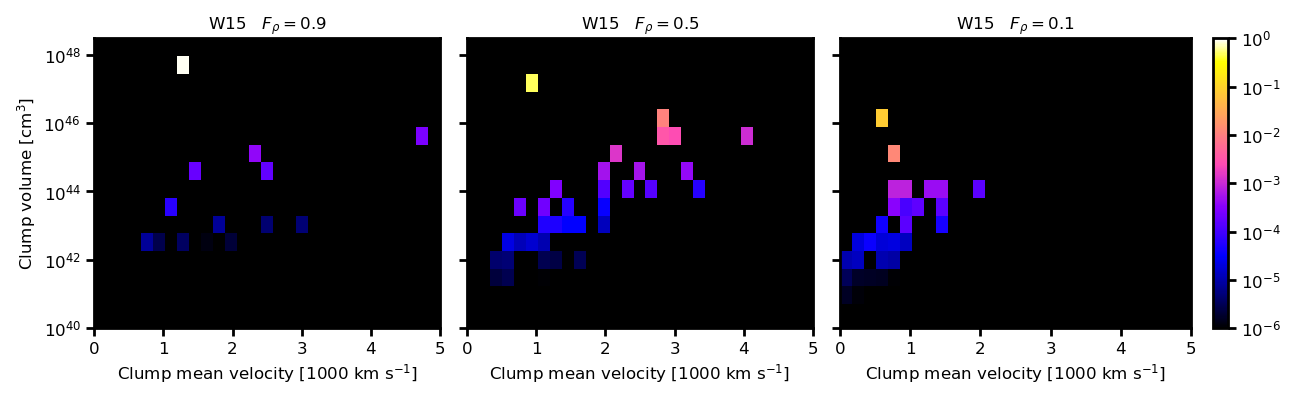}
\caption{Pseudocolour plots of the mass in individual clumps 
normalized by the total mass of NiCoFeX as function of the clump velocity 
(x-axis), and clump volume (y-axis) for different models. We 
assume that the clumps contain the densest NiCoFeX material 
and their mass sums up to different fractions of the total mass of NiCoFeX: 
$90\%$  ({\it left column}), $50\%$ ({\it central column}), and $10\%$ ({\it 
right column}). If more than one clump falls within the same velocity and 
volume bin, we add up the normalized masses of these clumps.
When considering the densest $90\%$ ($F_\rho=0.9$), all models 
have a dominant clump (the central bubble) in volume and fractional mass around 
$\bar v\sim1000\,$km/s, visible as the white or yellow squares in the top left 
corner of the panels. Model N20 (second row) has the fewest number of 
clumps followed by model B15 (top row). The two RSG models L15 (third row) and 
W15 
(bottom row) have more clumps with a wider spread in velocity. Considering 
$F_\rho=0.5$, the large central bubble fragments and smaller 
clumps get disconnected for all models. This fragmentation results in a wide 
distribution of clump sizes and velocities. This also happens when 
lowering $F_\rho$ to $0.1$. However, in this case many clumps also 
disappear, because their densities fall below the corresponding threshold (see 
in particular model W15 central and right lower panels). The largest dispersion 
of sizes and clumps is found for model L15 (third row). The binning in velocity 
is $167\,$km/s, the logarithmic bin width in volume is 
$0.5\log(\mathrm{cm}^3)$, and the colour scale of the normalized mass in the 
clumps is logarithmic.} 
\label{fig_mass_velvol}
\end{figure*} 

The fastest clumps of model N20 do not exceed velocities $\bar 
v_\mathrm{clump}\lesssim1500\,$km/s. The fastest clumps with $\bar 
v_\mathrm{clump}>4000\,$km/s are found for models L15 and W15. However, 
these fast clumps contain very little mass ($<10^{-3}$ of the total mass). 
Comparing the normalized masses of the clumps of models L15 (bottom 
left panel) and W15 (bottom right panel), we find that for 
$F_\rho=0.9$ (black lines) the distributions look 
quite similar, with a main peak around $1000\mathrm{~km/s}<\bar 
v_\mathrm{clump}<1500\mathrm{~km/s}$, a wide 
spread of mean clump velocities up to $\bar 
v_\mathrm{clump}\lesssim3000\,$km/s, and  one very fast clump. Also, 
$\bar v_\mathrm{clump}$ of the densest $50\%$ of the NiCoFeX ejecta decrease 
in both cases (blue lines). However, considering only the densest $10\%$,
the distribution of normalized clump mass as a function of velocity of model 
W15 is very narrow and constrained to low velocities $0<\bar 
v_\mathrm{clump}<2000\,$km/s. In contrast, model L15 still has a very broad 
distribution of velocities for its densest clumps.

In Fig.\,\ref{fig_mass_velvol}, we combine the information about the size of 
the clumps (y-axis), the mean clump velocities (x-axis) and the mass contained
in the clumps. The colour scale represents the NiCoFeX mass of a clump at the 
given velocity and for a given volume, normalized by the total mass of NiCoFeX 
of the simulation.  If more than one clump falls within the same velocity 
and 
volume bin, we add up the normalized masses of these clumps.
As expected, the central ejecta bubble (white or yellow 
squares in the top left corner of the panels) is always dominant in volume and 
mass fraction. For $F_\rho=0.9$, we see only a few clumps 
apart from the central ejecta. In general, models W15 and L15 have more 
clumps than B15, and model N20 has the smallest number of clumps (this actually 
holds for all density thresholds). The spread in clump sizes is comparable in 
all models, while the spread in velocity is significantly smaller for model N20 
compared to the other models. As noted before, for the $10\%$ of the densest 
NiCoFeX clumps, model L15 has the largest velocity spread, while at 
$F_\rho=0.9$ the spreads of models B15, L15, and W15 are 
comparable. Apart from model N20, the clump volume is correlated with the 
mean clump velocity as is most apparent in the third and fourth rows of the 
central column. This means that there is a general trend that the biggest clumps 
(apart from the central bubble) have the highest velocities, which makes 
sense because the fastest clumps have expanded most. However, due to 
their lower densities the fastest clumps do not necessarily have the highest 
clump masses.
 
\begin{figure*}
\includegraphics[width=.49\textwidth]{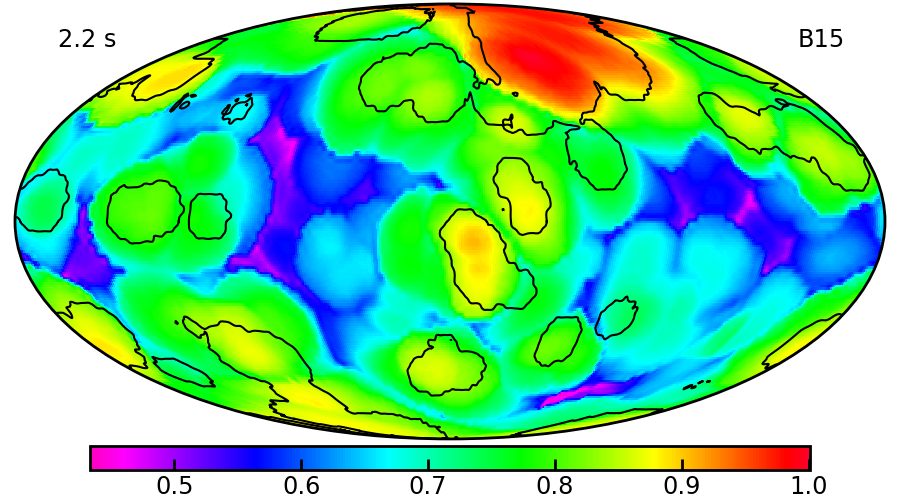}
\includegraphics[width=.49\textwidth]{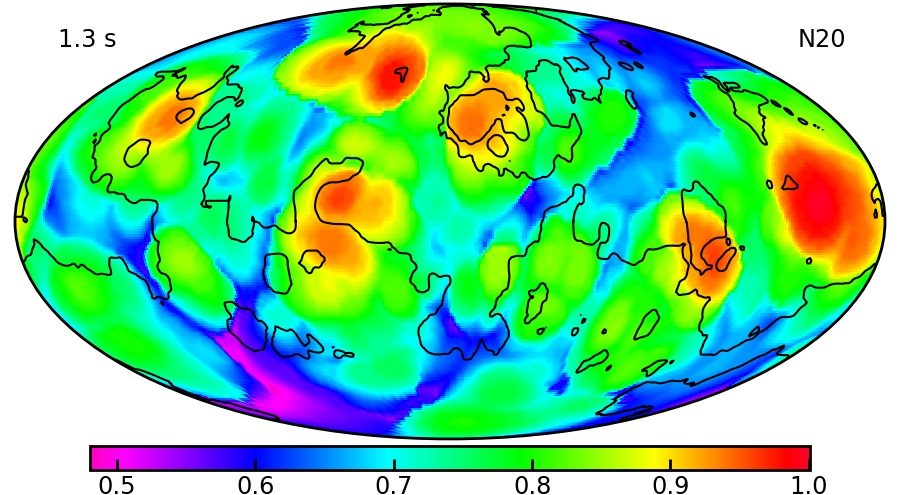}\\
\includegraphics[width=.49\textwidth]{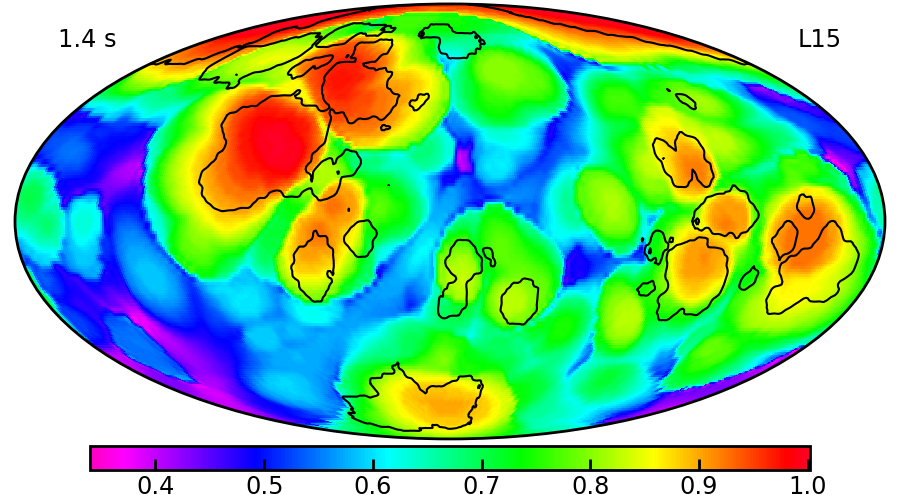}
\includegraphics[width=.49\textwidth]{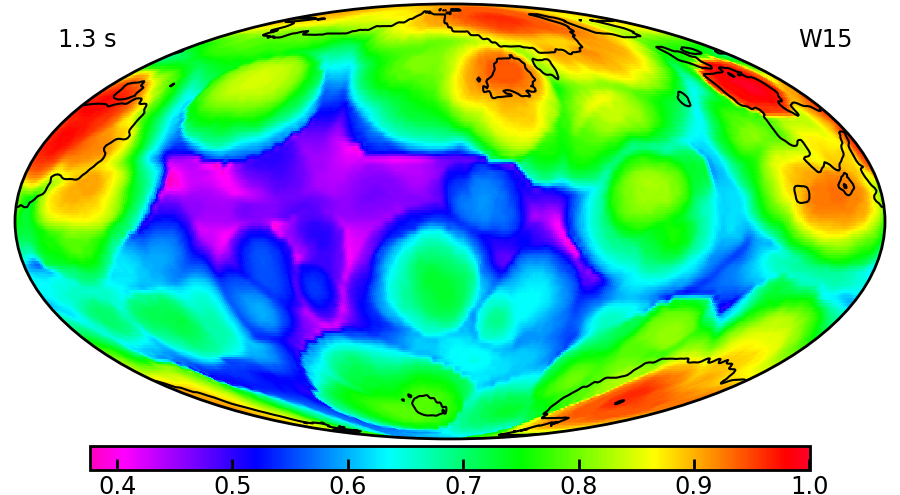}
\caption{Colour coded are the Mollweide projections of the 
normalized maximal radii $R^{56}_\mathrm{max}(\theta,\phi) / 
R^\mathrm{tot}_\mathrm{max}$ at $t\sim1-2\,$s, where 
$R^{56}_\mathrm{max}(\theta,\phi)$ is defined at $X_\mathrm{NiFoFeX}>10^{-3}$  
and $R^\mathrm{tot}_\mathrm{max} = 
\max\left(R^{56}_\mathrm{max}(\theta,\phi)\right)$ of all directions 
$(\theta,\phi)$. The black 
contour lines are the locations where $R^{56}_\mathrm{max}(\theta,\phi)=0.5 
R^\mathrm{tot}_\mathrm{max}$ at $t\sim1\,$yr. The 
most prominent structures of models B15, L15, and W15 are in the same 
directions at early and late times. Model N20 is an exception and does not 
exhibit any tight correlation between the biggest initial and final structures, 
because its initial nickel plumes are efficiently decelerated by the reverse 
shock and therefore not able to penetrate deep into the hydrogen envelope. 
Instead, they fragment into smaller-scale structures growing from RT instability 
at the He/H interface.}
\label{fig_moll}
\end{figure*}

\subsubsection{Comparison with initial explosion  asymmetries}
\label{sec_moll}

The final asymmetries, in particular the biggest structures, in the iron 
distribution are clearly linked to the biggest initial bubbles of 
neutrino-heated ejecta created at the onset of the explosion. This can be seen 
by comparing the isosurface plots of the NiCoFe-rich ejecta at 
early times in the left column of panels of figure 7 of 
\cite{Wongwathanarat2015} to the late times in Figs.\,\ref{fig_iso_B15} and 
\ref{fig_iso_all}. The fastest plumes shaped by the buoyant rise of 
neutrino-heated matter in the initial moments of the explosion end up as the 
biggest extended fingers or clumps at late times.

To emphasize the correlation between the initial and final asymmetries further, 
we plot the Mollweide projections of the maximal radius 
$R^{56}_\mathrm{max}(\theta,\phi)$ in a given direction $(\theta,\phi)$ at the 
time of shock revival compared to the corresponding radius at $t\sim1\,$yr in 
Fig.\,\ref{fig_moll}. 
Remember that $R^{56}_\mathrm{max}(\theta,\phi)$ was defined as the outermost 
radius where $X_\mathrm{NiFoFeX}>10^{-3}$. For the color-coding we rescaled the 
$R^{56}_\mathrm{max}(\theta,\phi)$ by the maximal 
$R^{56}_\mathrm{max}(\theta,\phi))$ of all directions $(\theta,\phi)$, 
$R^\mathrm{tot}_\mathrm{max} = 
\max\left(R^{56}_\mathrm{max}(\theta,\phi)\right)$, at the 
indicated times $t\sim1-2\,s$. The black contour lines 
are the locations where $R^{56}_\mathrm{max}(\theta,\phi)=0.5 
R^\mathrm{tot}_\mathrm{max}$ at $t\sim1\,$yr. These contour lines coincide 
almost perfectly with the highest amplitudes represented by yellow and red 
colours for 
models B15, L15, and W15. This correlation demonstrates that the most extended 
structures at $t\sim1-2\,$s are in the same directions as the most prominent 
features at 
$t\sim1\,$yr. Model N20, which does not have very extended structures, shows no 
clear correlations between the initial and the late time asymmetries.

\subsubsection{Spherical harmonics decomposition}
\label{sec_sphhar}

\begin{figure*}
\includegraphics[width=.62\textwidth]
{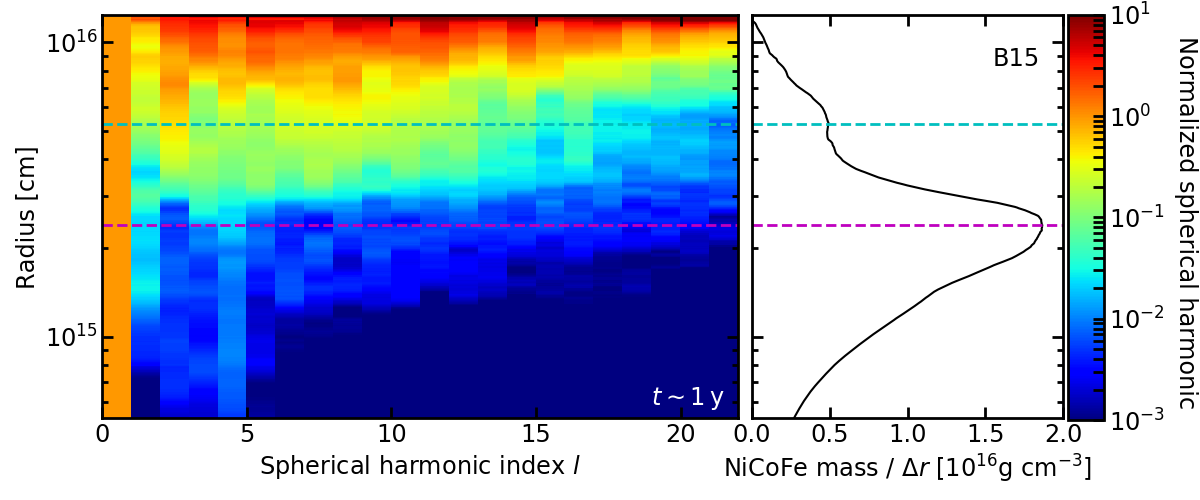}
\includegraphics[width=.37\textwidth]{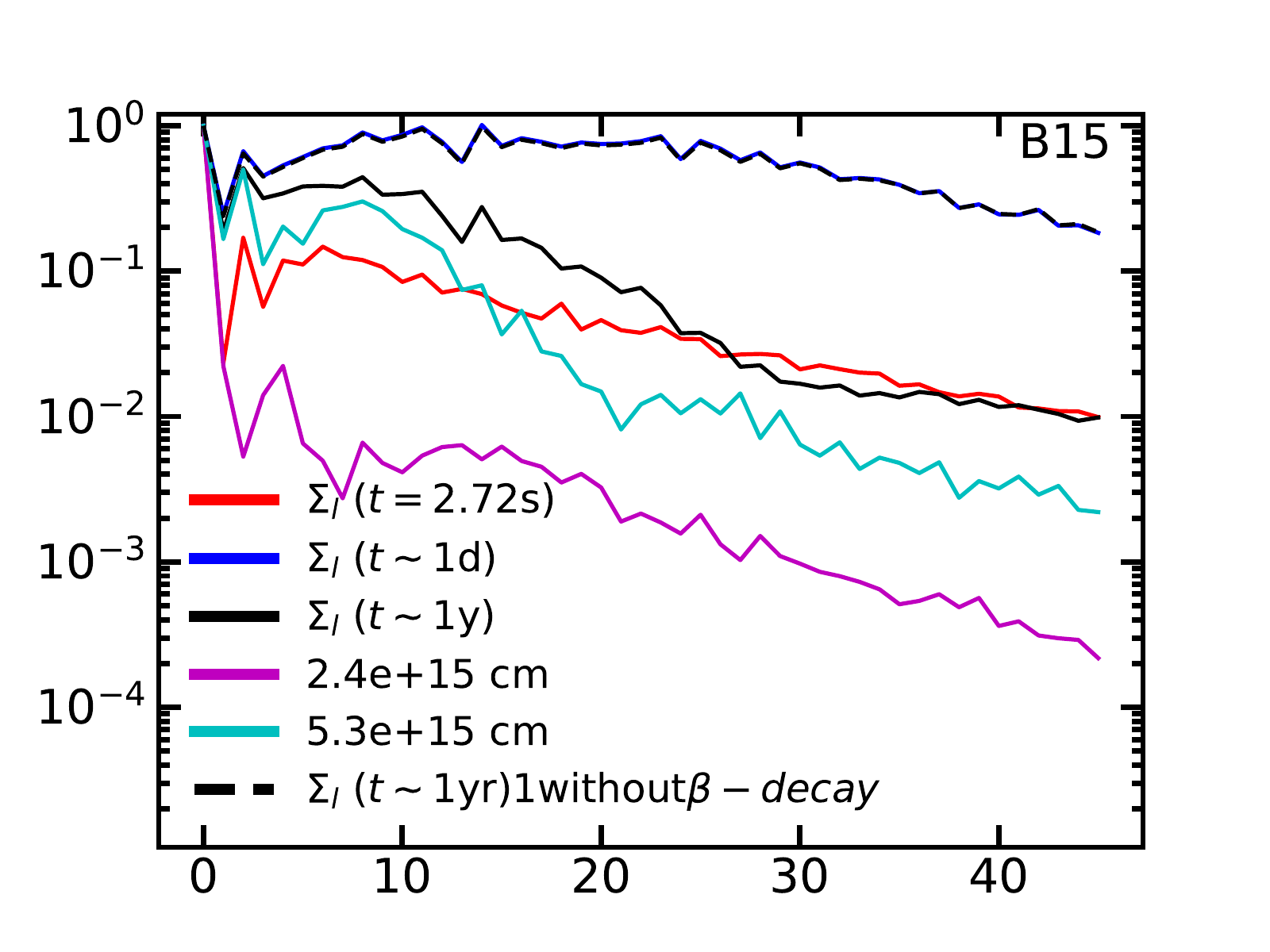}\\
\includegraphics[width=.62\textwidth]
{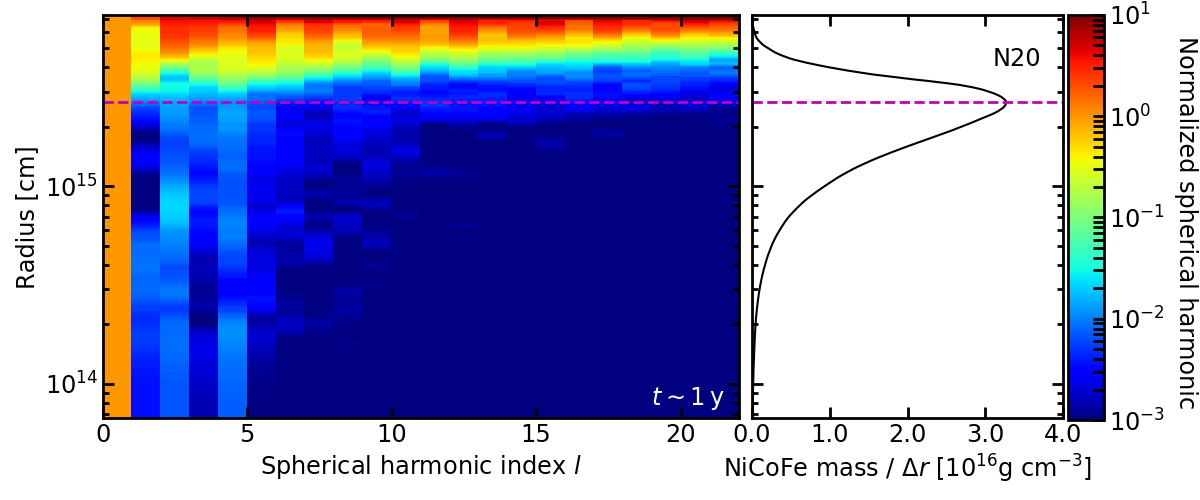}
\includegraphics[width=.37\textwidth]{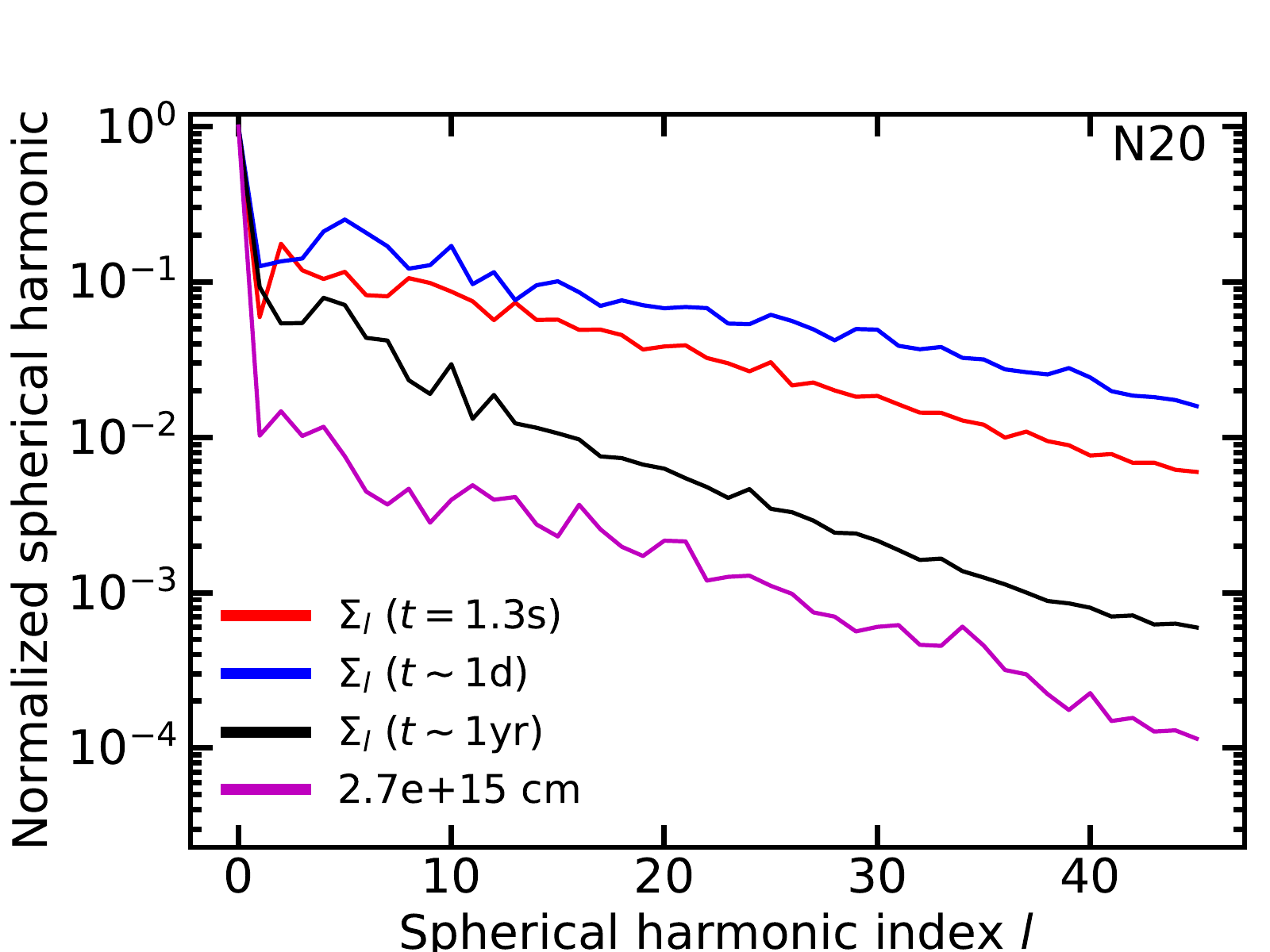}\\
\includegraphics[width=.62\textwidth]
{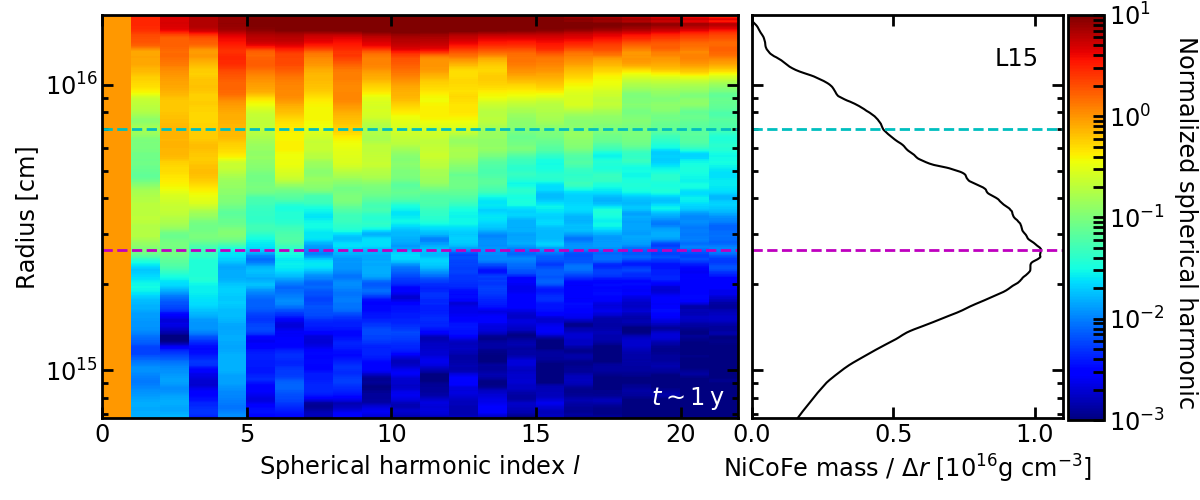}
\includegraphics[width=.37\textwidth]{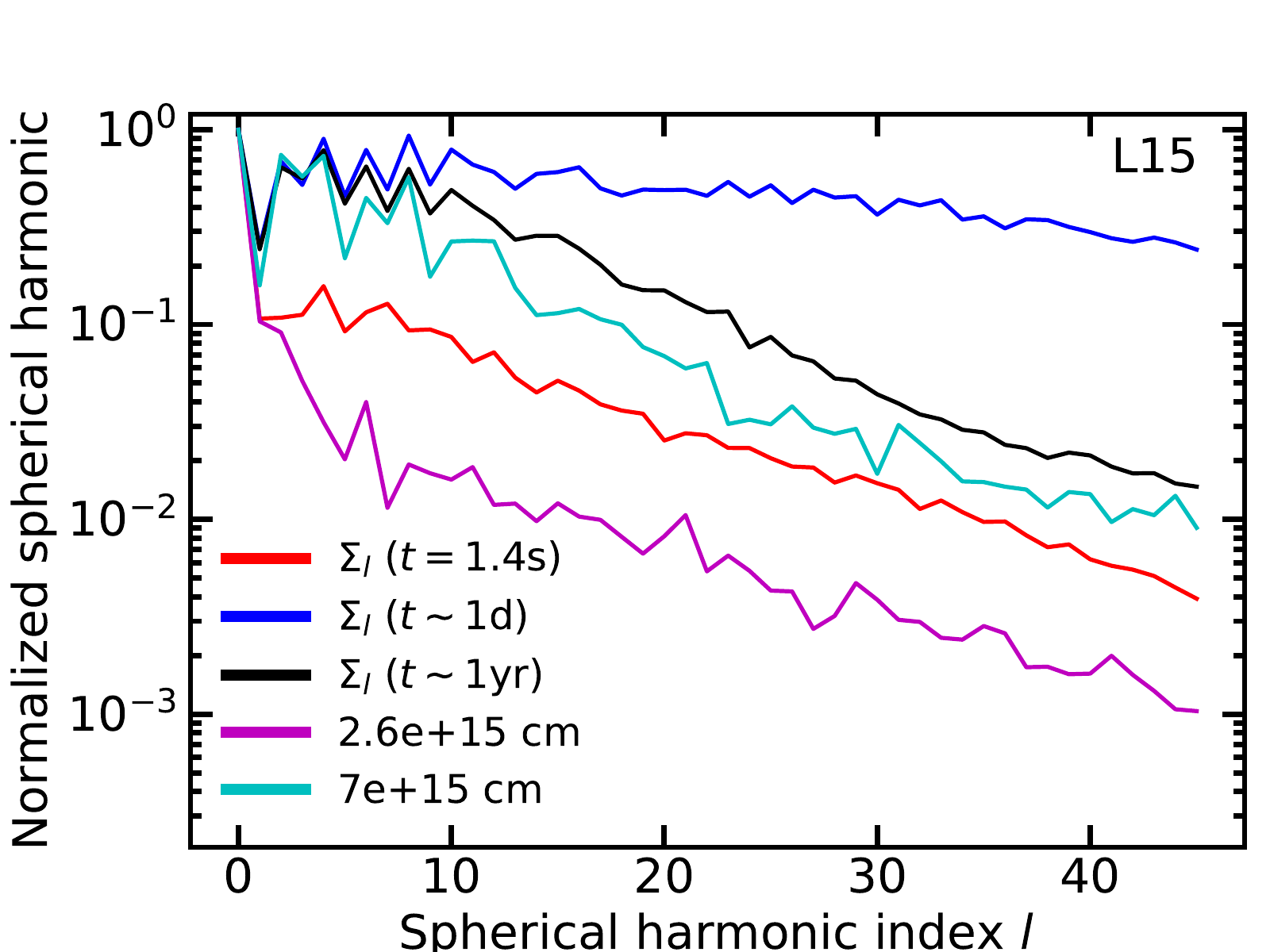}\\
\includegraphics[width=.62\textwidth]
{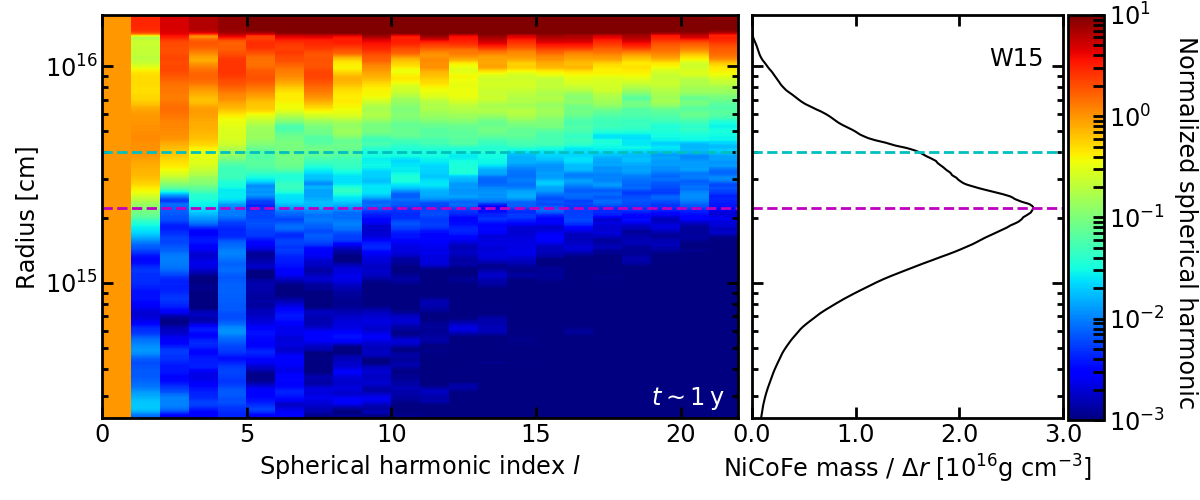}
\includegraphics[width=.37\textwidth]{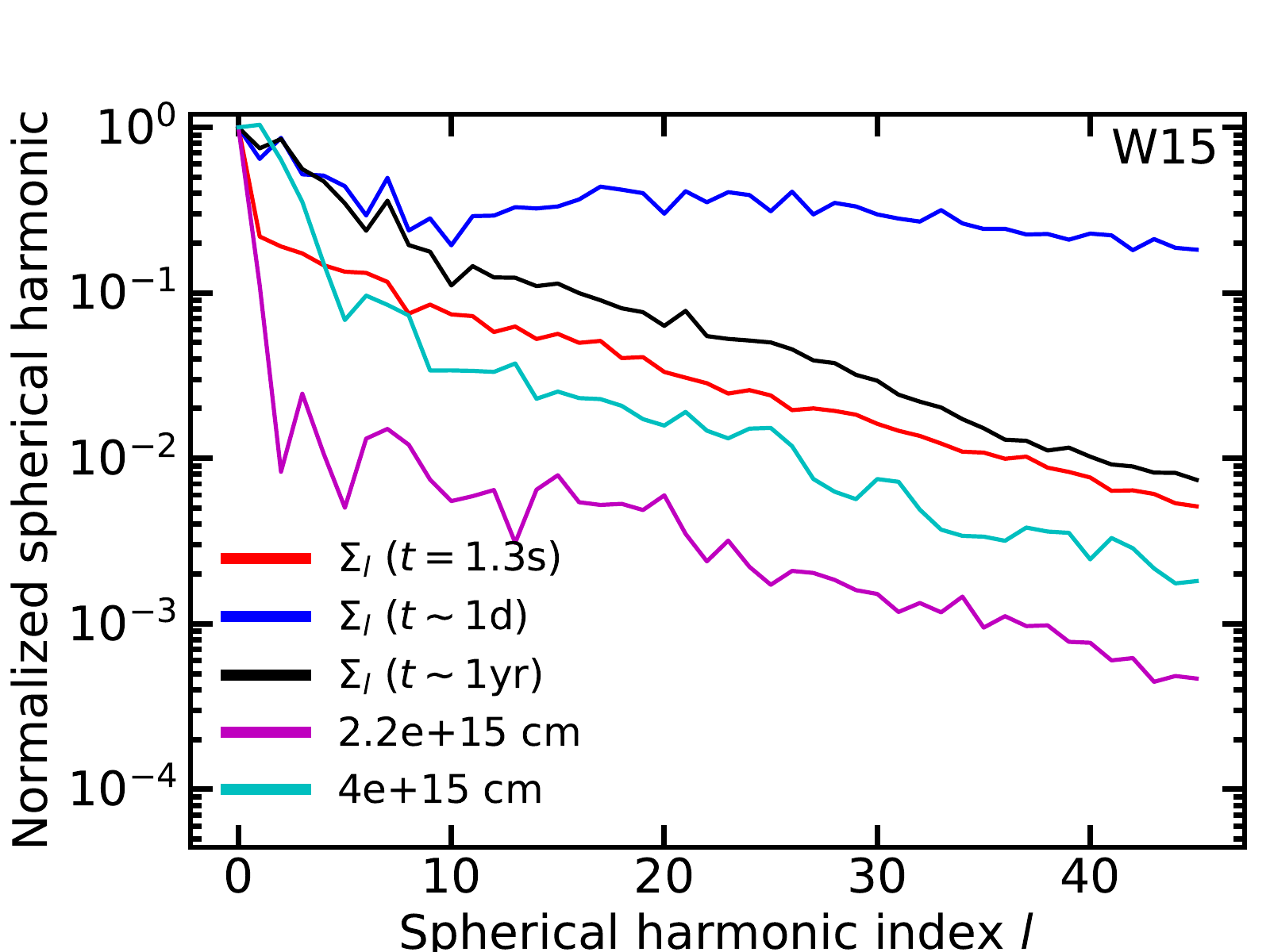}
\caption{Spherical harmonics decomposition of the linear momentum of 
NiCoFe-rich matter in radial direction of the 
clumps containing $90\%$ of the densest material of NiCoFe for different models 
B15 (top row), N20 (second row), L15 (third row), and W15 (bottom row). {\it 
Left column:} Spherical harmonic coefficients at a given radius normalized to 
the monopole as a function of $r$ and multipole order $l$ at $t\sim1\,$yr. {\it 
Central column:} Mass of NiCoFe per unit length scale as a function of radius. 
{\it 
Right column:} Normalized 
spherical harmonics coefficient at the radii given by the horizontal dashed 
lines in the left and central panels with the same colour. The blue and 
black lines represent the sum of the coefficients $c_l$ over all radial cells 
$\Sigma_l$ at $t\sim1\,$yr, $t\sim1\,$d  and at the indicated times 
{\protect \citep[which are approximately the first times given in figure 7 of 
][]{Wongwathanarat2015}}, respectively.  
All models have a dominant monopole representing the 
bulk of the material that is ejected nearly spherically. The most spherical 
model N20 has very weak higher multipoles and is dominated by the monopole 
everywhere, where most of the mass of NiCoFe is located 
($r\lesssim5\times10^{15}\,$cm). Model B15 has many, 
almost isotropically distributed RT fingers extending to large 
radii, explaining the high contribution of the spherical harmonics around 
$l=10$ (top right panel). The RSG models have only a 
few, but quite extended fingers leading to a strong quadrupole ($l=2$) and 
hexadecapole ($l=4$) asymmetry for model L15 (third row, right panel) and a 
strong dipole ($l=1$) and quadrupole ($l=2$) asymmetry for W15 (bottom right 
panel).}
\label{fig_sph_har}
\end{figure*}
To further analyse the spatial distribution of the NiCoFe-rich ejecta, 
we decompose the linear momentum of NiCoFe-rich matter in the radial 
direction of the models into spherical harmonics. The 
corresponding plots of the decomposition normalized to the monopole and for 
each radius of our numerical grid are displayed in the left column of panels 
in Fig.\,\ref{fig_sph_har}. To obtain a measure where most of the mass of the 
NiCoFe-rich ejecta is located we plot the mass of NiCoFe inside shells divided 
by the radial width of the corresponding shells, in the central column of 
panels. We mark the maximum with a magenta dashed line in each panel of 
the left and central columns. For all of our models, the monopole dominates at 
small radii $r\lesssim\mathrm{few}\times10^{15}\,$cm, including the radius 
containing the mass maximum. The dominating monopole at low radii can also be 
seen clearly in the right column of panels, where we plot the spherical 
harmonics decomposition at the radius of the mass maximum (magenta solid 
line). In all models the monopole is at least one order of magnitude larger 
than the dipole component.

When increasing the radius above $r\gtrsim\mathrm4\times10^{15}\,$cm, 
we find that higher multipoles become more and more important and start to 
dominate close to the fastest moving NiCoFe-rich ejecta at the top of each of 
the left panels. However, the mass in the topmost region in the plots is 
negligible. We thus plot an additional line at large, but not too large 
radii, which contains the main asymmetries of the explosions in the 
NiCoFe-rich fingers (dashed and solid cyan lines in the 
central and right panels of Fig.\,\ref{fig_sph_har}). We choose the radius at 
the locations where we see a slight bump of the $M_\mathrm{NiCoFe} / 
\Delta r$ distributions in the corresponding plots in the central panels.
Note that for model N20 due to the absence of extended RT fingers, there is no 
interesting region apart from the maximum of the $M_\mathrm{NiCoFe}/\Delta r$ 
distribution. For large radii the decompositions for the different models 
have different characteristics. Model N20 is dominated by the monopole at all 
radii where significant mass is located, while the other models start to have 
also significant contributions from higher multipoles. For model W15 (bottom 
row, right panel), the dipole at $r\sim4\times10^{15}\,$cm is even stronger 
than the monopole ($l=0$), and model L15 (third row) has very strong 
quadrupole ($l=2$) and hexadecapole ($l=4$) components. These structures are 
also confirmed in the corresponding plots at late times in 
Fig.\,\ref{fig_iso_all} (third and fourth 
row, central and right panels). In Fig.\,\ref{fig_iso_B15}, we see that model 
B15 has many more NiCoFe-rich fingers, and this also reflects in the spherical 
harmonics decomposition. We find a strong quadrupole contribution, but in 
contrast to all other models, we also find a local maximum of the spherical 
harmonic components around $l\lesssim10$ (first row, right panel in 
Fig.\,\ref{fig_sph_har}). 

To characterize the ejecta as a whole, we sum up the spherical harmonics 
coefficients $c_{l,i}$ over all radial zones with indices $i$, and normalize 
them to the respective monopole
\begin{equation}
 \Sigma_l = \frac{\sum_i c_{l,i} }{\sum_i c_{0,i}}\,.
\end{equation}
These summed up spherical harmonics are given by the black 
($t\sim1\,$yr), blue ($t\sim1\,$d) and red (times as indicated) lines in the 
right column of panels of Fig.\,\ref{fig_sph_har}. 

The black curves for all models are qualitatively 
similar to the cyan lines indicating the same trends of the spherical 
harmonics decomposition:
i) All models have a strong monopole ($l=0$). This holds in particular for 
model N20 (second row, right panel) which is completely dominated by it.
ii) Model B15 (top right panel) has a subdominant quadrupole ($l=2$), followed 
by a plateau of almost equally strong multipoles up to $l\sim10$, indicating a 
large number of individual NiCoFe-rich clumps and fingers.
iii) Model W15 (bottom, right panel) has subdominant dipole ($l=1$) and 
quadrupole ($l=2$) components, the former being significantly larger than the 
corresponding component of any other model.
iv) Compared to the other models, model L15 (third row, right panel) has more 
power at higher spherical harmonic coefficients $l>2$. In particular, the 
coefficients of the even indices $l=[2,4,6,8]$ and $l=3$ stick out. 
Also, all higher order multipoles are significantly larger than in any other 
model.

We now compare the final structures at $t\sim1\,$yr with the 
decomposition in spherical harmonics 
 at different times. The red 
curves show $\Sigma_l$ at the onset of the explosion and the blue curves
at $t\sim1\,$d. For low $l\lesssim20$, the black and red curves show similar 
trends for each of the models, i.e. the decomposition is dominated by the same 
coefficients at late and early times. However, there are significant 
differences, 
in particular, between the $\Sigma_l$ at explosion onset and $t\sim1\,$yr on the 
one side and $t\sim1\,$d on the other. The black and blue curves at 
$t\sim1\,$yr and $t\sim1\,$d, respectively, for models L15 and W15 seem to be 
upscaled from the red curves at the early times. This also holds for the red 
and blue curves of model B15, but the $\Sigma_l$ for $l\gtrsim20$ for 
$t\sim1\,$yr are of similar amplitude compared to those at the onset of the 
explosion.
The visible increase of the relative weight of the high $l$ 
amplitudes 
is related to the growth of small-scale RTIs during the expansion of the 
ejecta. In addition, the already fast moving, low-$l$ fingers get further 
accelerated. 
Consequently, the asymmetries seeded initially by the hydrodynamic instabilities 
during the onset of the explosion grow and lead to increasing $\Sigma_l$ also 
for $1<l\lesssim20$. This increase lasts until about $t\sim1\,$d (blue 
curves). Comparing the red and blue curves, we note that the relative weight of 
the high-$l$ amplitudes increases more strongly than the low-$l$ amplitudes. 
After $t\sim1\,$d the small NiCoFeX-rich fingers inflate due to 
$^{56}$Ni decay and many merge to form bigger structures. Therefore, the 
coefficients at $t\sim1\,$yr with $l\gtrsim20$ become relatively weaker 
compared with the lower coefficients ($l\lesssim20$). In addition, the major 
part of the energy deposition of the released $\beta$-decay energy occurs in 
the central bulk part of the NiCoFeX-rich ejecta. This leads to a faster 
acceleration of these central ejecta (compare also discussion in 
Sections\,\ref{sec_radial_vel} and\,\ref{sec_density_dist}).  Consequently, the 
central NiCoFeX-rich bubble inflates and the amplitude of the monopole of the 
spherical harmonics decomposition increases. In the case of model B15, the 
black curve ($t\sim\,1$yr) even crosses the red one, i.e. the small-scale 
structures become less dominant more efficiently. This is probably caused 
by the relatively stronger inflation of the NiCoFeX-rich structures as already 
discussed in Section\,\ref{sec_volume_increase}.

The interpretation of attributing the increase of relative weight of the low 
$l$ spherical harmonics to the $\beta$-decay caused inflation is supported by 
the following comparison to the simulation of model B15$_0$, which does not 
include the $\beta$ decay. The corresponding curve at $t\sim\,1$yr is given in 
the top right panel of Fig.\,\ref{fig_sph_har} as the dashed black curve. It 
differs only marginally from the blue curve at $t\sim1\,$d and at the given 
scale of the plot no differences are visible. In absence of $\beta$ decay, the 
$\Sigma_l$ for $l>1$ stay at very high amplitudes compared to spherical harmonic 
amplitude at $l=1$.

The major exception to the previous considerations is model N20, for which the 
asymmetries only slightly increase between the initial time and $t\sim1\,$d and 
then significantly decrease. This decrease results in a lower black 
curve compared to 
the red one of the second panel of the right column in 
Fig.\,\ref{fig_sph_har}. As we have seen before \citep[see for example 
Fig.\,\ref{fig_iso_all} and the discussion of N20 in][]{Wongwathanarat2015}, 
this model does not have strong RTIs and after $t\sim1\,$d it becomes more 
spherical during the evolution. 
Model N20 is particular, because the growth of the initial 
asymmetries is suppressed due to the interaction with the strong reverse 
shock from the He/H interface, and the structures caused by RTIs are generally 
much smaller. Their low growth factors and short growth times do not allow for 
a strong growth of RTIs. Therefore, the normalized amplitudes for $l>1$ only 
increase slightly until $t\sim1\,$d and later they decrease significantly as a 
consequence of the inflation due to the $\beta$ decay. At the late times, the 
latter leads to a more dominant $l=1$ component. The small 
extended structures outside of the spherical bulk NiCoFeX-rich ejecta merge and 
form a surface with a low level of corrugation. Additionally, the bulk ejecta 
expand due to the $\beta$-decay energy input and this central bubble `swallows' 
some of the slightly more extended fingers at least partially (compare second 
and fourth panel in the left column of Fig.\,\ref{fig_iso_all}).

\section{Conclusions}\label{sec_conclusions}
We studied the long-time evolution of supernova explosions for four progenitor 
models (B15, N20, L15, and W15) starting from the shock breakout and continued 
until the phase of homologous expansion was reached. B15 and N20 were based on 
BSG progenitors, and L15 and W15 on RSG progenitors. For one of the models, we 
performed simulations without (B15$_0$) and with enhanced (B15$_\mathrm{X}$) 
$\beta$ decay. 
In our standard treatment we consider only the radioactive decay of the 
network-produced $^{56}$Ni in shock-heated and neutrino-heated ejecta. This, 
however, is a lower bound of the $^{56}$Ni yield in the explosion, because some 
uncertain fraction of the slightly neutron-rich ($Y_e \le 0.49$) ejecta, whose 
heavy-element content we denote as `tracer-material' or `X-material', 
may actually end up as $^{56}$Ni. We tested the effects of a higher production 
of $^{56}$Ni compared to what we call `standard' $\beta$ decay in B15. To this 
end we added all the heavy nuclei ejected in neutrino heated matter as 
X-material to the mass of $^{56}$Ni in model B15$_\mathrm{X}$. 
\cite{Utrobin2019} used this 
maximal and also a `representative' mass of radioactive $^{56}$Ni, which they 
defined as all network-produced nickel plus $50$\% of the tracer mass. 
Therefore, the results we present here for the standard treatment of $^{56}$Ni 
decay heating are only conservative estimates of the effects of the $\beta$ 
decay, and the final velocity increase as well as the inflation of the volumes 
containing $^{56}$Ni-rich matter may be somewhat larger than found in our 
standard cases. Model B15$_\mathrm{X}$ provides the upper extreme.

Previous simulations until shock 
breakout, which were the starting point of our investigation, describe 
self-consistently the hydrodynamical instabilities that shape the ejecta 
structures from the onset of the explosion to the breakout 
\citep{Wongwathanarat2013, Wongwathanarat2015}. At later times $t\gtrsim1\,$d 
simulated here, there are mainly two new effects that shape the structures: 
the reverse shock that forms at the He/H shell interface and gets self 
reflected at the stellar centre, and the energy input from $\beta$-decaying 
$^{56}$Ni. The reverse shock first propagates backwards in the fluid frame, and 
slows down the expanding ejecta. Later, this shock reaches the innermost and 
densest ejecta. There, it increases the pressure and the temperature, which 
leads 
to the creation of a new outward moving shock, which we call the 
`self-reflected' reverse shock. 
Once this shock propagates outward, it accelerates mainly the central ejecta. 
The interaction of the shock waves with the ejecta depends sensitively on the 
progenitor structure \citep{Wongwathanarat2015}. 

The $\beta$ decay also contributes to the acceleration of the NiCoFe-rich 
($^{56}$Ni+$^{56}$Co+$^{56}$Fe) ejecta that consequently inflate compared to 
their surroundings. This inflation leads to the conversion of 
initially overdense $^{56}$Ni-rich clumps into underdense NiCoFe-rich fingers 
with high-density walls sourrounding and in between individual fingers. The 
corresponding density contrast between the underdense interior, which has up to 
one third of the ambient density for model 
B15$_\mathrm{X}$, and the overdense wall, which can be up to ten times denser 
than the ambient density, can thus be larger than one order of magnitude. 
However, the density contrast is typically less than a factor of $100$ for 
our most extreme model B15$_\mathrm{X}$ and significantly less for model B15 
with standard $\beta$ decay. The effects of the $\beta$ decay on 
the ejecta have been described before for artificial initial explosion 
asymmetries, which were not able to reproduce the high NiCoFe velocities 
required to explain the 
lightcurve of SN~1987A \citep[see e.g.][]{Herant1992, Benz1994, Blondin2001, 
wang2005}. Here, we extended previous discussions by providing a 
quantitative analysis in 3D of the properties of NiCoFe-rich clumps
and for self-consistent explosion models. 
The velocity increase due to the combined action of the self-reflected reverse 
shock and the $\beta$ decay between $t\sim1\,$d ($t\sim10\,$d) and $t\sim1\,$yr 
for the BSGs (RSGs) is about $100\,$km/s ($150\,$km/s) (see 
Table\,\ref{tab_vel_final}). The enhanced $\beta$ decay in model 
B15$_\mathrm{X}$ leads to a much stronger acceleration of up to about 
$350\,$km/s. The gain in velocity is less than the $~\sim30\%$ increase 
found by \cite{Herant1992,Benz1994} or several hundred km/s by 
\cite{Basko1994}. However, their initial velocity distributions were based 
on spherical explosions lacking the initial explosion asymmetries 
which led to much lower maximal velocities than the one 
we obtain from the self-consistent explosion models.
Unfortunately, \cite{Orlando2019,Orlando2020} and \cite{Ono2020}, who 
started their explosion models for SN~1987A with aspherical but 
still parameterized perturbations, do not discuss in detail the 
effect of the $\beta$ decay on the velocity distribution or the inflation of 
$^{56}$Ni-rich structures.

Depending on the progenitor and on the explosion dynamics, the structures of 
the NiCoFeX-rich ejecta at shock breakout can be described by (i) many, 
almost isotropically distributed NiCoFeX-rich fingers in model B15, (ii) many 
pronounced fingers grouped together in some preferred directions in L15 and 
W15 or (iii) no particularly elongated structures in N20. These fingers or 
clumps are 
related to the initial asymmetries arising due to hydrodynamic instabilities 
during the shock revival phase and are fragmenting during the propagation 
through the progenitor 
\citep{Wongwathanarat2015}. After about $t\sim1\,$yr of evolution, many of the 
fine structures have merged back to fewer large-scale structures that 
resemble the initial asymmetries. We characterize these structures by means of a 
spherical harmonics decomposition and find that the slow ejecta of all models 
are dominated by a spherical component. For the faster ejecta, where pronounced 
NiCoFeX fingers are present, we find different morphologies. In model N20 
pronounced RT fingers are absent and consequently the monopole 
dominates the entire ejecta. 
Models L15 and W15 have some subdominant multipoles $l=\{2,3,4,6,8\}$ and 
$l=\{1,2\}$, respectively. These large-scale asymmetries have their origin in a 
small number of elongated NiCoFe-rich fingers. In contrast, model B15 has a 
plateau of multipoles of similar amplitude around $0< l\lesssim10$. Model L15 
has the highest power in higher multipole degrees $l\geq1$ for the averaged 
spherical harmonics. The large magnitude of 
these multipoles indicates that this model is the most asymmetric in our 
model sample (compare also to Fig.\,\ref{fig_iso_all}). In addition, with $\bar 
v\sim 4000\,$km/s, it also has the fastest-moving clumps of the highest-density 
clumps ($F_\rho=0.1$), while the densest clumps of other models only 
reach velocities up to $\bar v\lesssim 2000\,$km/s (see right column of panels 
in Fig.\,\ref{fig_mass_velvol}). Note that 
the morphology of the final structures is sensitive to the chosen threshold in 
mass fraction above which we consider a fluid element to belong to 
NiCoFeX-rich structures. For example, for model B15, when choosing to plot the 
$10$\% or $25$\% of the total NiCoFeX mass with the highest mass 
fractions only, the structures 
appear elongated into a particular direction, with a few additional clumps 
(see Fig.\,\ref{fig_iso_B15_fractions}). Only for lower mass fraction 
thresholds, corresponding to more than $50$\% of the total NiCoFe 
mass, more and more fingers or clumps appear which are more isotropically 
distributed. When comparing to observations this should be kept in mind.

With our analysis of the Mollweide projection of the 
$R^{56}_\mathrm{max}(\theta,\phi)$ (Section\,\ref{sec_moll}) and the spherical 
harmonic decomposition (Section\,\ref{sec_sphhar}), we further 
demonstrated how the initial asymmetries are correlated with the biggest 
structures at late times. In particular for models B15, L15, and W15, 
the fastest plumes 
shaped by the buoyant rise of neutrino-heated matter in the initial moments of 
the explosion end up as the biggest extended fingers or clumps at late times.
Since the conditions of model N20 for the growth of the RTI are less 
favourable compared to the other models, the initial asymmetries do not 
lead to large-scale clumps and one single central bubble dominates the ejecta.

In Table\,\ref{tab_clumps}, we provide quantitative data of the clumps 
of the different models, such as the number of clumps, and their volume and 
surface filling factors. Because the threshold above which density a clump is 
defined ($\rho_\mathrm{NiCoFeX}>\rho^\mathrm{min}_\mathrm{NiCoFeX}$) is 
somewhat arbitrary, we give the data for different choices of the fraction of 
the total NiCoFeX mass inside the clumps, $F_\rho$. As expected, the filling 
factors of the models decrease with increasing threshold density. The number of 
clumps follows the opposite trend for small 
$\rho^\mathrm{min}_\mathrm{NiCoFeX}$, but when the threshold density increases 
too much, 
more NiCoFeX-rich material has too low densities and some of the clumps 
disappear leading to an overall decrease in clump number. Among the models with 
standard $\beta$ decay, model L15 has the largest volume filling factors. The 
large factors are related to the fastest moving clumps of model L15 compared to 
all other considered models. In contrast, model B15 has the largest surface 
filling factors. Its NiCoFeX-rich fingers are distributed almost 
isotropically, while model L15 has fingers in some preferred directions. The 
 filling factors obtained for clumps containing $90\%$ of the NiCoFeX mass for 
the models with standard $\beta$ decay span ranges 
$V_\mathrm{NiCoFeX}^\mathrm{1500} 
=0.766\mathrm{\,(N20)}\dots2.534\mathrm{\,(L15)}$, 
$V_\mathrm{NiCoFeX}^\mathrm{2500}=0.163\mathrm{\,(N20)}\dots0.527\mathrm{\,(L15)
}$, and 
$V_\mathrm{NiCoFeX}^\mathrm{fastest}=0.028\mathrm{\,(W15)}\dots0.138\mathrm{\,
(N20) }
$. These values can be very different from e.g. \cite{Basko1994} who studied a 
single spherical bubble that expands due to $\beta$ decay. They found that 
due to the mixing NiCoFe-rich bubble material with ambient NiCoFe-poor matter, 
the fraction of the volume occupied by NiCoFe relative to the total volume of 
the bubble is $f_n=0.3\dots0.9$. We cannot compare these values straight 
forwardly to ours, because we don't have a well defined bubble surface which we 
can use as a reference. However, when assuming a radial velocity of the 
outermost shell of the spherical bubble between $1500\,$km/s 
and $2500\,$km/s our values are consistent with those of \cite{Basko1994}. From 
observations of SN~1987A, \cite{Li1993} estimated the filling factor 
$V_\mathrm{NiCoFeX}^\mathrm{2500}\gtrsim0.3$ for this SN. Assuming that most of 
the radioactive ejecta material was observed, we can compare to our results for 
$F_\rho=0.9$ in Table\,\ref{tab_clumps}: Models 
B15$_0$ and N20 are not compatible with this estimate, and model W15 is 
marginally compatible only. However, our estimates for L15, B15, and 
B15$_\mathrm{X}$ seem very reasonable
$V_\mathrm{NiCoFeX}^\mathrm{2500}=0.32\mathrm{(B15)}\dots0.64(\mathrm{B15}_X)$

As already pointed out by \cite{Blondin2001} and \cite{wang2005} for spherical 
shells, we find highly overdense material at the walls between neighbouring 
NiCoFeX-rich clumps/fingers. The density contrast between underdense, inflated 
clumps and overdense, compressed walls depends sensitively on the amount of 
initially synthesized $^{56}$Ni. For example for model B15 we obtain a factor of 
a few, while for B15$_\mathrm{X}$ the density in the clump borders can be 
several ten times the density inside the clump. Unfortunately, these numbers 
cannot directly be compared to the values in \cite{Blondin2001} and 
\cite{wang2005}, because they investigated the shell of a big central bubble, 
which we do not observe in our simulations due to the significant asymmetries 
arising during the explosion. 

With only two members of each class, we avoid a detailed assessment of 
differences between supernovae from RSG and BSG progenitors here and leave this 
task for future studies. However, one possibly generic feature revealed by of 
our analysis is the 
stronger density contrast of the iron-rich fingers and their surrounding 
material in BSG explosions compared to RSG explosions. Since  hydrogen envelopes 
of BSGs are more compact than those of RSGs (the radius of the former is roughly 
one tenth of that of the latter),
but the mass of synthesized $^{56}$Ni is comparable within a factor of two 
($0.034\dots0.56\,$M$_\odot$), the $\beta$-decay energy is deposited in a 
smaller volume. This higher energy per volume deposition leads to a stronger 
relative inflation of the NiCoFe-rich fingers/bubbles (see 
Fig.\,\ref{fig_expansion}). Thus the density contrast in BSG explsions is 
expected to be stronger than explosions of RSGs. With the same reasoning we 
also can explain the higher 
number of clumps obtained for the RSGs cases: due to the weaker inflation there 
is a 
weaker tendency of the transiently fragmented clumps to merge again and 
recombine to bigger structures.

A natural extension of this work is a more detailed study of 
differences between RSG and BSG explosions. In future studies, we will also
investigate how some of the described structures and morphological features 
may be connected to observations of known young supernova remnants like 
SN~1987A, Cassiopeia A, or the Crab Nebula. These investigations require 
longer-time simulations of the transition from the infant stage of the remnants 
to their fully developed state as, for example, recently conducted by 
\cite{Orlando2020b}. There the authors 
studied the gaseous ejecta of the explosion of a stripped progenitor to better 
understand the morphology of Cassiopeia A. In these kind of long-time 
simulations, one needs to consider how different physical 
effects like the faster cooling of the extended iron-rich ejecta material may 
lead to a slower expansion of the latter. Furthermore, one has to 
understand the effects of the interaction of the clumps with the reverse shock 
formed by the interaction with the circumstellar (CSM) or interstellar medium 
(ISM).

\section*{Acknowledgements}
We thank the anonymous referee for valuable questions and comments,
Anders Jerkstrand for fruitful discussions, 
Alexandra Gessner for providing the code version of 
{\protect\textsc{Prometheus-HotB}}, and Margarita Petkova and Alexey Krukau of 
the Computational Center for Particle and Astrophysics
(C2PAP) for developing a new parallel version of the code. MG 
acknowledges support through the Generalitat Valenciana via the grant 
CIDEGENT/2019/031. This research was supported by the Deutsche 
Forschungsgemeinschaft through  Excellence Cluster Universe (EXC 
153 http://www.universe-cluster.de/) and Sonderforschungsbereich SFB 
1258 ``Neutrinos and Dark Matter in Astro- and Particle Physics'', and by the 
European Research Council through grant ERC-AdG No. 341157-COCO2CASA. The 
computations were carried out on Hydra of the Max Planck Computing and Data 
Facility (MPCDF) Garching and on the Cluster of the Computational Center for 
Particle and Astrophysics (C2PAP) Garching at the Leibnitz Supercomputing Centre 
(LRZ) Garching. 

{\it Software}: Prometheus-HOTB \citep{Fryxell1991, Mueller1991b, 
Mueller1991, Kifonidis2003, Scheck2006, Arcones2007, Wongwathanarat2013, 
Wongwathanarat2015, Ertl2016}, Numpy and SciPy \citep{Jones2001}, IPython 
\citep{Perez2007}, Matplotlib \citep{Hunter2007}, VisIt \citep{Childs2012}.

{\it Data availability}: The data underlying this article will beshared on 
reasonable request to the corresponding author.
\appendix
\section{Energy deposition due to {\protect $\beta$}-decay}\label{app_beta}
\begin{figure}
\includegraphics[width=.46\textwidth]{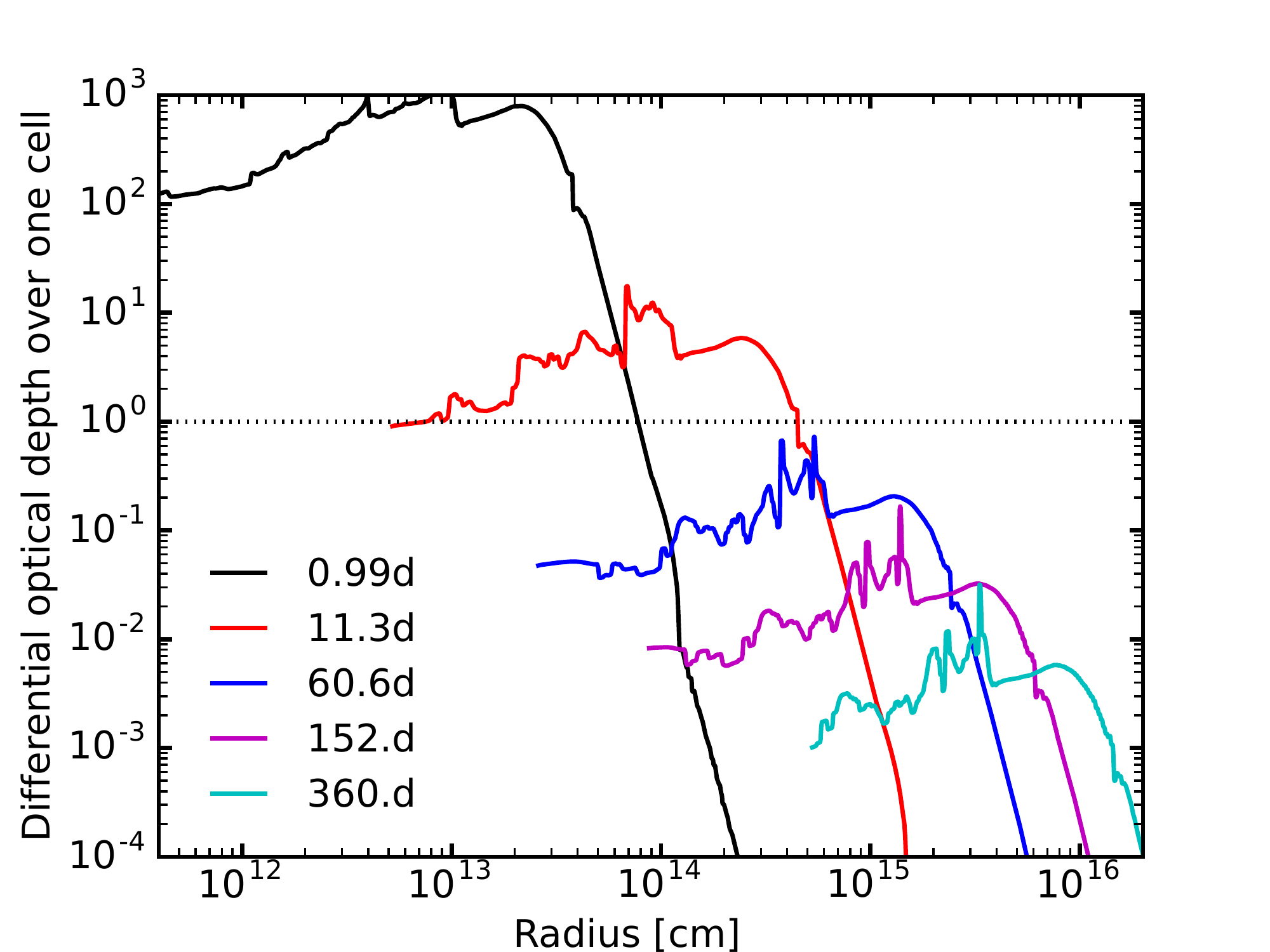}
\caption{ Differential optical depth of individual grid cells $\Delta 
\tau_\gamma$ of model B15 at different times for a randomly selected direction 
at $\theta=1.1$ and $\varphi=-2.0$. The dashed line indicates  
$\Delta\tau_\gamma=1$. Up to about $10\,$d the radiation is approximately 
trapped even within each cell of the numerical grid.}
\label{fig_local_opt_depth}
\end{figure}
\begin{figure}  
\includegraphics[width=.46\textwidth]{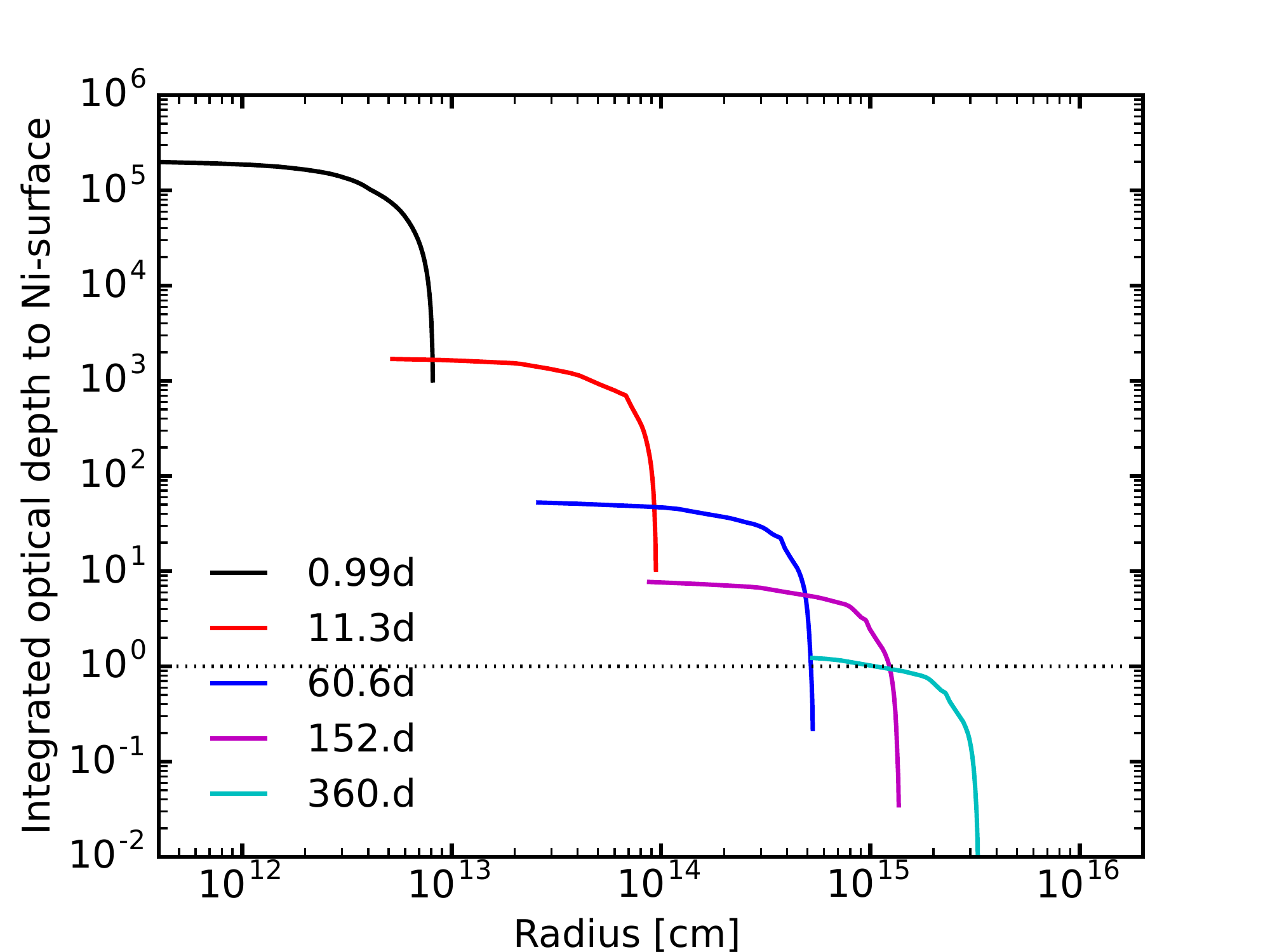}
\caption{ Radially integrated optical depth to the surface of the NiCoFe-rich 
ejecta $\tau_\gamma$ of model B15 at different times for a randomly selected 
direction at $\theta=1.1$ and $\varphi=-2.0$. The reference surface is defined 
by the outermost radii where the mass fraction of $X_\mathrm{NiCoFe}$ drops 
below $10^{-3}$. The dashed line 
indicates $\tau_\gamma=1$. Only later than $t\sim150\,$d, the optical 
depth to the NiCoFe surface drops below $\tau_\gamma<1.0$ for most of the 
material.}
\label{fig_int_opt_depth}
\end{figure}
In this Appendix, we test our implementation of the (local) energy deposition 
due to $\beta$ decay. As described in Section \ref{sec_beta}, we deposit only 
a fraction of the decay energy locally. This energy fraction depends on the 
interaction probability of the radiation with the ejecta, which is determined 
by the optical depth of the photons. The corresponding differential optical 
depth of individual numerical cells $\Delta \tau_\gamma$ is a measure for the 
radial optical depth of the local structures we can resolve within our 
simulations. $\Delta \tau_\gamma$ is plotted for model B15 and for a randomly 
chosen 
direction in Fig.\,\ref{fig_local_opt_depth}. Up to about $t\gtrsim10\,$d, this 
{\it local} 
optical depth is larger than one, meaning that until this time most energy is 
deposited locally inside the corresponding numerical cell. In 
Fig.\,\ref{fig_int_opt_depth}, we show the radially integrated optical depth up 
to the NiCoFe surface, which is defined as the outermost radial location where 
the mass fraction of NiCoFe $X_\mathrm{NiCoFe}$ exceeds $10^{-3}$. Up to 
$t\sim150\,$d, the optical 
depth $\tau_\gamma$ of most of the material is still significantly larger than 
$1$, and only after that time, the bulk of the material becomes transparent to 
the released $\gamma$ rays. Note that we always consider local deposition of  
the energy of the positron emitted during the $^{56}$Co decay. 

\begin{figure}  
\includegraphics[width=.46\textwidth]{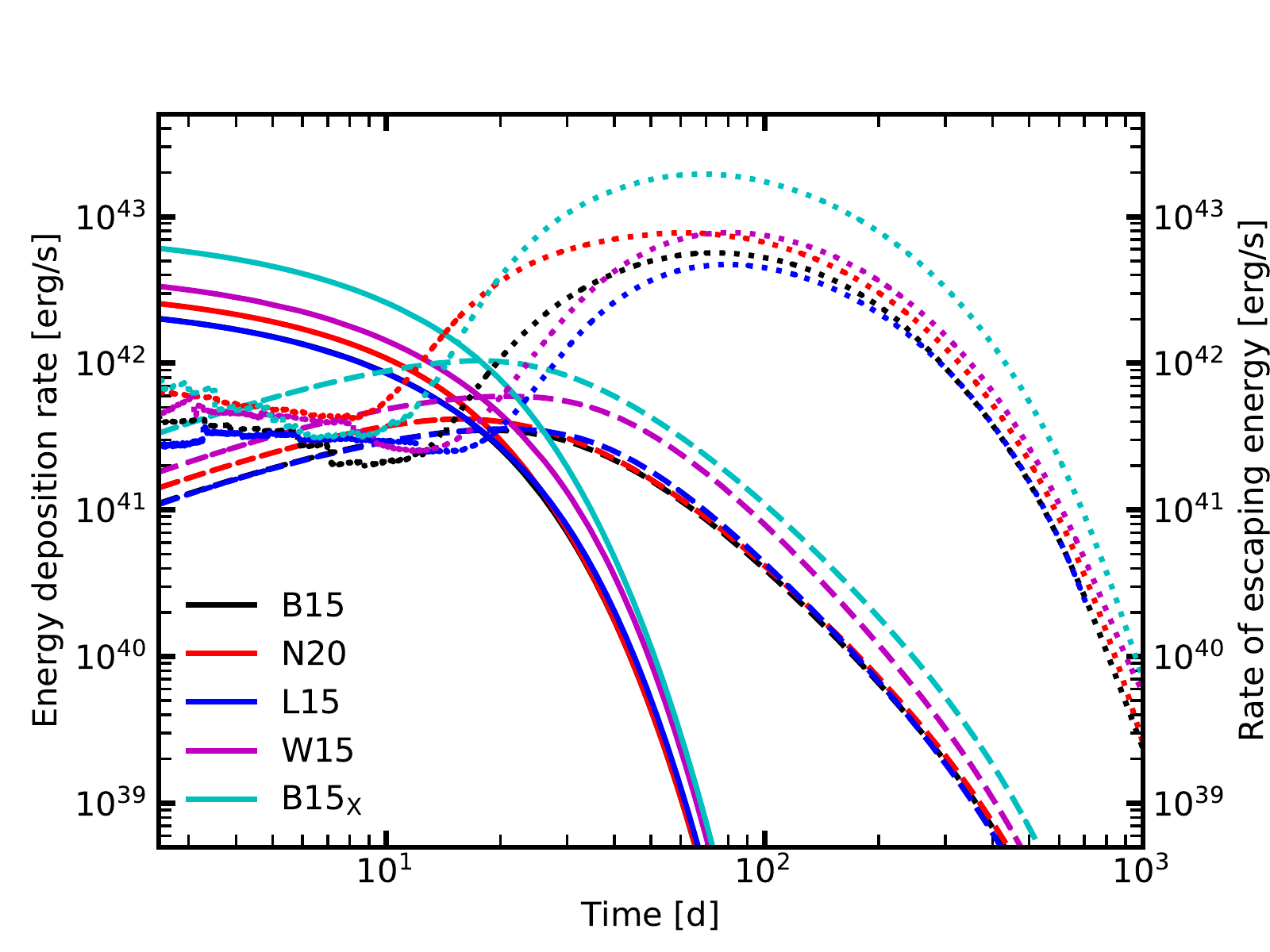}
\caption{Volume integrated local $\beta$-decay energy deposition rates from 
$^{56}$Ni (solid lines) and from $^{56}$Co (dashed lines) compared to the rate 
of the energy which escapes from the NiCoFe-rich regions (dotted lines) for 
different models. After a few tens of days, more energy escapes from the 
NiCoFe-rich regions and around $t\sim200\,$d the energy deposition rate inside 
the NiCoFe-rich volume drops to about one percent of the initial rate. Note 
that part of the escaping energy is deposited in the volume close to the 
NiCoFe-rich clumps.}
\label{fig_beta_losses}
\end{figure}
In Fig.\,\ref{fig_beta_losses}, we show the total energy deposition rates of 
the decaying $^{56}$Ni (solid lines) and $^{56}$Co (dashed lines) which is 
deposited locally for our different models. The dotted lines give the loss rate 
of the energy that depending on $\tau_\gamma^\mathrm{mean}$ is not deposited 
locally, but instead homogeneously or leaves the NiCoFe-rich region completely.

The total energy per second $dQ^\text{tot} / dt$ released due to the radioactive 
decay of $^{56}$Ni and $^{56}$Co is:
\begin{eqnarray}
\frac{dQ_\text{Ni}^\text{tot}}{dt}&=&-\frac{\ln{2}}{\tau^\text{Ni}_{1/2}} 
N_\text{Ni} (t) Q_\text{Ni}\label{eq_Qdot_Ni}\,\\
\frac{dQ_\text{Co}^\text{tot}}{dt}&=&-\frac{\ln{2}}{\tau^\text{Co}_{1/2}}
N_\text{Co}(t) 
Q_\text{Co}\label{eq_Qdot_Co}\,,
\end{eqnarray}
where
\begin{eqnarray}
 N_\text{Ni}(t)&=&N_\text{Ni}^0 \exp\left({-\frac{t \ln{2}}{ 
\tau^\text{Ni}_{1/2}}}\right) \,,\\
 N_\text{Co}(t)&=&N_\text{Ni}^0 \frac{\tau^\text{Co}_{1/2}}{\tau^
\text{Ni}_{1/2}-\tau^\text{Co}_{1/2}} \times \nonumber\\
&& \left[\exp\left(-\frac{t \ln{2}}{ \tau^\text{Ni}_{1/2}}\right) - 
\exp\left(-\frac{t \ln{2}}{ \tau^\text{Co}_{1/2}}\right) \right]\,.
 \end{eqnarray}

Initially, only a small fraction of the $\beta$-decay energy, which is emitted 
interior to the NiCoFe-surface, leaves the volume enclosed by this surface and 
most of the energy is deposited in the ejecta rich in NiCoFe. After about a few 
tens of days, $^{56}$Co decay liberates more energy than $^{56}$Ni decay, and 
the losses to the surroundings of the NiCoFe-rich ejecta exceed the locally 
deposited energy. At about $200\,$d the combined local deposition rate of 
$^{56}$Ni and $^{56}$Co is only one percent of the initial one.

\begin{table*}
\begin{tabular}{c c c c c c}
Model& 
$E^\mathrm{tot}_\mathrm{Ni}=\int \Delta\epsilon_\text{Ni}dt$&
$Q^\mathrm{tot}_\mathrm{Ni}=\int-\frac{dQ^\text{tot}_\text{Ni}}{dt} dt$&
$E^\mathrm{tot}_\mathrm{Co}=\int \Delta\epsilon_\text{Co} dt$&
$Q^\mathrm{tot}_\mathrm{Co}=\int -\frac{dQ^\text{tot}_\text{Co}}{dt} dt$ &
$(E^\mathrm{tot}_\mathrm{Ni}+E^\mathrm{tot}_\mathrm{Co})\times(Q^\mathrm{tot}
_\mathrm{ Ni} +Q^\mathrm{tot}_\mathrm{Co})^{-1}$\\
&[$10^{48}$erg]&[$10^{48}$erg]&[$10^{48}$erg]&[$10^{48}$erg]\\\hline
B15 & 1.99& 2.02& 1.66& 4.40&  0.57\\
N20 & 2.47& 2.59& 1.83& 5.63&  0.52\\
L15 & 1.51& 1.54& 1.78& 4.40&  0.55\\
W15 & 3.31& 3.30& 3.10& 7.21&  0.61\\
B15$_\text{X}$ & 5.95& 6.07& 4.71& 13.2&  0.55\\
\end{tabular}
\caption{Integrated energies $E^\mathrm{tot}_\mathrm{Ni}$ (second column) and 
$E^\mathrm{tot}_\mathrm{Co}$ (fourth column) absorbed directly at 
the $\beta$-decay production sites. These values are compared to the 
theoretically expected total energy budget $Q^\mathrm{tot}_\mathrm{Ni}$ (third 
column) and $Q^\mathrm{tot}_\mathrm{Co}$ (fifth column) from
integrating Eqs.\,\ref{eq_Qdot_Ni}  and \ref{eq_Qdot_Co}, respectively. The 
last column gives the fraction of the total decay 
energy deposited directly at the production sites in the 
NiCoFeX-rich ejecta compared to the total energy budget of the respective 
decay.}
\label{tab_decay_energies}
\end{table*}

As can be seen also in Fig.\ref{fig_beta_losses}, almost the entire decay 
energy from $^{56}$Ni is deposited locally. We give the integrated energies 
deposited locally and produced during the different decays in 
Table\,\ref{tab_decay_energies} for the $^{56}$Ni and $^{56}$Co decays for all 
our models. Since the maximum of the $^{56}$Co-decay occurs at times when the 
matter becomes transparent to $\gamma$ rays, only roughly $30-40\%$ of the 
energy 
of this decay is absorbed by the ejecta locally at the $\beta$-decay sites. 
This means in total, depending on the model $52-61\%$ 
of the total available energy budget of the $^{56}$Ni-decay chain is transformed 
into the internal energy of the NiCoFe-rich ejecta.
\bibliographystyle{mnras}
\bibliography{SN}
\end{document}